\documentclass[%prl,
aps,pra,
 ,twocolumn
]{revtex4}
\usepackage{graphicx}
\usepackage{subfigure}
\usepackage{amsmath}
\usepackage{amssymb}
\usepackage{amsthm}
\usepackage{color}
\usepackage{microtype}
\usepackage{tikz}
\usepackage{makecell}

\usepackage{hyperref}
\hypersetup{colorlinks,linkcolor={blue},citecolor={red},urlcolor={blue}}

\newcommand{\op}[1]{{\hat{#1}}}

\newcommand{\ket}[1]{| #1 \rangle}
\newcommand{\bra}[1]{\langle #1 |}

\newcommand{\vt}[1]{{\boldsymbol{#1}}}
\theoremstyle{plain}

\theoremstyle{plain}

\begin{document}

\title{Universal quantum computation with optical four-component cat qubits}

\author{Daiqin Su$^{1}$}
\author{Ish Dhand$^{1, 2}$}
\author{Timothy C. Ralph$^{3}$}

\affiliation{$^{1}$Xanadu, Toronto, Ontario, M5G 2C8, Canada}
\affiliation{$^{2}$Institut f{\"u}r Theoretische Physik, Albert-Einstein-Allee 11, Universit{\"a}t Ulm, 89069 Ulm, Germany}
\affiliation{$^{3}$Centre for Quantum Computation and Communication Technology, School of Mathematics and Physics, The University of Queensland, St. Lucia, QLD 4072, Australia}

\date{\today}

%%\tableofcontents
\begin{abstract}
We propose a teleportation-based scheme to implement a universal set of quantum gates with a four-component cat code, assisted by appropriate entangled resource states and 
photon number resolving detection. The four-component cat code features the ability to recover from single photon loss.
Here, we propose a concrete procedure to correct the single photon loss, including detecting the single photon loss event and recovering the initial states. By concatenating with
standard qubit error correcting codes, we estimate the loss threshold for fault-tolerant quantum computation and obtain a significant improvement over the 
two-component cat code. 
\end{abstract}

\maketitle

{\it Introduction.}--Quantum computers promise to solve some problems faster than classical computers by utilizing 
quantum entanglement and coherence~\cite{365700, 10.1145/237814.237866, nielsen2002quantum, ladd2010quantum}.
However, noise can easily destroy the entanglement and coherence of a quantum system, 
dramatically degrading the performance of a quantum computer~\cite{PhysRevA.51.992}. 
The standard procedure to deal with noise is to introduce quantum error correcting codes, in particular, to encode quantum information in the subspace of a 
multi-qubit system and correct the errors by using the redundant subspace
~\cite{PhysRevA.52.R2493, PhysRevLett.77.793, PhysRevA.54.3824, PhysRevLett.77.198, PhysRevLett.77.2585, PhysRevA.55.900, KITAEV20032, PhysRevA.86.032324}. 
An alternative is to encode the quantum information in an
infinite-dimensional bosonic system, e.g., a single harmonic oscillator. These sorts of codes are called bosonic codes. Examples include the Gottesman-Kitaev-Preskill (GKP) codes~\cite{PhysRevA.64.012310}, cat codes~\cite{PhysRevA.59.2631, PhysRevA.68.042319, PhysRevLett.111.120501, Mirrahimi_2014}, 
binomial codes~\cite{PhysRevX.6.031006} and rotation-symmetric codes~\cite{PhysRevX.10.011058}. 

A bosonic code is usually introduced to correct a specific kind of error, e.g., the GKP code corrects random phase and amplitude displacements~\cite{PhysRevA.64.012310}. 
However, a bosonic code itself is not sufficient for fault-tolerant quantum computation. The strategy is to concatenate the bosonic codes with qubit codes, which potentially 
reduces the resource overhead. It has been shown that a finite-squeezing GKP code can allow for fault tolerance~\cite{PhysRevLett.112.120504}, 
and the squeezing threshold can be reduced to around 10 dB~\cite{PhysRevX.8.021054, PhysRevA.101.012316, PhysRevA.99.032344}. 
Another important bosonic code is the two-component cat code, which can protect itself from phase rotation error but not photon loss, 
leading to a biased error model. The photon loss can be corrected by
concatenating the cat code with some simple qubit codes, e.g., repetition code, giving more feasible protocols for fault-tolerant quantum computing~\cite{PhysRevX.9.041053, puri2020bias}. 
In optical platforms where gates 
are probabilistic, additional located erasure errors need to be addressed~\cite{PhysRevLett.100.030503}, which can also be handled by concatenating with qubit codes. 
Along this line, the photon loss threshold for the two-component optical cat codes has been estimated~\cite{PhysRevLett.100.030503, Lee2013near}. 

Another way to mitigate the photon loss is to introduce a cat code with more components~\cite{bergmann2016quantum}, e.g., the four-component cat code~\cite{Mirrahimi_2014}. The four-component cat code is introduced to 
correct the single photon loss and its performance has been studied in Ref.~\cite{PhysRevA.97.032346}. 
The preparation and manipulation of the four-component cat code have been proposed for platforms
with strong nonlinearity~\cite{Mirrahimi_2014}. However for optical platforms, techneques to implement the universal set of gates and to correct the single photon loss have not 
been explored. 

In this work we propose a scheme to implement the universal set of gates for the four-component cat code via teleportation, with appropriate entangled resource states and 
photon-number-resolving (PNR) detection. This is particularly appealing due to the rapid development of PNR detectors in the optical platform~\cite{PhysRevA.71.061803, PhysRevA.99.043822, morais2020precisely}. 
The advantage is twofold: the PNR detectors not only can be used for teleportation, but also can be used to generate the required entangled resource states~\cite{PhysRevA.100.012326, PhysRevA.100.052301, magana2019multiphoton}. We remark that this scheme is 
also promising for other platforms provided PNR detection and entangled resource states are available. Also based on teleportation, we then develop a 
concrete procedure to correct the single photon loss, which accomplishes the syndrome measurement and recovery operation simultaneously. 
By further concatenating with qubit codes, we estimate the 
photon loss threshold for fault-tolerant quantum computation.

{\it Encoding.}--In the four-component cat code, the state of a qubit is encoded in the superposition of four coherent states $\ket{\pm \alpha}$ and $\ket{\pm i \alpha}$, 
where $\alpha$ is assumed to be real. 
Specifically, the encoded Pauli $Z$ eigenstates are defined as $\ket{0_L} = \ket{\mathcal{C}^{+}_{\alpha}} = \mathcal{N}_+ (\ket{\alpha} + \ket{- \alpha})$ 
and $\ket{1_L} = \ket{\mathcal{C}^{+}_{i \alpha}} = \mathcal{N}_+ (\ket{i \alpha} + \ket{- i \alpha})$, where 
$\mathcal{N}_+ = 1/\sqrt{2(1 + e^{-2 \alpha^2})}$ is the normalization factor. Note that $\ket{0_L}$ and $\ket{1_L}$ are not orthogonal, 
$\langle 0_L \ket{1_L} = \langle \mathcal{C}^{+}_{\alpha} \ket{\mathcal{C}^{+}_{i \alpha}} = \cos \alpha^2/ \cosh \alpha^2$,
however the overlap is small for large $\alpha$. Since $\ket{\mathcal{C}^{+}_{\alpha}}$ and $\ket{\mathcal{C}^{+}_{i \alpha}}$ are related by a %$\frac{\pi}{2}$ 
$\pi/2$ phase rotation, 
the encoded Pauli $X$ operation can thus be implemented deterministically. This is similar to the situation for the two-component cat code where the logical
states $\ket{\alpha}$ and $\ket{-\alpha}$ are non-orthogonal and are related by a $\pi$ phase rotation. 

{\it Teleportation and universal set of gates.}--We choose the universal set of quantum gates as Pauli $X$, arbitrary rotation about $Z$, Hadamard gate and 
controlled-$Z$ gate. The Pauli $X$ gate can be implemented deterministically while other gates are implemented probabilistically 
via gate teleportation. Before studying the gate teleportation, we first consider the state teleportation which simply propagates the input 
state to the output, and can be considered as an identity gate or a memory. 

\begin{figure}
\includegraphics[width=0.85 \columnwidth]{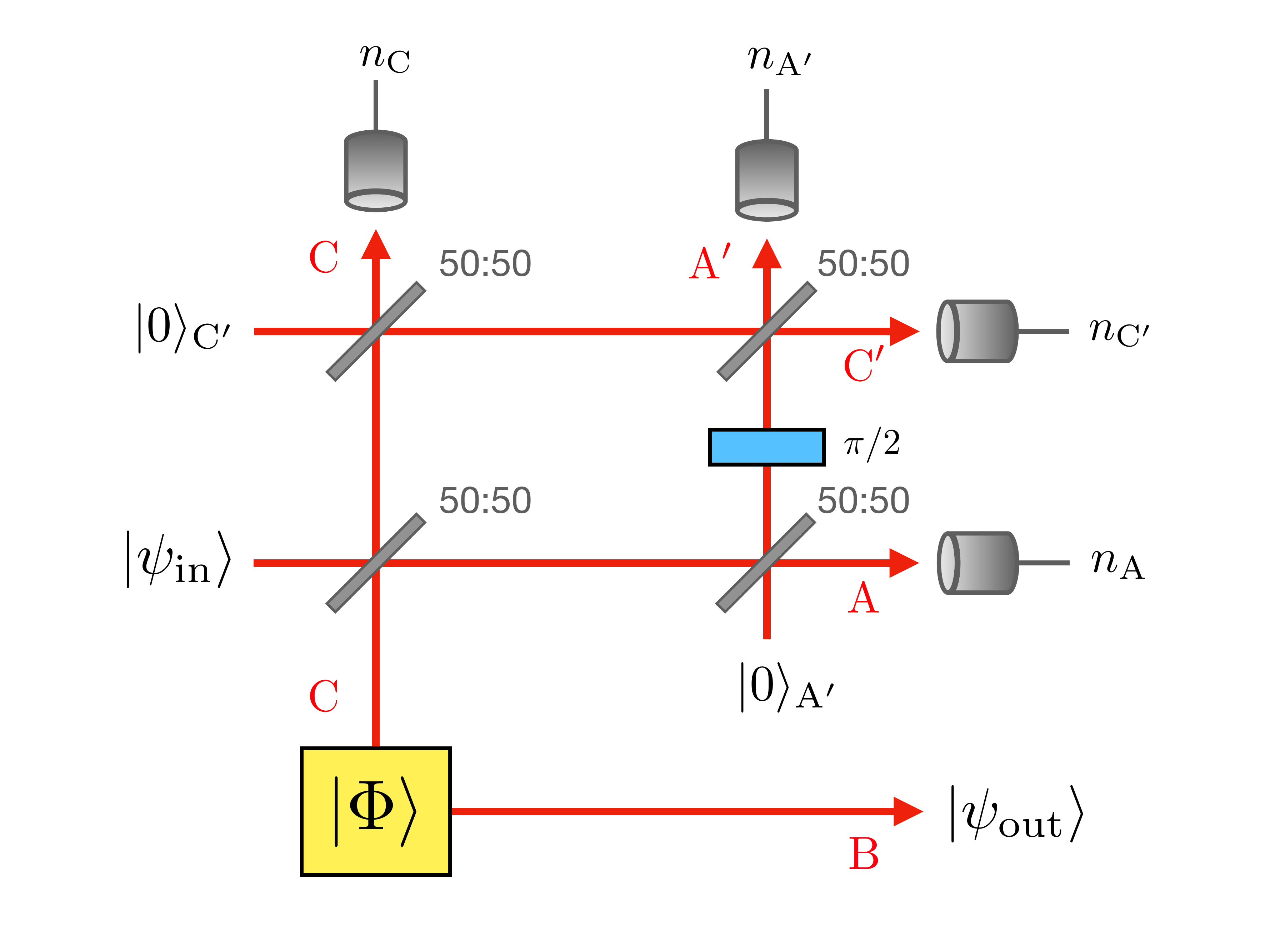}
\caption{ A schematic of the four-port teleporter, which consists of four 50:50 beam splitters, one $\pi/2$ phase shifter, four photon-number-resolving detectors
and an entangled resource state $\ket{\Phi}$. The entangled resource state has to be appropriately chosen depending on the gate to implement. }
\label{fig:4-port-TP}
\end{figure}

The state teleportation requires an entangled resource state and the Bell measurement. Here we choose the entangled resource state as the encoded Bell state, 
\begin{eqnarray}\label{eq:bell}
\ket{\Phi_0} = \mathcal{N} \big(\ket{0_L} \ket{0_L} + \ket{1_L} \ket{1_L} \big),
\end{eqnarray}
where $\mathcal{N} = \cosh \alpha^2/\sqrt{2(\cosh^2 \alpha^2 + \cos^2 \alpha^2)}$ is the normalization factor. The Bell measurement is performed by appropriately
arranging four $50:50$ beam splitters, one phase shifter and four PNR detectors, as shown in Fig.~\ref{fig:4-port-TP}. A similar circuit has been used to amplify
coherent states and teleport four-component cat states~\cite{Neergaard-Nielsen_2013}. 
The novelty of our circuit is that we use PNR detectors and a different entangled resource state, which are essential for implementing quantum gates and correcting single 
photon loss. 

\begin{widetext}

\begin{table}[h]
\caption{ Measurement outcomes of four PNR detectors and the corresponding Pauli corrections required in order to recover the input state in state teleportation. 
 It is explicitly indicated if a detector detects no photons, otherwise the detector detects at least one photon. Here $N = n_A + n_C + n_{A'} + n_{C'}$ is the total number of 
 detected photon and $k$ is a positive integer. } %title of the table
\centering % centering table
\begin{tabular}{@{}c|c|c|c|c}%\toprule
\hline \hline
  {\bf No correction} & {\bf $Z$ correction}  &  {\bf $X$ correction}  & {\bf $ZX$ correction} & {\bf Failure} \\
    \hline
\makecell{ $[0, n_C, n_{A'}, n_{C'}]$ \\ $[n_A, 0, n_{A'}, n_{C'}]$ \\ \\ $[0, 0, n_{A'}, n_{C'}]$ \\ $[n_A, n_C, 0, 0]$ \\ \\ $( N = 4k )$}
& \makecell{ $[0, n_C, n_{A'}, n_{C'}]$ \\ $[n_A, 0, n_{A'}, n_{C'}]$ \\ \\ $[0, 0, n_{A'}, n_{C'}]$ \\ $[n_A, n_C, 0, 0]$ \\ \\  $[n_A, 0, 0, 0]$ \\ $[0, n_C, 0, 0]$ \\ \\ $( N = 4k-2 )$}  
& \makecell{ $[n_A, n_C, 0, n_{C'}]$ \\ $[n_A, n_C, n_{A'}, 0]$ \\ \\ $( N = 4k )$}
& \makecell{ $[n_A, n_C, 0, n_{C'}]$ \\ $[n_A, n_C, n_{A'}, 0]$ \\ \\ $[0, 0, n_{A'}, 0]$ \\ $[0, 0, 0, n_{C'}]$ \\ \\ $( N = 4k-2 )$}
& \makecell{$[n_A, 0, n_{A'}, 0]$ \\ $[n_A, 0, 0, n_{C'}]$ \\  $[0, n_C, n_{A'}, 0]$ \\ $[0, n_C, 0, n_{C'}]$ \\ \\ $( N = 2k )$}  ~ \makecell{ $[4k, 0, 0, 0]$ \\ $[0, 4k, 0, 0]$ \\ $[0, 0, 4k, 0]$ \\$[0, 0, 0, 4k]$ \\ \\ $[0, 0, 0, 0]$} \\
\hline \hline
\end{tabular}
\label{tab:ClickPatterns}
\end{table}

\end{widetext}

The four PNR detectors count the photon number at each output, giving click patterns $[n_A, n_C, n_{A'}, n_{C'}]$. If all components in the circuit are ideal, i.e., no photon loss,
then the total number of detected photons $N = n_A + n_C + n_{A'} + n_{C'}$ is even. This is because both the encoded states and the entangled resource state consist of even numbers
of photons. The click patterns are divided into two categories, one for success of teleportation and the other for failure of teleportation. When the teleportation was successful, one 
needs to perform appropriate encoded Pauli corrections, depending on the click patterns, to recover the input state. These are summarized in Table~\ref{tab:ClickPatterns}. 

To implement the gates via teleportation, the entangled resource states need to be modified accordingly. For rotation about the $Z$ direction, one needs
an entangled state $\ket{\Phi_R} \sim e^{i \theta/2} \ket{0_L} \ket{0_L} + e^{- i \theta/2} \ket{1_L} \ket{1_L} $ with $\theta$ the rotation angle; 
for the Hadamard gate, one needs an entangled state
$\ket{\Phi_H} \sim \ket{0_L} \ket{0_L} + \ket{0_L} \ket{1_L} + \ket{1_L} \ket{0_L} - \ket{1_L} \ket{1_L} $; and for the controlled-$Z$ gate, one needs a four-qubit entangled state 
$\ket{\Phi_{CZ}} \sim \ket{0_L} \ket{0_L} \ket{0_L} \ket{0_L} + \ket{0_L} \ket{0_L} \ket{1_L} \ket{1_L} + \ket{1_L} \ket{1_L} \ket{0_L} \ket{0_L} - \ket{1_L} \ket{1_L}\ket{1_L} \ket{1_L}$,
which is used as the shared entanglement of two teleportation circuits. Depending on the gate to implement, the Pauli correction needs to be adjusted accordingly, for example, for the 
Hadamard gate the $Z$ correction is converted to $X$ correction and the $X$ correction is converted to $Z$ correction~\cite{PhysRevLett.100.030503}.

{\it Generating entangled resource states.}--We now consider a strategy for creating the  entangled resource state required for the teleportation. We take the single mode encoded superposition state, $\ket{+_L} \sim \ket{0_L} + \ket{1_L}$, with amplitude $\sqrt{2} \alpha$ as our ``free" resource state and seek a non-deterministic, heralded, linear optics method for producing the teleportation resource state given in Eq.~\eqref{eq:bell}.

\begin{figure}
\includegraphics[width=0.75 \columnwidth]{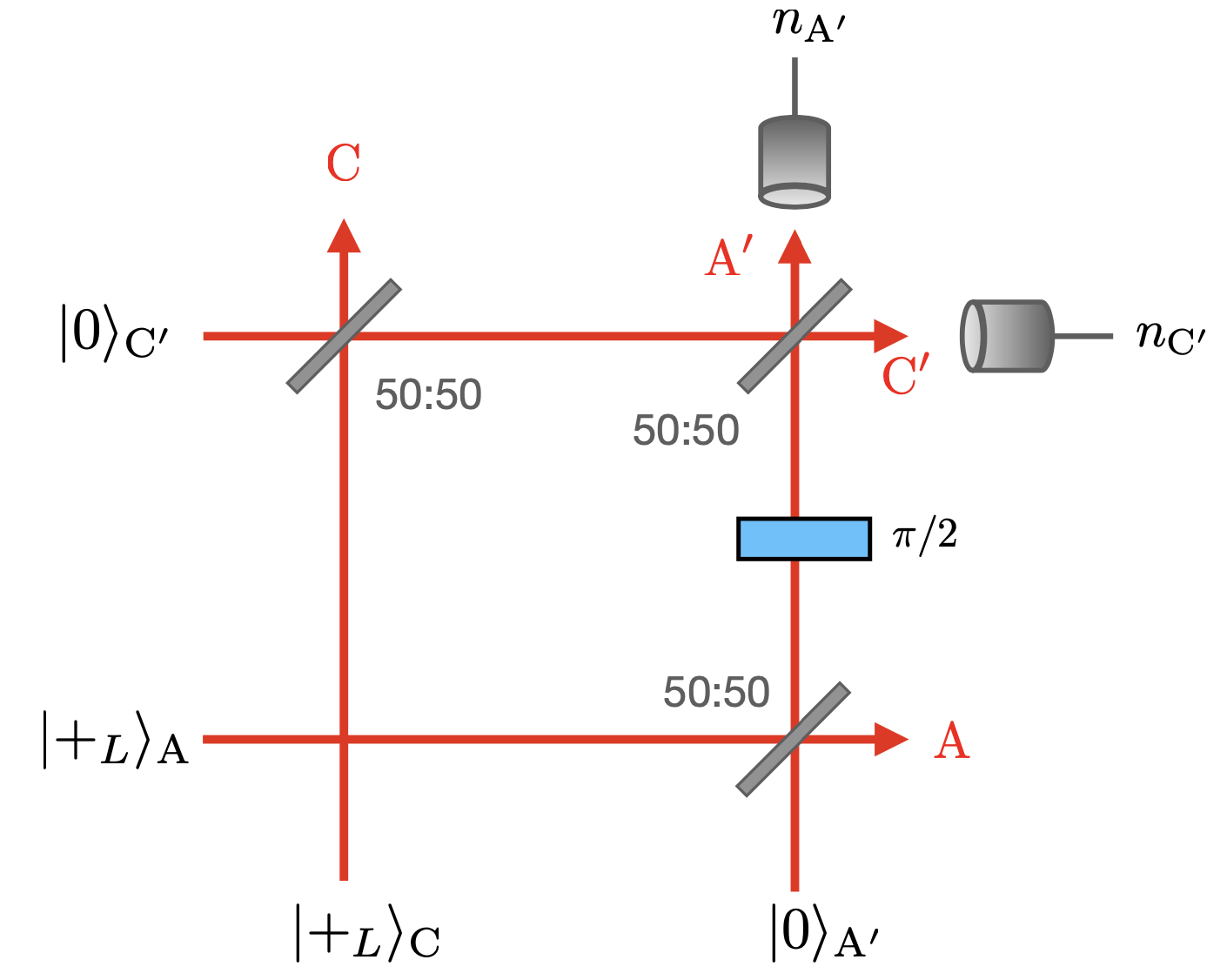}
\caption{ An optical circuit to generate the entangled resource state $\ket{\Phi_0}$. The states $\ket{+_L}_A$ and $\ket{+_L}_C$ are encoded superposition states with amplitude $\sqrt{2} \alpha$. }
\label{fig:resoure-state}
\end{figure}

Consider an optical circuit depicted in Fig.~\ref{fig:resoure-state} which mixes two superposition states and two vacuum inputs together on a series of 50:50 beam splitters. 
When we obtain equal counts on the two detectors, $n_{{\mathrm A}^{\prime}} = n_{{\mathrm C}^{\prime}}$, 
and $n_{{\mathrm A}^{\prime}}$ is even then the projected state is the teleportation resource state given in Eq.~\eqref{eq:bell}. If $n_{{\mathrm A}^{\prime}} = n_{{\mathrm C}^{\prime}}$, 
but $n_{{\mathrm A}^{\prime}}$ is odd we obtain the phase flipped Bell state
$\mathcal{N} \big(\ket{0_L} \ket{0_L} - \ket{1_L} \ket{1_L} \big)$, which is still useful for teleportation, just requiring a different Pauli operator correction.

The entangled resource states used to implement the universal set of gates for the four-component cat code are non-Gaussian states. For optical platforms, these non-Gaussian 
states can be probabilistically generated by measuring multimode Gaussian states using PNR detectors and post selecting certain measurement outcomes~\cite{PhysRevA.100.012326, PhysRevA.100.052301, magana2019multiphoton}. For platforms where
strong nonlinearity is available, these non-Gaussian states can be generated deterministically.

{\it Correcting single photon loss.}--The standard error correction procedure consists of two steps: the first is to detect whether an error has occurred without disturbing the state, 
known as the syndrome measurement; the second is to recover the original state according to the results of the syndrome measurement, known as the recovery operation.
The performance of the four-component cat code in correcting single photon loss has been explored in Ref.~\cite{PhysRevA.97.032346}, however 
a concrete error correcting procedure is still absence. 
Here we propose a concrete procedure to correct single photon loss via teleportation, which realizes the syndrome measurement and recovery operation simultaneously. 
 
The four-component cat code is designed to correct single photon loss using the redundancy in the Fock basis states. 
The states of a qubit are encoded in the subspace with even numbers of photons.
If there is single photon loss, the states are shifted to the subspace with odd numbers of photons. Therefore, a syndrome measurement is basically a parity measurement. 
When detecting an even number of photon, one concludes that no photons have been lost and no further operation is needed; 
when detecting an odd number of photon, one concludes that a photon has been lost and a photon needs to be added to the output state to recover the original state. 
If more than one photons have been lost, the above procedure results in uncorrectable errors. 

Define two new cat states $\ket{\mathcal{C}^{-}_{\alpha} } = \mathcal{N}_- (\ket{\alpha} - \ket{- \alpha})$ and 
$\ket{\mathcal{C}^{-}_{i \alpha} } = \mathcal{N}_- (\ket{i \alpha} - \ket{- i \alpha})$, with $\mathcal{N}_- = 1/\sqrt{2(1 - e^{-2 \alpha^2})}$ the normalization factor. 
The states $\ket{\mathcal{C}^{-}_{\alpha} }$  and $\ket{\mathcal{C}^{-}_{i \alpha} }$ consist of odd numbers of photons, therefore are not in the code subspace.
We refer to them as being in the loss subspace.  
Together with $\ket{\mathcal{C}^{+}_{\alpha} }$ and $\ket{\mathcal{C}^{+}_{i \alpha} }$, one can form four states that are superposition of four coherent states:
$\ket{\psi_{\alpha}^{(0)}} = c_0 \ket{\mathcal{C}^{+}_{\alpha}} + c_1 \ket{\mathcal{C}^{+}_{i \alpha}}$, 
$\ket{\psi_{\alpha}^{(1)}} = c_0 \ket{\mathcal{C}^{-}_{\alpha}} + i c_1 \ket{\mathcal{C}^{-}_{i \alpha}}$,
$\ket{\psi_{\alpha}^{(2)}} = c_0 \ket{\mathcal{C}^{+}_{\alpha}} - c_1 \ket{\mathcal{C}^{+}_{i \alpha}}$, and 
$\ket{\psi_{\alpha}^{(3)}} = c_0 \ket{\mathcal{C}^{-}_{\alpha}} - i c_1 \ket{\mathcal{C}^{-}_{i \alpha}}$, where $c_0$ and $c_1$ are complex coefficients. 
Note that $\ket{\psi_{\alpha}^{(0)}}$ and $\ket{\psi_{\alpha}^{(2)}}$ are in the code subspace and are related by a Pauli $Z$ transformation, while 
$\ket{\psi_{\alpha}^{(1)}}$ and $\ket{\psi_{\alpha}^{(3)}}$ are not in the code subspace. When a photon is lost, these states are transformed as
$\ket{\psi_{\alpha}^{(k)}} \rightarrow  \ket{\psi_{\alpha}^{[(k+1) \, {\rm mod} \, 4]}}$, up to a normalization constant. 

Consider an encoded cat qubit with initial state $\ket{\psi_{\alpha}^{(0)}}$ going through a lossy channel which is modelled by a beam splitter with transmission $\eta = 1 - \epsilon$. If there is no photon loss, the state becomes $\ket{\psi_{\alpha'}^{(0)}}$ with $\alpha' = \sqrt{\eta} \alpha$; if there is single photon loss, the state becomes $\ket{\psi_{\alpha'}^{(1)}}$ (see Appendix for details). The state after the lossy channel is then fed into the teleportation circuit shown in Fig.~\ref{fig:4-port-TP}, namely, it acts as the input state. Note that the amplitude of the cat code decreases from $\alpha$ to $\sqrt{\eta} \alpha$ regardless of whether a photon is lost or not, therefore the amplitude of the entangled resource state has to be adjusted such that the amplitude of arm $C$ is $\alpha'$, whilst that of arm $B$ remains $\alpha$. In the case where there is no photon loss, the state is simply teleported to the output and the total number of detected photon is even. The teleportation corrects the amplitude: $\alpha' \rightarrow \alpha$. In the case where a photon is lost, the total number of detected photon is odd. This is because the input couples with one of the modes of the encoded Bell state Eq.~\eqref{eq:bell}, which consists of even numbers of photon. Therefore, the parity of the total number of detected photon tells us whether there is single photon loss or not when the encoded cat qubit goes through the lossy channel. This accomplishes the syndrome measurement. 

Interestingly, the recovery operation can also be implemented via teleportation. From Eq.~\eqref{eq:bell} and Fig.~\ref{fig:4-port-TP} it is evident that the output state of the teleportation (in the other mode of the encoded Bell state) consists of even numbers of photon and is in the code subspace. In order to recover the initial state of the encoded cat qubit, one only needs to apply appropriate encoded unitary transformations. In the case of state teleportation and no photon loss, these required unitary transformations are encoded Pauli operators, as shown in Table~\ref{tab:ClickPatterns}. We find that one can still recover the encoded initial state by applying appropriate encoded Pauli operators when there is single photon loss in the encoded cat qubit. The encoded Pauli operator that needs to be applied depends on the click pattern and the total number of detected photon (see Supplementary Material for details), as summarized in Table~\ref{tab:RecoveryLoss}. The Pauli correction in the teleportation thus plays the role as recovery operation.

\begin{widetext}

\begin{table}[h]
\caption{ {\bf Pauli corrections (recovery operations) with single photon loss}. The measurement patterns that allow for initial state recovery are listed. Here $I$ is identity operator and $X, Y, Z$ are three encoded Pauli operators, $N$ is the total number of detected photon and $k$ is an integer.  } %title of the table
\centering % centering table
\begin{tabular}{@{} l |c|c|c|c|c|c}%\toprule
\hline \hline
   & ~~$[0, n_C, n_{A'}, n_{C'}]$~~ & ~~$[n_A, 0, n_{A'}, n_{C'}]$~~  & ~~$[n_A, n_C, 0, n_{C'}]$~~ & ~~$[n_A, n_C, n_{A'}, 0]$~~ & ~~$[0, 0, n_{A'}, n_{C'}]$~~ & ~~$[n_A, n_C, 0, 0]$~~\\
    \hline
   $N = 4k - 1$ & $I$ & $I$ & $X$ & $X$ & $I$ & $X$ \\
   \hline
   $N = 4k + 1$ & $Z$ & $Z$ & $ZX$ & $ZX$ & $Z$ & $ZX$  \\
\hline %\hline
\end{tabular}
\label{tab:RecoveryLoss}
\end{table}

\end{widetext}

Other than photon loss in the encoded cat qubit, there are various sources of photon loss in the teleportation circuit, including the encoded Bell state, beam splitters, phase shifter and PNR detectors. An important question is whether the proposed error correction procedure can also correct single photon loss within the teleportation circuit. The answer is affirmative. In the situation where a single photon is lost in one of the modes of the entangled Bell state, we show that the initial state can be recovered (see Supplementary Material for details). The output states after photon number detection are the same as those where a single photon is lost in the encoded cat qubit, except for a possible global phase that is irrelevant. This implies the Pauli corrections that need to be applied are also given by Table~\ref{tab:RecoveryLoss}. 

Single photon loss in the beam splitters, phase shifter and photon number detectors can also be mitigated. Assume that the optical elements and the PNR detectors are not perfect, but the rate of photon loss along each path (the modes $A, C, A'$ and $C'$) is the same. Under this balanced loss assumption, one can combine all lossy channels along each path into one effective lossy channel. Since the loss rate in each path is the same, the effective lossy channels commute with the optical elements. One can therefore move them to the front of the teleportation circuit, giving a lossy channel in the input mode $A$, which may cause single photon loss in the encoded input state, and another lossy channel in one of the arms of the Bell state, which may cause singe photon loss in the Bell state. As a result, the photon loss from the optical elements and PNR detectors can be mitigated via the proposed error correction procedure.

{\it Estimating loss threshold.}--Since the four-component cat code can protect the encoded qubit from single photon loss, it is expected that the loss threshold will be higher 
if it is used for fault-tolerant quantum computation, as compared to codes which cannot correct single photon loss like the two-component cat code. 
Here we estimate the loss threshold based on the results of Ref.~\cite{PhysRevLett.100.030503}, where concatenation of the two-component cat code with the Steane code or 
Golay code is used to estimate the loss threshold. The purpose of this preliminary estimation is to show that the four-component cat code is a promising candidate for fault-tolerant computation. To obtain an accurate
loss threshold value one needs to perform a numerical simulation. By concatenating the cat codes with some more advanced qubit codes, e.g., surface code, 
which promises a much higher threshold, one expects the loss threshold would be higher. We leave this for future work.

There are two kinds of error one needs to deal with: the located error due to the failure of the teleportation and the phase flip error coming from the photon loss. 
If the teleportation failed, the gate implementation was not successful and one simply removes the relevant qubits. For the two-component cat code, the teleportation fails when
two PNR detectors detect no photons. For the four-component cat code, there are more click patterns that result in failure of teleportation, as shown in Table~\ref{tab:ClickPatterns}. 
The failure probability is obtained by summing the probabilities of detecting all those failure click patterns. 
%Figure~\ref{fig:FailProb} shows the teleportation failure probability (located error rate) for the four-component cat code, also shown is the failure probability for the two-component cat code. 
We find that for a fixed cat amplitude, the failure probability for the two-component cat code is lower than that for the four-component cat code. 
If the failure probability is fixed, the amplitude of the four-component cat code is approximately twice as big as that of the two-component cat code (see supplementary materials). 

The phase flip error rate for the two-component cat code is the probability of losing an odd number of photons and is dominated by the probability of losing one photon. 
Since the four-component cat code can correct a single photon loss, so the phase flip error rate is dominated by the probability of losing two and three photons. 
The phase flip error rate depends on the cat amplitude and the photon loss rate. In order to obtain a fair comparison of loss threshold between the two cat codes, 
we adjust the amplitudes of the two codes so as to set the located error rate of the two codes equal. By fixing the phase flip error rate, we can compare the 
loss value between the two cat codes. 

\begin{figure}
\includegraphics[width= 0.95\columnwidth]{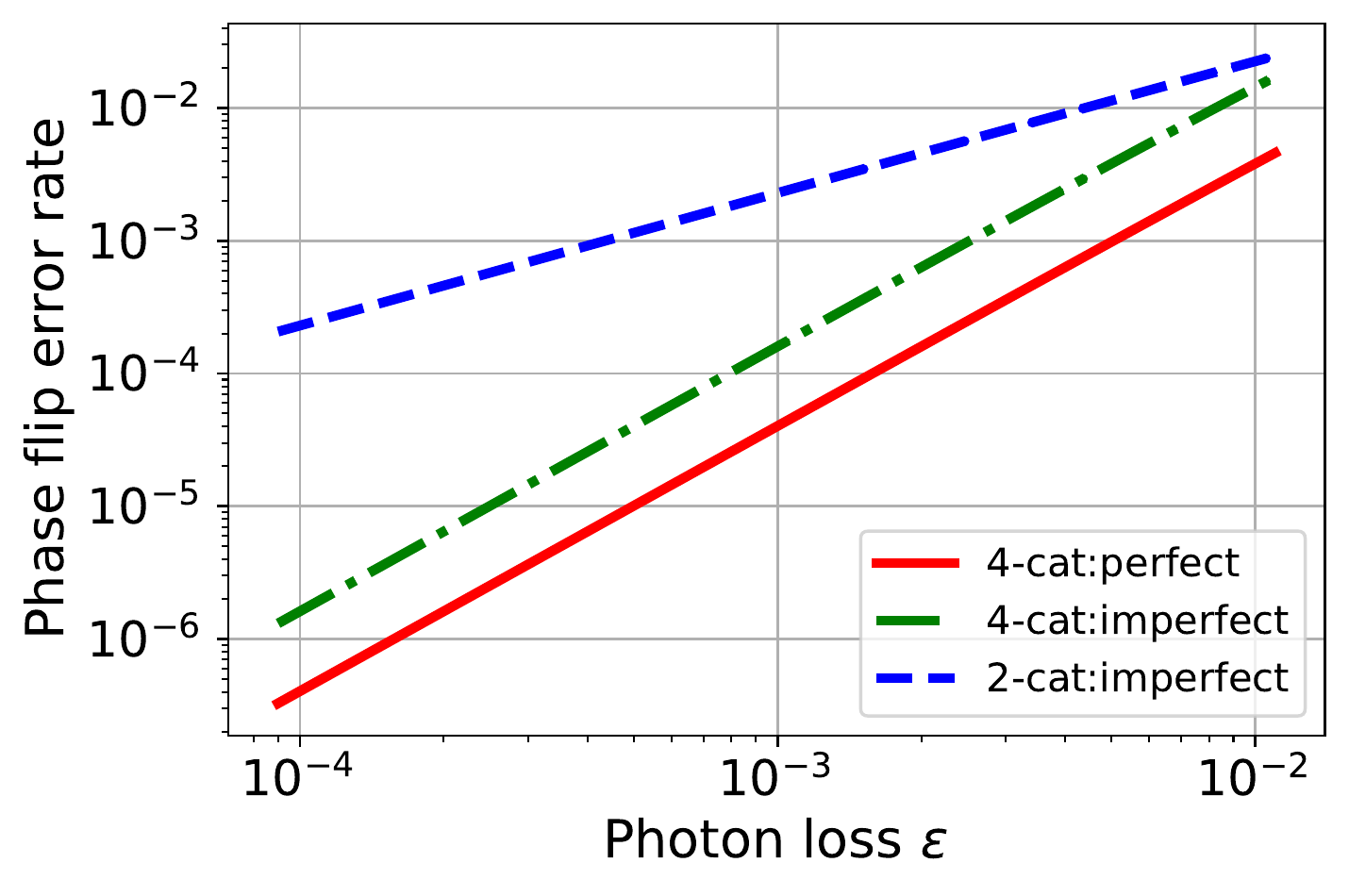}
\caption{ Relation between phase flip error rate and photon loss for the two-component and four-component cat codes under the condition that the teleportation failure probabilities
(located error rates) are the same. The blue (dashed) line is for the two-component cat code, the green (dashed-dotted) and red (solid) lines are for the four-component cat code with lossy and perfect teleportation circuit, respectively. The effective loss rate in the teleportation circuit is chosen to be the same as that in the encoded input state. 
The amplitudes of the two-component and four component cat codes are chosen as $\alpha_2 = 1.5$ and $\alpha_4 = 3.0$, respectively. }
\label{fig:MapLossThreshold}
\end{figure}

It is shown in Ref.~\cite{PhysRevLett.100.030503} that if the amplitude of the two-component cat code $\alpha_2 > 1.2$, the loss threshold is 
around $\epsilon = 5 \times 10^{-4}$. As an example, we choose the amplitude of the two-component cat code as $\alpha_2  = 1.5$, so a four-component cat code with 
amplitude $\alpha_4 = 3.0$ would produce approximately the same located error rate. Once the cat amplitudes are given, the relation between the phase flip error rate and the loss 
is known, as plotted in Fig.~\ref{fig:MapLossThreshold}. It is evident that a significant improvement in the loss threshold can be achieved by employing the four-component cat code. For a phase flip error rate that corresponds to a loss value around $5 \times 10^{-4}$ for the 
two-component cat code, the corresponding loss value for the four-component cat code is about $5 \times 10^{-3}$ with perfect teleportation circuit and $3 \times 10^{-3}$ with lossy teleportation circuit.

{\it Conclusion.}--We developed a scheme to implement the universal set of quantum gates for the four-component cat code via gate teleportation. 
This is the first proposal for optical implementation  
and is particularly promising due to the fast development of PNR detectors.  It is also suitable for other platforms provided the PNR detection is available. Also based on the 
teleportation, a concrete procedure to correct the single photon loss was proposed. Furthermore, we estimated the loss threshold for the four-component cat code when it is 
concatenated with qubit codes, and found a significant improvement as compared to the two-component cat code. 
This shows that the four-component cat code is a promising candidate for fault-tolerant quantum computation. 

\textit{Note}: In the preparation of this work, Ref.~\cite{hastrup2021all} was posted, which like this current work addresses the task of all-optical error correction with a cat code albeit using a different and complementary approach.

We thank Austin Lund for providing some valuable references. 

\

\bibliography{ref_4CatCode}

\appendix
\vspace{0.5cm}

\begin{widetext}

%\section{Supplementary materials}
\section{State teleportation}\label{sec:GoodTeleportation}

In this section we discuss the detailed implementation of the state teleportation. 

For convenience we explicitly write down four two-component cat states, 
$\ket{\mathcal{C}^{\pm}_{\alpha}}$ and $\ket{\mathcal{C}^{\pm}_{i\alpha}}$, which are the
superpositions of coherent states $\ket{\pm \alpha}$ or $\ket{\pm i \alpha}$, namely, 
\begin{eqnarray}\label{eq:four-cat-state}
\ket{\mathcal{C}^{+}_{\alpha} } &=& \mathcal{N}_+ (\ket{\alpha} + \ket{- \alpha}), ~~~~~~ \ket{\mathcal{C}^{+}_{i \alpha} } = \mathcal{N}_+ (\ket{i \alpha} + \ket{- i \alpha}), \nonumber\\
\ket{\mathcal{C}^{-}_{\alpha} } &=& \mathcal{N}_- (\ket{\alpha} - \ket{- \alpha}), ~~~~~~ \ket{\mathcal{C}^{-}_{i \alpha} } = \mathcal{N}_- (\ket{i \alpha} - \ket{- i \alpha}),
\end{eqnarray}
where $\mathcal{N}_\pm = 1/\sqrt{2(1 \pm e^{-2 \alpha^2})}$ are the normalization constants. Note that $\ket{\mathcal{C}^{+}_{\alpha}}$ and $\ket{\mathcal{C}^{-}_{\alpha}}$ are the even and odd two-component cat states, respectively. The logical eigenstates of the Pauli $Z$ operator are encoded as
\begin{eqnarray}\label{eq:NewEncode}
\ket{0_L} = \ket{\mathcal{C}^{+}_{\alpha}}, ~~~~~~~~ \ket{1_L} = \ket{\mathcal{C}^{+}_{i \alpha}}. 
\end{eqnarray}
Note that $\ket{0_L}$ and $\ket{1_L}$ are not orthogonal, $\langle 0_L \ket{1_L} = \langle \mathcal{C}^{+}_{\alpha} \ket{\mathcal{C}^{+}_{i \alpha}} = \cos \alpha^2/ \cosh \alpha^2$,
however, the overlap is small for sufficiently large $\alpha$. A single-qubit logical state is a superposition of $\ket{0_L}$ and $\ket{1_L}$, namely, $\ket{\psi} = c_0 \ket{0_L} + c_1 \ket{1_L}$, 
which is in general a superposition of four coherent states. This is the reason why we call it four-component cat code. 
An essential feature of the four-component cat code is that the logical states are encoded in the subspace with even number of photons, which is crucial for correcting single photon loss. 

The teleportation circuit is shown in Fig.~\ref{fig:4-port-TP}. An input mode (mode $A$) couples with one of the modes of the entangled resource 
state (mode $C$) and another two modes (modes $A'$ and $C'$) with vacuum inputs via a series of 50:50 beam splitters. The optical fields in these four modes are finally detected by four PNR detectors, 
leaving the other mode (mode $B$) of the entangled resource state as the output. For state teleportation, 
the entangled resource state is chosen as the logical Bell state $\ket{\Phi_0}$, which can be written as
%the logical Bell state $\ket{\Phi_0}$ can be written as
\begin{eqnarray}
\ket{\Phi_0}_{BC} &=& \mathcal{N} \left(\ket{0_L}_{B} \ket{0_L}_{C} + \ket{1_L}_{B} \ket{1_L}_{C} \right) 
= \mathcal{N} \left(\ket{\mathcal{C}^{+}_{\alpha}}_{B} \ket{\mathcal{C}^{+}_{\alpha}}_{C} + \ket{\mathcal{C}^{+}_{i \alpha}}_{B} \ket{\mathcal{C}^{+}_{i \alpha}}_{C} \right)
\end{eqnarray}
by using Eqs.~\eqref{eq:four-cat-state} and \eqref{eq:NewEncode}, 
where we have explicitly written down the subscripts ``$B$" and ``$C$" to indicate the corresponding modes. In the case of no photon loss, the input state is in the code subspace, which we 
assume as
\begin{eqnarray}
\ket{\psi_{\rm in}}_A = c_0 \ket{0_L}_A + c_1 \ket{1_L}_A = c_0 \ket{\mathcal{C}^{+}_{\alpha}}_A + c_1 \ket{\mathcal{C}^{+}_{i \alpha}}_A,
\end{eqnarray}
where the subscript ``$A$" indicates that the input is in mode $A$, and $c_0$ and $c_1$ are coefficients that normalize the input state. Then the overall input state before teleportation is
\begin{eqnarray}
\ket{\Psi_{\rm in}} &=& \ket{\psi_{\rm in}}_A \otimes \ket{\Phi_0}_{BC} \otimes \ket{0, 0}_{A' C'}
\nonumber\\
&=&
\mathcal{N} \left[ \left( c_0 \ket{\mathcal{C}^{+}_{\alpha}}_A \ket{\mathcal{C}^{+}_{\alpha}}_C 
+ c_1 \ket{\mathcal{C}^{+}_{i \alpha}}_A \ket{\mathcal{C}^{+}_{\alpha}}_C \right) \ket{\mathcal{C}^{+}_{\alpha}}_B 
+ \left( c_0 \ket{\mathcal{C}^{+}_{\alpha}}_A \ket{\mathcal{C}^{+}_{i \alpha}}_C 
+ c_1 \ket{\mathcal{C}^{+}_{i \alpha}}_A \ket{\mathcal{C}^{+}_{i \alpha}}_C \right) \ket{\mathcal{C}^{+}_{i \alpha}}_B \right] \otimes \ket{0, 0}_{A' C'}. 
\end{eqnarray}

The teleportation circuit consists of four $50:50$ beam splitters and a $\frac{\pi}{2}$ phase shifter. We denote their action as unitary operators 
$\op{U}_{AC}$, $\op{U}_{AA'}$, $\op{U}_{CC'}$, $\op{U}_{A'C'}$ and $\op{U}_{A'}(\pi/2)$, respectively. Here $\op{U}_{AC}$ represents the unitary transformation of the beam splitter located
at the intersection between modes $A$ and $C$, and notations are similar for other unitary operators. 
By considering the action of a 50:50 beam splitter upon two input coherent states and the action of a phase shifter, it can be shown that 
\begin{eqnarray*}
 \ket{\mathcal{C}^{+}_{\alpha}}_A \ket{\mathcal{C}^{+}_{\alpha}}_C \ket{0}_{A'} \ket{0}_{C'} 
 &\longrightarrow&
\bigg| 0, \alpha, -\frac{\alpha}{\sqrt{2}}, \frac{\alpha}{\sqrt{2}} \bigg\rangle + \bigg| \alpha, 0, \frac{i \alpha}{\sqrt{2}}, \frac{i \alpha}{\sqrt{2}} \bigg\rangle
+ \bigg| - \alpha, 0, - \frac{i \alpha}{\sqrt{2}}, - \frac{i \alpha}{\sqrt{2}} \bigg\rangle + \bigg| 0, - \alpha, \frac{\alpha}{\sqrt{2}}, - \frac{\alpha}{\sqrt{2}} \bigg\rangle,
\nonumber\\
 \ket{\mathcal{C}^{+}_{i \alpha}}_A \ket{\mathcal{C}^{+}_{i \alpha}}_C \ket{0}_{A'} \ket{0}_{C'} 
 &\longrightarrow&
\bigg| 0, i \alpha, -\frac{i \alpha}{\sqrt{2}}, \frac{i \alpha}{\sqrt{2}} \bigg\rangle + \bigg| i \alpha, 0, - \frac{\alpha}{\sqrt{2}}, - \frac{\alpha}{\sqrt{2}} \bigg\rangle
+ \bigg| - i \alpha, 0, \frac{\alpha}{\sqrt{2}}, \frac{\alpha}{\sqrt{2}} \bigg\rangle + \bigg| 0, -i \alpha, \frac{i \alpha}{\sqrt{2}}, - \frac{i \alpha}{\sqrt{2}} \bigg\rangle,
\nonumber\\
 \ket{\mathcal{C}^{+}_{\alpha}}_A \ket{\mathcal{C}^{+}_{i \alpha}}_C \ket{0}_{A'} \ket{0}_{C'} 
 &\longrightarrow&
\bigg| \frac{\beta^*}{\sqrt{2}}, \frac{\beta}{\sqrt{2}},  0, \beta \bigg\rangle+ \bigg| \frac{\beta}{\sqrt{2}}, \frac{\beta^*}{\sqrt{2}}, - \beta^*, 0 \bigg\rangle
+ \bigg| - \frac{\beta}{\sqrt{2}}, - \frac{\beta^*}{\sqrt{2}}, \beta^*, 0 \bigg\rangle + \bigg| - \frac{\beta^*}{\sqrt{2}}, - \frac{\beta}{\sqrt{2}}, 0, - \beta \bigg\rangle,
\nonumber\\
\ket{\mathcal{C}^{+}_{i \alpha}}_A \ket{\mathcal{C}^{+}_{\alpha}}_C \ket{0}_{A'} \ket{0}_{C'} 
 &\longrightarrow&
\bigg| - \frac{\beta^*}{\sqrt{2}}, \frac{\beta}{\sqrt{2}},  - \beta, 0 \bigg\rangle + \bigg| \frac{\beta}{\sqrt{2}}, - \frac{\beta^*}{\sqrt{2}}, 0, - \beta^* \bigg\rangle
+ \bigg| - \frac{\beta}{\sqrt{2}}, \frac{\beta^*}{\sqrt{2}}, 0, \beta^* \bigg\rangle+ \bigg| \frac{\beta^*}{\sqrt{2}}, - \frac{\beta}{\sqrt{2}}, \beta, 0 \bigg\rangle,
\end{eqnarray*}
where have defined $\beta = \alpha e^{i \pi/4}$, and omitted the subscripts to simplify the notation and chosen the order of the modes as $``ACA'C' "$. 
By combining these results, we find that the output state after the teleportation circuit (before photon number measurement) is 
\begin{eqnarray}\label{eq:OutPutNoLoss}
&& \ket{\Psi}_{BAA'CC'} = \op{U}_{A'C'} \op{U}_{A'}(\pi/2) \op{U}_{AA'} \op{U}_{CC'} \op{U}_{AC} \ket{\Psi_{\rm in}}
\nonumber\\
&\propto&
\bigg[ \bigg( \bigg| 0, \alpha, - \frac{\alpha}{\sqrt{2}}, \frac{\alpha}{\sqrt{2}} \bigg\rangle 
+ \bigg| 0, - \alpha, \frac{\alpha}{\sqrt{2}}, - \frac{\alpha}{\sqrt{2}} \bigg\rangle \bigg) c_0 \ket{\mathcal{C}^{+}_{\alpha}}_B
%\nonumber\\
%&&
+ \bigg( \bigg| 0, i \alpha, - \frac{i \alpha}{\sqrt{2}}, \frac{i \alpha}{\sqrt{2}} \bigg\rangle 
+ \bigg|0, - i \alpha, \frac{i \alpha}{\sqrt{2}}, - \frac{i \alpha}{\sqrt{2}} \bigg\rangle \bigg) c_1 \ket{\mathcal{C}^{+}_{i \alpha}}_B \bigg]
\nonumber\\
&&
+ \bigg[ \bigg( \bigg| \alpha, 0, \frac{i \alpha}{\sqrt{2}}, \frac{i \alpha}{\sqrt{2}} \bigg\rangle 
+ \bigg| - \alpha, 0, - \frac{i \alpha}{\sqrt{2}}, - \frac{i \alpha}{\sqrt{2}} \bigg\rangle \bigg) c_0 \ket{\mathcal{C}^{+}_{\alpha}}_B 
%\nonumber\\
%&&
+ \bigg( \bigg| i \alpha, 0, - \frac{\alpha}{\sqrt{2}}, - \frac{\alpha}{\sqrt{2}} \bigg\rangle 
+ \bigg| - i \alpha, 0, \frac{\alpha}{\sqrt{2}}, \frac{\alpha}{\sqrt{2}} \bigg\rangle \bigg) c_1 \ket{\mathcal{C}^{+}_{i \alpha}}_B \bigg]
\nonumber\\
&&
+ \bigg[ \bigg( \bigg| \frac{\beta}{\sqrt{2}}, - \frac{\beta^*}{\sqrt{2}}, 0, - \beta^* \bigg\rangle 
+ \bigg| - \frac{\beta}{\sqrt{2}}, \frac{\beta^*}{\sqrt{2}}, 0, \beta^* \bigg\rangle \bigg) c_1 \ket{\mathcal{C}^{+}_{\alpha}}_B 
%\nonumber\\
%&&
+ \bigg( \bigg| \frac{\beta^*}{\sqrt{2}}, \frac{\beta}{\sqrt{2}},  0, \beta \bigg\rangle 
+ \bigg| - \frac{\beta^*}{\sqrt{2}}, - \frac{\beta}{\sqrt{2}}, 0, - \beta \bigg\rangle \bigg) c_0 \ket{\mathcal{C}^{+}_{i \alpha}}_B \bigg]
\nonumber\\
&&
+ \bigg[ \bigg( \bigg| - \frac{\beta^*}{\sqrt{2}}, \frac{\beta}{\sqrt{2}}, - \beta, 0 \bigg\rangle 
+ \bigg| \frac{\beta^*}{\sqrt{2}}, - \frac{\beta}{\sqrt{2}}, \beta, 0 \bigg\rangle \bigg) c_1 \ket{\mathcal{C}^{+}_{\alpha}}_B 
%\nonumber\\
%&&
+ \bigg( \bigg| \frac{\beta}{\sqrt{2}}, \frac{\beta^*}{\sqrt{2}}, - \beta^*, 0 \bigg\rangle 
+ \bigg| - \frac{\beta}{\sqrt{2}}, - \frac{\beta^*}{\sqrt{2}}, \beta^*, 0 \bigg\rangle \bigg) c_0 \ket{\mathcal{C}^{+}_{i \alpha}}_B \bigg],
\nonumber\\
\end{eqnarray}
where the proportionality factor is $\mathcal{N} \mathcal{N}_+^{\, 2}$. 

Four PNR detectors are used to count the number of photons at modes $A, C, A'$ and $C'$, giving click patterns
$[n_A, n_C, n_{A'}, n_{C'}]$. From the output state, Eq.~\eqref{eq:OutPutNoLoss}, we see that at least one of the detectors registers no photons. 
We will derive the conditional output state and its occurring probability for a particular click pattern, from which one can explicitly see whether the teleportation is
successful and what Pauli correction needs to be applied. The results are summarized in Tables~\ref{tab:OneZeroNoLoss}, 
\ref{tab:TwoZeroNoLoss} and \ref{tab:ThreeZeroNoLoss}.
%\textcolor{blue}{When two or more detectors detect vacuum, the teleportation fails}. 

\begin{table}[h]
\caption{ {\bf Pauli corrections for state teleportation}: click patterns with one vacuum output, no photon loss.  } %title of the table
\centering % centering table
\begin{tabular}{@{} l |c|c|c|c}%\toprule
\hline %\hline
   & ~~~$[0, n_C, n_{A'}, n_{C'}]$~~~ & ~~~$[n_A, 0, n_{A'}, n_{C'}]$~~~  & ~~~$[n_A, n_C, 0, n_{C'}]$~~~ & ~~~$[n_A, n_C, n_{A'}, 0]$~~~ \\
    \hline
   $N = 4k$ & $I$ & $I$ & $X$ & $X$ \\
   \hline
   $N = 4k + 2$ & $Z$ & $Z$ & $ZX$ & $ZX$ \\
\hline %\hline
\end{tabular}
\label{tab:OneZeroNoLoss}
\end{table}

\begin{table}[h]
\caption{ {\bf Pauli corrections for state teleportation}: click patterns with two vacuum outputs, no photon loss.   } %title of the table
\centering % centering table
\begin{tabular}{@{} l |c|c|c|c|c|c}%\toprule
\hline %\hline
   & ~~$[0, 0, n_{A'}, n_{C'}]$~~ & ~~$[n_A, n_C, 0, 0]$~~  & ~~$[n_A, 0, n_{A'}, 0]$~~ & ~~$[n_A, 0, 0, n_{C'}]$~~ & ~~$[0, n_C, n_{A'}, 0]$~~ & ~~$[0, n_C, 0, n_{C'}]$~~ \\
    \hline
   $N = 4k$ & $I$ & $X$ & failed & failed & failed & failed \\
   \hline
   $N = 4k - 2$ & $Z$ & $ZX$ & failed & failed & failed & failed \\
\hline %\hline
\end{tabular}
\label{tab:TwoZeroNoLoss}
\end{table}

\begin{table}[h]
\caption{ {\bf Pauli corrections for state teleportation}: click patterns with three vacuum outputs, no photon loss.  } %title of the table
\centering % centering table
\begin{tabular}{@{} l |c|c|c|c}%\toprule
\hline %\hline
   & ~~~$[n_A, 0, 0, 0]$~~~ & ~~~$[0, n_C, 0, 0, 0]$~~~  & ~~~$[0, 0, n_{A'}, 0]$~~~ & ~~~$[0, 0, 0, n_{C'}]$~~~ \\
    \hline
   $N = 4k$ & failed & failed & failed & failed \\
   \hline
   $N = 4k - 2$ & $Z$ & $Z$ & $ZX$ & $ZX$ \\
\hline %\hline
\end{tabular}
\label{tab:ThreeZeroNoLoss}
\end{table}

\subsection{Absence of photon at one detector}

Consider situations where only one detector detects no photons and each of the other three detectors registers at least one photon. 
This includes four sorts of click pattern: 
\begin{eqnarray*}
[0, n_C, n_{A'}, n_{C'}], ~~~~~~ [n_A, 0, n_{A'}, n_{C'}], ~~~~~~ [n_A, n_{C}, 0, n_{C'}], ~~~~~~ [n_A, n_{C}, n_{A'}, 0]. 
\end{eqnarray*}
We find that the encoded input state can be successfully teleported to the output for all these click patterns. The detailed analysis is given as follows. 

(i) $n_{A} =0$, no click at mode $A$.  

When the detector at mode $A$ detects no photons, namely, the measurement outcome is $\vt{n} = [0, n_C, n_{A'}, n_{C'}]$, the conditional output state is
\begin{eqnarray}
\langle0, n_C, n_{A'}, n_{C'} \ket{\Psi}_{BAA'CC'} =
\mathcal{N} \mathcal{N}_+^{\, 2} \frac{e^{- \alpha^2}}{\sqrt{n_C ! \, n_{A'} ! \, n_{C'} !}} \frac{ (-1)^{n_{A'}} \alpha^{N}}{(\sqrt{2})^{n_{A'} + n_{C'}}} 
\big[1 + (-1)^N \big] \bigg( c_0 \ket{\mathcal{C}^{+}_{\alpha}}_B + i^N c_1 \ket{\mathcal{C}^{+}_{i \alpha}}_B \bigg),
\end{eqnarray} 
where $N = n_A + n_C + n_{A'}  + n_{C'}$ is the total number of detected photons. 
This implies the (unnormalized) output state at mode $B$ is 
\begin{eqnarray}\label{eq:Case1output}
\ket{\psi_1} =  c_0 \ket{\mathcal{C}^{+}_{\alpha}}_B + i^N c_1 \ket{\mathcal{C}^{+}_{i \alpha}}_B, 
\end{eqnarray}
and the measurement probability is %\textcolor{red}{ [missing a normalization factor of the output state] }
\begin{eqnarray}
P([0, n_C, n_{A'}, n_{C'}]) = \mathcal{N}^{\, 2} \mathcal{N}_+^{\, 4} \frac{e^{- 2 \alpha^2}}{ n_C ! \, n_{A'} ! \, n_{C'} !} \frac{\alpha^{2 N}}{2^{n_{A'} + n_{C'}}} \big[1 + (-1)^N \big]^2 
|\langle \psi_1 \ket{\psi_1}|^2. 
\end{eqnarray}

(ii) $n_C = 0$, no click at mode $C$. 

The detector at mode $C$ detects no photons, namely, the measurement outcome is $\vt{n} = [n_A, 0, n_{A'}, n_{C'}]$, the conditional output state is
\begin{eqnarray}
\langle n_A, 0, n_{A'}, n_{C'} \ket{\Psi}_{BAA'CC'} =
\mathcal{N} \mathcal{N}_+^{\, 2} \frac{e^{- \alpha^2}}{\sqrt{n_A ! \, n_{A'} ! \, n_{C'} !}} \frac{i^{n_{A'} + n_{C'}} \alpha^{N}}{(\sqrt{2})^{n_{A'} + n_{C'}}} 
\big[1 + (-1)^N \big] \bigg( c_0 \ket{\mathcal{C}^{+}_{\alpha}}_B + i^N c_1 \ket{\mathcal{C}^{+}_{i \alpha}}_B \bigg). 
\end{eqnarray} 
%where $N = n_A + n_{A'}  + n_{C'}$ is the total number of detected photons. 
This implies the (unnormalized) output state at mode $B$ is
\begin{eqnarray}\label{eq:Case1output}
\ket{\psi_2} =  c_0 \ket{\mathcal{C}^{+}_{\alpha}}_B + i^N c_1 \ket{\mathcal{C}^{+}_{i \alpha}}_B,
\end{eqnarray}
and the measurement probability is 
\begin{eqnarray}
P([n_A, 0, n_{A'}, n_{C'}]) = \mathcal{N}^{\, 2} \mathcal{N}_+^{\, 4} \frac{e^{- 2 \alpha^2}}{n_{A} ! \, n_{A'} ! \, n_{C'} !} \frac{\alpha^{2N}}{2^{n_{A'} + n_{C'}}} \big[1 + (-1)^N \big]^2
|\langle \psi_2 \ket{\psi_2}|^2. 
\end{eqnarray}
It is evident that the output states are the same for cases (i) and (ii). 

(iii) $n_{A'} = 0$, no click at mode $A'$.

The detector at mode $A'$ detects no photons, namely, the measurement outcome is $\vt{n} = [n_A, n_C, 0, n_{C'}]$, the conditional output state is
\begin{eqnarray}
\langle n_A, n_C, 0, n_{C'} \ket{\Psi}_{BAA'CC'} =
\mathcal{N} \mathcal{N}_+^{\, 2} \frac{e^{- \alpha^2}}{\sqrt{n_A ! \, n_{C} ! \, n_{C'} !}} \frac{ e^{- i \pi (n_A - n_{C} - n_{C'})/4 } \alpha^{N}}{(\sqrt{2})^{n_{A} + n_{C}}} 
\big[1 + (-1)^N \big] \bigg( c_0 \ket{\mathcal{C}^{+}_{i \alpha}}_B + i^N c_1 \ket{\mathcal{C}^{+}_{\alpha}}_B \bigg).
%\nonumber\\
\end{eqnarray} 
%where $N = n_A + n_{C}  + n_{C'}$ is the total number of detected photons. 
This implies the (unnormalized) output state at mode $B$ is
\begin{eqnarray}\label{eq:Case1output}
\ket{\psi_3} =  c_0 \ket{\mathcal{C}^{+}_{i \alpha}}_B + i^N c_1 \ket{\mathcal{C}^{+}_{\alpha}}_B,
\end{eqnarray}
and the measurement probability is 
\begin{eqnarray}
P([n_A, n_{C}, 0, n_{C'}]) = \mathcal{N}^{\, 2} \mathcal{N}_+^{\, 4} \frac{e^{- 2 \alpha^2}}{n_{A} ! \, n_{C} ! \, n_{C'} !} \frac{\alpha^{2N}}{2^{n_{A} + n_{C}}} \big[1 + (-1)^N \big]^2 
|\langle \psi_3 \ket{\psi_3}|^2.
\end{eqnarray}

(iv) $n_{C'} = 0$, no click at mode $C'$.

The detector at mode $C'$ detects no photons, namely, the measurement outcome is $\vt{n} = [n_A, n_C, n_{A'}, 0]$, the conditional state is
\begin{eqnarray}
\langle n_A, n_C, n_{A'}, 0 \ket{\Psi}_{BAA'CC'} &=&
\mathcal{N} \mathcal{N}_+^{\, 2} \frac{e^{- \alpha^2}}{\sqrt{n_A ! \, n_{C} ! \, n_{A'} !}} \frac{(-1)^{n_{A'}} e^{i \pi (n_A - n_{C} - n_{A'})/4 } \alpha^{N}}{(\sqrt{2})^{n_{A} + n_{C}}} 
%\nonumber\\
%&&
%\times 
\big[1 + (-1)^N \big]  \bigg( c_0 \ket{\mathcal{C}^{+}_{i \alpha}}_B + i^N c_1 \ket{\mathcal{C}^{+}_{\alpha}}_B \bigg). 
\nonumber\\
\end{eqnarray} 
%where $N = n_A + n_{C}  + n_{A'}$ is the total number of detected photons. 
This implies the (unnormalized) output state at mode $B$ is
\begin{eqnarray}\label{eq:Case1output}
\ket{\psi_4} =  c_0 \ket{\mathcal{C}^{+}_{i \alpha}}_B + i^N c_1 \ket{\mathcal{C}^{+}_{\alpha}}_B,
\end{eqnarray}
and the measurement probability is 
\begin{eqnarray}
P([n_A, n_{C}, n_{A'}, 0]) = \mathcal{N}^{\, 2} \mathcal{N}_+^{\, 4} \frac{e^{- 2 \alpha^2}}{n_{A} ! \, n_{C} ! \, n_{A'} !} \frac{\alpha^{2N}}{2^{n_{A} + n_{C}}} \big[1 + (-1)^N \big]^2
|\langle \psi_4 \ket{\psi_4}|^2.
\end{eqnarray}
It is evident that the output states are the same for cases (iii) and (iv).

For all these cases, the probability of detecting an odd number of photon is zero. This is consistent with the fact that the entangled resource state, input state and 
output state all consist of even numbers of photon. For cases (i) and (ii), the output state is exactly the same as the input state when $N = 4 k$ with $k$ a positive integer, indicating a
successful teleportation without further corrections. When $N = 4 k + 2$, the output state is 
\begin{eqnarray}
\ket{\psi_{\rm out}} = c_0 \ket{\mathcal{C}^{+}_{\alpha}}_B - c_1 \ket{\mathcal{C}^{+}_{i \alpha}}_B, 
\end{eqnarray}
and can be converted to the input state by applying a Pauli $Z$ correction. 
For cases (iii) and (iv), the output state is 
\begin{eqnarray}
\ket{\psi_{\rm out}} = c_0 \ket{\mathcal{C}^{+}_{i \alpha}}_B + c_1 \ket{\mathcal{C}^{+}_{\alpha}}_B
\end{eqnarray}
when $N = 4 k$, and it can be converted to the input state by applying a Pauli $X$ correction, which is simply a $\frac{\pi}{2}$ phase shift and is easy to implement. 
When $N = 4 k + 2$, the output state is 
\begin{eqnarray}
\ket{\psi_{\rm out}} = c_0 \ket{\mathcal{C}^{+}_{i \alpha}}_B - c_1 \ket{\mathcal{C}^{+}_{\alpha}}_B, 
\end{eqnarray}
and it can be converted to the input state by first applying a Pauli $X$ correction and then a Pauli $Z$ correction. This is summarized in Table~\ref{tab:OneZeroNoLoss}.

\subsection{Absence of photons at two detectors }

When two detectors detect no photons and each of the other two detectors registers at least one photon, the teleportation succeeds for click patterns:
\begin{eqnarray*}
[0, 0, n_{A'}, n_{C'}], ~~~~~~ [n_A, n_{C}, 0, 0]; 
\end{eqnarray*}
and fails for click patterns:
\begin{eqnarray*}
[n_A, 0, n_{A'}, 0], ~~~~~~ [n_A, 0, 0, n_{C'}], ~~~~~~ [0, n_C, n_{A'}, 0], ~~~~~~ [0, n_C, 0, n_{C'}]. 
\end{eqnarray*}
The results are summarized in Table~\ref{tab:TwoZeroNoLoss} and the detailed analysis is given as follows.

(i) $n_A = n_C = 0$, no click at modes $A$ and $C$.

The detectors at modes $A$ and $C$ detect no photons, namely, the measurement outcome is $\vt{n} = [0, 0, n_{A'}, n_{C'}]$. The conditional output state is
\begin{eqnarray}
\langle0, 0, n_{A'}, n_{C'} \ket{\Psi}_{BAA'CC'} =
\mathcal{N} \mathcal{N}_+^{\, 2} \frac{e^{- \alpha^2}}{\sqrt{n_{A'} ! \, n_{C'} !}} \frac{i^N \alpha^{N}}{(\sqrt{2})^{N}} 
\big[1 + (-1)^N \big] \big(1 + i^{n_{A'}  - n_{C'}} \big) \bigg( c_0 \ket{\mathcal{C}^{+}_{\alpha}}_B + i^N c_1 \ket{\mathcal{C}^{+}_{i \alpha}}_B \bigg).
\end{eqnarray} 
%where $N = n_{A'}  + n_{C'}$ is the total number of detected photons. 
This implies the (unnormalized) output state at mode $B$ is 
\begin{eqnarray}\label{eq:Case1output}
\ket{\psi_5} =  c_0 \ket{\mathcal{C}^{+}_{\alpha}}_B + i^N c_1 \ket{\mathcal{C}^{+}_{i \alpha}}_B,
\end{eqnarray}
and the measurement probability is 
\begin{eqnarray}
P([0, 0, n_{A'}, n_{C'}]) = \mathcal{N}^{\, 2} \mathcal{N}_+^{\, 4} \frac{e^{- 2 \alpha^2}}{n_{A'} ! \, n_{C'} !} \frac{\alpha^{2N}}{2^{N}} \big[1 + (-1)^N \big]^2 \big|1 + i^{n_{A'}  - n_{C'}} \big|^2
|\langle \psi_5 \ket{\psi_5}|^2.
\end{eqnarray}
The measurement probability is nonzero only when $N$ is even and $n_{A'}  - n_{C'} = 4 \ell$ with $\ell$ an integer.

(ii) $n_{A'} = n_{C'} = 0$,  no click at modes $A'$ and $C'$. 

The detectors at mode $A'$ and $C'$ detect no photons, namely, the measurement outcome is $\vt{n} = [n_A, n_C, 0, 0]$. The conditional output state is
\begin{eqnarray*}
\langle n_A, n_C, 0, 0 \ket{\Psi}_{BAA'CC'} =
\mathcal{N} \mathcal{N}_+^{\, 2} \frac{e^{- \alpha^2}}{\sqrt{n_A ! \, n_{C} ! }} \frac{e^{- i \pi (n_A - n_{C})/4 }  \alpha^{N}}{(\sqrt{2})^{N}} 
\big[1 + (-1)^N \big] \big(1 + i^{n_A - n_C} \big) \bigg( c_0 \ket{\mathcal{C}^{+}_{i \alpha}}_B + i^N c_1 \ket{\mathcal{C}^{+}_{\alpha}}_B \bigg).
\end{eqnarray*} 
%where $N = n_A + n_{C} $ is the total number of detected photons. 
This implies the (unnormalized) output state at mode $B$ is
\begin{eqnarray}\label{eq:Case1output}
\ket{\psi_6} =  c_0 \ket{\mathcal{C}^{+}_{i \alpha}}_B + i^N c_1 \ket{\mathcal{C}^{+}_{\alpha}}_B,
\end{eqnarray}
and the measurement probability is 
\begin{eqnarray}
P([n_A, n_{C}, 0, 0]) = \mathcal{N}^{\, 2} \mathcal{N}_+^{\, 4} \frac{e^{- 2 \alpha^2}}{n_{A} ! \, n_{C} ! } \frac{\alpha^{2N}}{2^{N}} \big[1 + (-1)^N \big]^2 \big|1 + i^{n_A - n_C} \big|^2
|\langle \psi_6 \ket{\psi_6}|^2.
\end{eqnarray}
The measurement probability is nonzero only when $N$ is even and $ n_A - n_C= 4 \ell$ with $\ell$ an integer.

The teleportation fails for the following click patterns. Here we explicitly write down the conditional output states, from which one can see clearly why the teleportation fails. 

(iii) $n_{C} = n_{C'} = 0$,  no click at modes $C$ and $C'$. 

The detectors at mode $C$ and $C'$ detect no photons, namely, the measurement outcome is $\vt{n} = [n_A, 0, n_{A'}, 0]$ with $n_A \ne 0$ and $ n_{A'} \ne 0$. The conditional output state is
\begin{eqnarray}
&& \langle n_A, 0, n_{A'}, 0 \ket{\Psi}_{BAA'CC'} 
\nonumber\\
&=&
\mathcal{N} \mathcal{N}_+^{\, 2} \frac{e^{-\alpha^2}}{\sqrt{n_A ! \, n_{A'} !}} \alpha^N \big[ 1 + (-1)^{N} \big] i^{n_{A'}}
\bigg[ \frac{1}{(\sqrt{2})^{n_{A'}}} \bigg( c_0 \ket{\mathcal{C}^{+}_{\alpha}}_B + i^{N} c_1 \ket{\mathcal{C}^{+}_{i \alpha}}_B, \bigg)
\nonumber\\
&&
+ \frac{ e^{i N \pi /4}}{(\sqrt{2})^{n_{A}}} \bigg( c_0 \ket{\mathcal{C}^{+}_{i \alpha}}_B + i^{N} c_1 \ket{\mathcal{C}^{+}_{\alpha}}_B, \bigg)
\bigg]
\nonumber\\
&=&
\mathcal{N} \mathcal{N}_+^{\, 2} \frac{ e^{-\alpha^2}}{\sqrt{n_A ! n_{A'} !}} \alpha^N \big[ 1 + (-1)^N \big] i^{n_{A'}}
\bigg\{ \bigg[ \frac{1}{(\sqrt{2})^{n_{A'}}} c_0 + \frac{ e^{i N \pi/4}}{(\sqrt{2})^{n_{A}}} i^N c_1 \bigg]  \ket{\mathcal{C}^{+}_{\alpha}}_B
\nonumber\\
&&
+ \bigg[ \frac{1}{(\sqrt{2})^{n_{A'}}} i^N c_1 + \frac{ e^{i N \pi/4}}{(\sqrt{2})^{n_{A}}} c_0 \bigg]  \ket{\mathcal{C}^{+}_{i \alpha}}_B
\bigg\}.
\end{eqnarray}

(iv) $n_{C} = n_{A'} = 0$,  no click at modes $C$ and $A'$. 

The detectors at mode $C$ and $A'$ detect no photons, namely, the measurement outcome is $\vt{n} = [n_A, 0, 0, n_{C'}]$ with $n_A \ne 0$ and $ n_{C'} \ne 0$. The conditional output state is
\begin{eqnarray}
&& \langle n_A, 0, 0, n_{C'} \ket{\Psi}_{BAA'CC'} 
\nonumber\\
&=&
\mathcal{N} \mathcal{N}_+^{\, 2} \frac{e^{-\alpha^2}}{\sqrt{n_A ! \, n_{C'} !}} \alpha^{N} \big[ 1 + (-1)^{N} \big] i^{n_{C'}}
\bigg[ \frac{1}{(\sqrt{2})^{n_{C'}}} \bigg( c_0 \ket{\mathcal{C}^{+}_{\alpha}}_B + i^{N} c_1 \ket{\mathcal{C}^{+}_{i \alpha}}_B, \bigg)
\nonumber\\
&&
+ \frac{ e^{- i N \pi/4}}{(\sqrt{2})^{n_{A}}} \bigg( c_0 \ket{\mathcal{C}^{+}_{i \alpha}}_B + i^{N} c_1 \ket{\mathcal{C}^{+}_{\alpha}}_B, \bigg)
\bigg]
\nonumber\\
&=&
\mathcal{N} \mathcal{N}_+^{\, 2} \frac{e^{-\alpha^2}}{\sqrt{n_A ! \, n_{C'} !}} \alpha^N \big[ 1 + (-1)^N \big] i^{n_{C'}}
\bigg\{ \bigg[ \frac{1}{(\sqrt{2})^{n_{C'}}} c_0 + \frac{ e^{- i N \pi/4}}{(\sqrt{2})^{n_{A}}} i^N c_1 \bigg]  \ket{\mathcal{C}^{+}_{\alpha}}_B
\nonumber\\
&&
+ \bigg[ \frac{1}{(\sqrt{2})^{n_{C'}}} i^N c_1 + \frac{e^{- i N \pi/4}}{(\sqrt{2})^{n_{A}}} c_0 \bigg]  \ket{\mathcal{C}^{+}_{i \alpha}}_B
\bigg\}.
\end{eqnarray}

(v) $n_{A} = n_{C'} = 0$,  no click at modes $A$ and $C'$. 

The detectors at mode $A$ and $C'$ detect no photons, namely, the measurement outcome is $\vt{n} = [0, n_C, n_{A'}, 0]$ with $n_C \ne 0$ and $ n_{A'} \ne 0$. The conditional output state is
\begin{eqnarray}
&& \langle 0, n_C, n_{A'}, 0 \ket{\Psi}_{BAA'CC'} 
\nonumber\\
&=&
\mathcal{N} \mathcal{N}_+^{\, 2} \frac{e^{-\alpha^2}}{\sqrt{n_C ! \, n_{A'} !}} \alpha^{N} \big[ 1 + (-1)^{N} \big] (-1)^{n_{A'}}
\bigg[ \frac{1}{(\sqrt{2})^{n_{A'}}} \bigg( c_0 \ket{\mathcal{C}^{+}_{\alpha}}_B + i^{N} c_1 \ket{\mathcal{C}^{+}_{i \alpha}}_B, \bigg)
\nonumber\\
&&
+ \frac{e^{- i N \pi/4}}{(\sqrt{2})^{n_{C}}} \bigg( c_0 \ket{\mathcal{C}^{+}_{i \alpha}}_B + i^{N} c_1 \ket{\mathcal{C}^{+}_{\alpha}}_B, \bigg)
\bigg]
\nonumber\\
&=&
\mathcal{N} \mathcal{N}_+^{\, 2} \frac{e^{-\alpha^2}}{\sqrt{n_C ! \, n_{A'} !}} \alpha^{N} \big[ 1 + (-1)^{N} \big] (-1)^{n_{A'}}
\bigg\{ \bigg[ \frac{1}{(\sqrt{2})^{n_{A'}}} c_0 + \frac{e^{- i N \pi /4}}{(\sqrt{2})^{n_{C}}} i^N c_1 \bigg]  \ket{\mathcal{C}^{+}_{\alpha}}_B
\nonumber\\
&&
+ \bigg[ \frac{1}{(\sqrt{2})^{n_{A'}}} i^N c_1 + \frac{ e^{- i N \pi/4}}{(\sqrt{2})^{n_{C}}} c_0 \bigg]  \ket{\mathcal{C}^{+}_{i \alpha}}_B
\bigg\}.
\end{eqnarray}

%The outcome probability is
%\begin{eqnarray}
%P([0, n_C, n_{A'}, 0]) = | \langle 0, n_C, n_{A'}, 0 \ket{\Psi}_{BAA'CC'} |^2. 
%\end{eqnarray}

(vi) $n_{A} = n_{A'} = 0$,  no click at modes $A$ and $A'$. 

The detectors at mode $A$ and $A'$ detect no photons, namely, the measurement outcome is $\vt{n} = [0, n_C, 0, n_{C'}]$ with $n_C \ne 0$ and $ n_{C'} \ne 0$. The conditional output state is
\begin{eqnarray}
&& \langle 0, n_C, 0, n_{C'} \ket{\Psi}_{BAA'CC'} 
\nonumber\\
&=&
\mathcal{N} \mathcal{N}_+^{\, 2} \frac{e^{-\alpha^2}}{\sqrt{n_C ! \, n_{C'} !}} \alpha^{N} \big[ 1 + (-1)^{N} \big] 
\bigg[ \frac{1}{(\sqrt{2})^{n_{C'}}} \bigg( c_0 \ket{\mathcal{C}^{+}_{\alpha}}_B + i^{N} c_1 \ket{\mathcal{C}^{+}_{i \alpha}}_B, \bigg)
\nonumber\\
&&
+ \frac{ e^{i N \pi /4}}{(\sqrt{2})^{n_{C}}} \bigg( c_0 \ket{\mathcal{C}^{+}_{i \alpha}}_B + i^{N} c_1 \ket{\mathcal{C}^{+}_{\alpha}}_B, \bigg)
\bigg]
\nonumber\\
&=&
\mathcal{N} \mathcal{N}_+^{\, 2} \frac{e^{-\alpha^2}}{\sqrt{n_C ! \, n_{C'} !}} \alpha^{N} \big[ 1 + (-1)^{N} \big] 
\bigg\{ \bigg[ \frac{1}{(\sqrt{2})^{n_{C'}}} c_0 + \frac{ e^{i N \pi /4}}{(\sqrt{2})^{n_{C}}} i^N c_1 \bigg]  \ket{\mathcal{C}^{+}_{\alpha}}_B
\nonumber\\
&&
+ \bigg[ \frac{1}{(\sqrt{2})^{n_{C'}}} i^N c_1 + \frac{ e^{i N \pi/4}}{(\sqrt{2})^{n_{C}}} c_0 \bigg]  \ket{\mathcal{C}^{+}_{i \alpha}}_B
\bigg\}.
\end{eqnarray}

%The outcome probability is
%\begin{eqnarray}
%P([0, n_C, 0, n_{C'}]) = | \langle 0, n_C, 0, n_{C'} \ket{\Psi}_{BAA'CC'} |^2. 
%\end{eqnarray}

\subsection{Absence of photons at three detectors }

There are four sorts of click pattern when only one of the detectors detects photons, namely, 
\begin{eqnarray}
[n_A, 0, 0, 0], ~~~~~~ [0, n_C,  0, 0], ~~~~~~ [0, 0, n_{A'}, 0], ~~~~~~ [0, 0, 0, n_{C'}]. 
\end{eqnarray}
The teleportation succeeds when the total number of detected photon $N = 4 k +2$ with $k$ a nonnegative integer. The results are summarized in Table~\ref{tab:ThreeZeroNoLoss} and
the detailed analysis is given as follows. 

(i) $n_A = N \ne 0$, clicks at mode $A$.

When the measurement outcome is $\vt{n} = [N, 0, 0, 0]$, the conditional output state is
\begin{eqnarray}
&& \langle N, 0, 0, 0 \ket{\Psi}_{BAA'CC'} 
\nonumber\\
&=&
\mathcal{N} \mathcal{N}_+^{\, 2} \big[ 1 + (-1)^N \big] e^{-\alpha^2}
\bigg[ \frac{\alpha^N}{\sqrt{N!}} \bigg( c_0 \ket{\mathcal{C}^{+}_{\alpha}}_B + i^N c_1 \ket{\mathcal{C}^{+}_{i \alpha}}_B \bigg)
%\nonumber\\
%&&
+ \frac{1}{\sqrt{N!}}\bigg( \frac{\beta^{*N}}{(\sqrt{2})^N} + \frac{\beta^N}{(\sqrt{2})^N} \bigg) \bigg( c_0 \ket{\mathcal{C}^{+}_{i \alpha}}_B + i^N c_1 \ket{\mathcal{C}^{+}_{\alpha}}_B \bigg)
\bigg]
\nonumber\\
&=&
\mathcal{N} \mathcal{N}_+^{\, 2} \big[ 1 + (-1)^N \big] e^{-\alpha^2} \frac{\alpha^N}{\sqrt{N!}}
\bigg[ \bigg( c_0 \ket{\mathcal{C}^{+}_{\alpha}}_B + i^N c_1 \ket{\mathcal{C}^{+}_{i \alpha}}_B \bigg)
+ \frac{e^{i N \pi/4} + e^{- i N \pi/4}} {(\sqrt{2})^N} \bigg( c_0 \ket{\mathcal{C}^{+}_{i \alpha}}_B + i^N c_1 \ket{\mathcal{C}^{+}_{\alpha}}_B \bigg)
\bigg]
\nonumber\\
&=&
\mathcal{N} \mathcal{N}_+^{\, 2} \big[ 1 + (-1)^N \big] e^{-\alpha^2} \frac{\alpha^N}{\sqrt{N!}}
\bigg\{ \bigg[c_0 + \frac{2 \cos(N \pi/4)}{(\sqrt{2})^N} i^N c_1 \bigg] \ket{\mathcal{C}^{+}_{\alpha}}_B
+ \bigg[ i^N c_1 + \frac{2 \cos(N \pi/4)}{(\sqrt{2})^N} c_0 \bigg] \ket{\mathcal{C}^{+}_{i \alpha}}_B
\bigg\}. 
\end{eqnarray}
%The measurement probability is
%\begin{eqnarray}
%P([N, 0, 0, 0]) = | \langle N, 0, 0, 0 \ket{\Psi}_{BAA'CC'} |^2. 
%\end{eqnarray}
Note that the measurement probability is nonzero only when $N$ is even. Assume that $N = 2m$ with $m \ge 1$, then the conditional output state can be written as 
\begin{eqnarray}
&& \langle 2m, 0, 0, 0 \ket{\Psi}_{BAA'CC'} 
\nonumber\\
&=&
2 \mathcal{N} \mathcal{N}_+^{\, 2} e^{-\alpha^2} \frac{\alpha^{2m}}{\sqrt{(2m)!}}
\bigg\{ \bigg[c_0 + \frac{ \cos(m \pi/2)}{2^{m-1}} (-1)^m c_1 \bigg] \ket{\mathcal{C}^{+}_{\alpha}}_B
+ \bigg[ (-1)^m c_1 + \frac{ \cos(m \pi/2)}{2^{m-1}} c_0 \bigg] \ket{\mathcal{C}^{+}_{i \alpha}}_B
\bigg\}
\nonumber\\
&=&
\left\{ 
\begin{array}{lc} 
2 \mathcal{N} \mathcal{N}_+^{\, 2} e^{-\alpha^2} \frac{\alpha^{2m}}{\sqrt{(2m)!}} 
\big( c_0 \ket{\mathcal{C}^{+}_{\alpha}}_B - c_1 \ket{\mathcal{C}^{+}_{i \alpha}}_B \big), & m ~ \text{is odd}; \\ 
2 \mathcal{N} \mathcal{N}_+^{\, 2} e^{-\alpha^2} \frac{\alpha^{2m}}{\sqrt{(2m)!}}
\bigg\{ \bigg[c_0 + \frac{(-1)^{m/2}}{2^{m-1}} c_1 \bigg] \ket{\mathcal{C}^{+}_{\alpha}}_B
+ \bigg[ c_1 + \frac{ (-1)^{m/2} }{2^{m-1}} c_0 \bigg] \ket{\mathcal{C}^{+}_{i \alpha}}_B
\bigg\}, & m ~ \text{is even}. 
\end{array} \right.
\end{eqnarray}
It is thus evident that the input state can be recovered by applying an encoded Pauli $Z$ correction when $m$ is odd, and the teleportation fails when $m$ is even.

(ii) $n_C = N \ne 0$, clicks at mode $C$.

When the measurement outcome is $\vt{n} = [0, N, 0, 0]$, the conditional state is
\begin{eqnarray}
&& \langle 0, N, 0, 0 \ket{\Psi}_{BAA'CC'} 
\nonumber\\
&=&
\mathcal{N} \mathcal{N}_+^{\, 2} \big[ 1 + (-1)^N \big] e^{-\alpha^2} \frac{\alpha^N}{\sqrt{N!}}
\bigg[ \bigg( c_0 \ket{\mathcal{C}^{+}_{\alpha}}_B + i^N c_1 \ket{\mathcal{C}^{+}_{i \alpha}}_B \bigg)
+ \frac{e^{i N \pi/4} + e^{- i N \pi/4}} {(\sqrt{2})^N} \bigg( c_0 \ket{\mathcal{C}^{+}_{i \alpha}}_B + i^N c_1 \ket{\mathcal{C}^{+}_{\alpha}}_B \bigg)
\bigg]
\nonumber\\
&=&
\langle N, 0, 0, 0 \ket{\Psi}_{BAA'CC'}. 
\end{eqnarray}
The conditional output state is the same as case (i). %when the PNR detector at mode $A$ detects $N$ photons. 

\

(iii) $n_{A'} = N \ne 0$, clicks at mode $A'$.

When the measurement outcome is $\vt{n} = [0, 0, N, 0]$, the conditional output state is
\begin{eqnarray}
&& \langle 0, 0, N, 0 \ket{\Psi}_{BAA'CC'} 
\nonumber\\
&=&
\mathcal{N} \mathcal{N}_+^{\, 2} \big[ 1 + (-1)^N \big] e^{-\alpha^2}  \frac{\alpha^N}{\sqrt{N!}}
\bigg[ \frac{(-1)^N + i^N}{(\sqrt{2})^N} \bigg( c_0 \ket{\mathcal{C}^{+}_{\alpha}}_B + i^N c_1 \ket{\mathcal{C}^{+}_{i \alpha}}_B \bigg)
+ (-1)^N e^{- i N \pi/4} \bigg( c_0 \ket{\mathcal{C}^{+}_{i \alpha}}_B + i^N c_1 \ket{\mathcal{C}^{+}_{\alpha}}_B \bigg)
\bigg]
\nonumber\\
&=&
\mathcal{N} \mathcal{N}_+^{\, 2} \big[ 1 + (-1)^N \big] e^{-\alpha^2}  \frac{\alpha^N}{\sqrt{N!}} i^N
\bigg[ \frac{1 + i^N}{(\sqrt{2})^N} \bigg( c_0 \ket{\mathcal{C}^{+}_{\alpha}}_B + i^N c_1 \ket{\mathcal{C}^{+}_{i \alpha}}_B \bigg)
+ e^{ i N \pi/4} \bigg( c_0 \ket{\mathcal{C}^{+}_{i \alpha}}_B + i^N c_1 \ket{\mathcal{C}^{+}_{\alpha}}_B \bigg)
\bigg]
\nonumber\\
&=&
\mathcal{N} \mathcal{N}_+^{\, 2} \big[ 1 + (-1)^N \big] e^{-\alpha^2}  \frac{\alpha^N}{\sqrt{N!}} i^N
\bigg\{ \bigg[ e^{ i N \pi/4} i^N c_1 +  \frac{1 + i^N}{(\sqrt{2})^N} c_0 \bigg]  \ket{\mathcal{C}^{+}_{\alpha}}_B
+ \bigg[ e^{ i N \pi/4} c_0 +  \frac{1 + i^N}{(\sqrt{2})^N} i^N c_1 \bigg]  \ket{\mathcal{C}^{+}_{i \alpha}}_B
\bigg\}. 
\end{eqnarray}
%The measurement probability is
%\begin{eqnarray}
%P([0, 0, N, 0]) = | \langle 0, 0, N, 0 \ket{\Psi}_{BAA'CC'} |^2. 
%\end{eqnarray}
Note that the probability is nonzero only when $N$ is even. Assume that $N = 2m$ with $m \ge 1$, then the conditional output state can be written as
\begin{eqnarray}
&& \langle 0, 0, 2m, 0 \ket{\Psi}_{BAA'CC'} 
\nonumber\\
&=&
2 \mathcal{N} \mathcal{N}_+^{\, 2} e^{-\alpha^2} \frac{(-1)^m \alpha^{2m} }{\sqrt{(2m)!}}
\bigg\{ \bigg[ i^m (-1)^m c_1 +  \frac{1 + (-1)^m}{2^m} c_0 \bigg]  \ket{\mathcal{C}^{+}_{\alpha}}_B
+ \bigg[ i^m c_0 +  \frac{1 + (-1)^m}{2^m} (-1)^m c_1 \bigg]  \ket{\mathcal{C}^{+}_{i \alpha}}_B
\bigg\}
\nonumber\\
&=&
\left\{ 
\begin{array}{lc} 
2 \mathcal{N} \mathcal{N}_+^{\, 2} e^{-\alpha^2} \frac{\alpha^{2m}}{\sqrt{(2m)!}} (-i)^m
\big( c_0 \ket{\mathcal{C}^{+}_{i \alpha}}_B - c_1 \ket{\mathcal{C}^{+}_{\alpha}}_B \big), & m ~ \text{is odd}; \\ 
2 \mathcal{N} \mathcal{N}_+^{\, 2} e^{-\alpha^2} \frac{\alpha^{2m}}{\sqrt{(2m)!}}
\bigg\{ \bigg[ i^m c_1 + \frac{1}{2^{m-1}} c_0 \bigg] \ket{\mathcal{C}^{+}_{\alpha}}_B
+ \bigg[ i^m c_0 + \frac{1}{2^{m-1}} c_1 \bigg] \ket{\mathcal{C}^{+}_{i \alpha}}_B
\bigg\}, & m ~ \text{is even}. 
\end{array} \right.
\end{eqnarray}
It is thus evident that the input state can be recovered by first applying an encoded Pauli $X$ correction and then a Pauli $Z$ correction when $m$ is odd, and the teleportation fails when 
$m$ is even.

(iv) $n_{C'} = N \ne 0$, clicks at mode $C'$.

When the measurement outcome is $\vt{n} = [0, 0, 0, N]$, the conditional output state is
\begin{eqnarray}
&& \langle 0, 0, 0, N \ket{\Psi}_{BAA'CC'} 
\nonumber\\
&=&
\mathcal{N} \mathcal{N}_+^{\, 2} \big[ 1 + (-1)^N \big] e^{-\alpha^2}  \frac{\alpha^N}{\sqrt{N!}}
\bigg[ \frac{1 + i^N}{(\sqrt{2})^N} \bigg( c_0 \ket{\mathcal{C}^{+}_{\alpha}}_B + i^N c_1 \ket{\mathcal{C}^{+}_{i \alpha}}_B \bigg)
+ e^{i N \pi/4} \bigg( c_0 \ket{\mathcal{C}^{+}_{i \alpha}}_B + i^N c_1 \ket{\mathcal{C}^{+}_{\alpha}}_B \bigg)
\bigg]
\nonumber\\
&=&
\mathcal{N} \mathcal{N}_+^{\, 2} \big[ 1 + (-1)^N \big] e^{-\alpha^2}  \frac{\alpha^N}{\sqrt{N!}}
\bigg\{ \bigg[ e^{i N \pi/4} i^N c_1 +  \frac{1 + i^N}{(\sqrt{2})^N} c_0 \bigg]  \ket{\mathcal{C}^{+}_{\alpha}}_B
+ \bigg[ e^{i N \pi/4} c_0 +  \frac{1 + i^N}{(\sqrt{2})^N} i^N c_1 \bigg]  \ket{\mathcal{C}^{+}_{i \alpha}}_B
\bigg\}. 
\end{eqnarray}
%The measurement probability is
%\begin{eqnarray}
%P([0, 0, 0, N]) = | \langle 0, 0, 0, N \ket{\Psi}_{BAA'CC'} |^2. 
%\end{eqnarray}
Note that the probability is nonzero only when $N$ is even. Assume that $N = 2m$ with $m \ge 1$, then the conditional output state can be written as
\begin{eqnarray}
&& \langle 0, 0, 0, 2m \ket{\Psi}_{BAA'CC'} 
\nonumber\\
&=&
2 \mathcal{N} \mathcal{N}_+^{\, 2} e^{-\alpha^2} \frac{\alpha^{2m}}{\sqrt{(2m)!}}
\bigg\{ \bigg[ i^m (-1)^m c_1 +  \frac{1 + (-1)^m}{2^m} c_0 \bigg]  \ket{\mathcal{C}^{+}_{\alpha}}_B
+ \bigg[ i^m c_0 +  \frac{1 + (-1)^m}{2^m} (-1)^m c_1 \bigg]  \ket{\mathcal{C}^{+}_{i \alpha}}_B
\bigg\}
\nonumber\\
&=&
\left\{ 
\begin{array}{lc} 
2 \mathcal{N} \mathcal{N}_+^{\, 2} e^{-\alpha^2} \frac{\alpha^{2m}}{\sqrt{(2m)!}} i^m
\big( c_0 \ket{\mathcal{C}^{+}_{i \alpha}}_B - c_1 \ket{\mathcal{C}^{+}_{\alpha}}_B \big), & m ~ \text{is odd}; \\ 
2 \mathcal{N} \mathcal{N}_+^{\, 2} e^{-\alpha^2} \frac{\alpha^{2m}}{\sqrt{(2m)!}}
\bigg\{ \bigg[ i^m c_1 + \frac{1}{2^{m-1}} c_0 \bigg] \ket{\mathcal{C}^{+}_{\alpha}}_B
+ \bigg[ i^m c_0 + \frac{1}{2^{m-1}} c_1 \bigg] \ket{\mathcal{C}^{+}_{i \alpha}}_B
\bigg\}, & m ~ \text{is even}. 
\end{array} \right.
\end{eqnarray}
It is thus evident that the input state can be recovered by first applying an encoded Pauli $X$ correction and then a Pauli $Z$ correction when $m$ is odd, and the teleportation fails  
when $m$ is even. 

In summary, when one of the PNR detectors detects $4k+2$ photons with $k$ an nonnegative integer, the teleportation succeeds; while when it detects $4k$ photons, the 
teleportation fails. 

Finally, when all PNR detectors detect vacuum outputs, the teleportation fails. The conditional output state is
\begin{eqnarray}
\langle 0, 0, 0, 0 \ket{\Psi}_{BAA'CC'} = 4 \mathcal{N} \mathcal{N}_+^{\, 2} e^{-\alpha^2} (c_0 + c_1) \big( \ket{\mathcal{C}^{+}_{\alpha}}_B + \ket{\mathcal{C}^{+}_{i \alpha}}_B \big),
\end{eqnarray}
showing that the normalized output state is independent of the input state. The measurement probability is
\begin{eqnarray}
P([0, 0, 0, 0]) = | \langle 0, 0, 0, 0 \ket{\Psi}_{BAA'CC'} |^2 
= 32 \mathcal{N}^{\, 2} \mathcal{N}_+^{\, 4} e^{- 2 \alpha^2} |c_0 + c_1|^2 \big(1 + \langle  \mathcal{C}^{+}_{\alpha} \ket{\mathcal{C}^{+}_{i \alpha}}  \big). 
\end{eqnarray}

\subsection{Teleportation failure probability and phase flip error rate}

\begin{table}[h]
\caption{ Click patterns resulting in failure of teleportation.  } %title of the table
\centering % centering table
\begin{tabular}{@{}c|c|c|c}%\toprule
\hline \hline
  & Detect four vacuum outputs  & \makecell{Detect three vacuum outputs \\ ($k \ge 1$) } & \makecell{Detect two vacuum outputs \\ ($n_1, n_2 \ge 1$, $n_1 + n_2$ is even) } \\
    \hline
\makecell{Click \\ pattern} & $[0, 0, 0, 0]$  & \makecell{$[4k, 0, 0, 0]$ \\ $[0, 4k, 0, 0]$ \\ $[0, 0, 4k, 0]$ \\$[0, 0, 0, 4k]$ }
&  \makecell{$[n_1, 0, n_2, 0]$ \\ $[n_1, 0, 0, n_2]$ \\  $[0, n_1, n_2, 0]$ \\ $[0, n_1, 0, n_2]$} \\
\hline \hline
\end{tabular}
\label{tab:FailurePatterns}
\end{table}

The click patterns that result in failure of teleportation are listed in Table~\ref{tab:FailurePatterns}. The failure probability is the sum of the measurement probabilities
of all these click patterns. Note that the failure probability depends on the input state. We plot the failure probability as a function of the cat-state amplitude 
$\alpha$ in Fig.~\ref{fig:FailProb} for several typical input states. We can see that for most input states the failure probability approaches to one when $\alpha \rightarrow 0$, however
it decreases exponentially when $\alpha \rightarrow \infty$. For comparison, we also plot the teleportation failure probability for the two-component cat code in Fig.~\ref{fig:FailProb}.
The relation between the failure probability and the cat-tate amplitude $\alpha$ for the two-component cat code is given by (without photon loss) \cite{PhysRevLett.100.030503}
\begin{eqnarray}
P_{\rm fail} = \frac{2}{1 + e^{2\alpha^2}}
\end{eqnarray}
for input state $\ket{+_L}$. 
It is evident that for a given amplitude $\alpha$, the failure probability for the four-component cat code is higher than that for the two-component cat code. To keep the failure probability
at the same level, the amplitude of the four-component cat code has to be approximately twice as big as that of the two-component cat code. 

\begin{figure}
\includegraphics[width=0.5 \columnwidth]{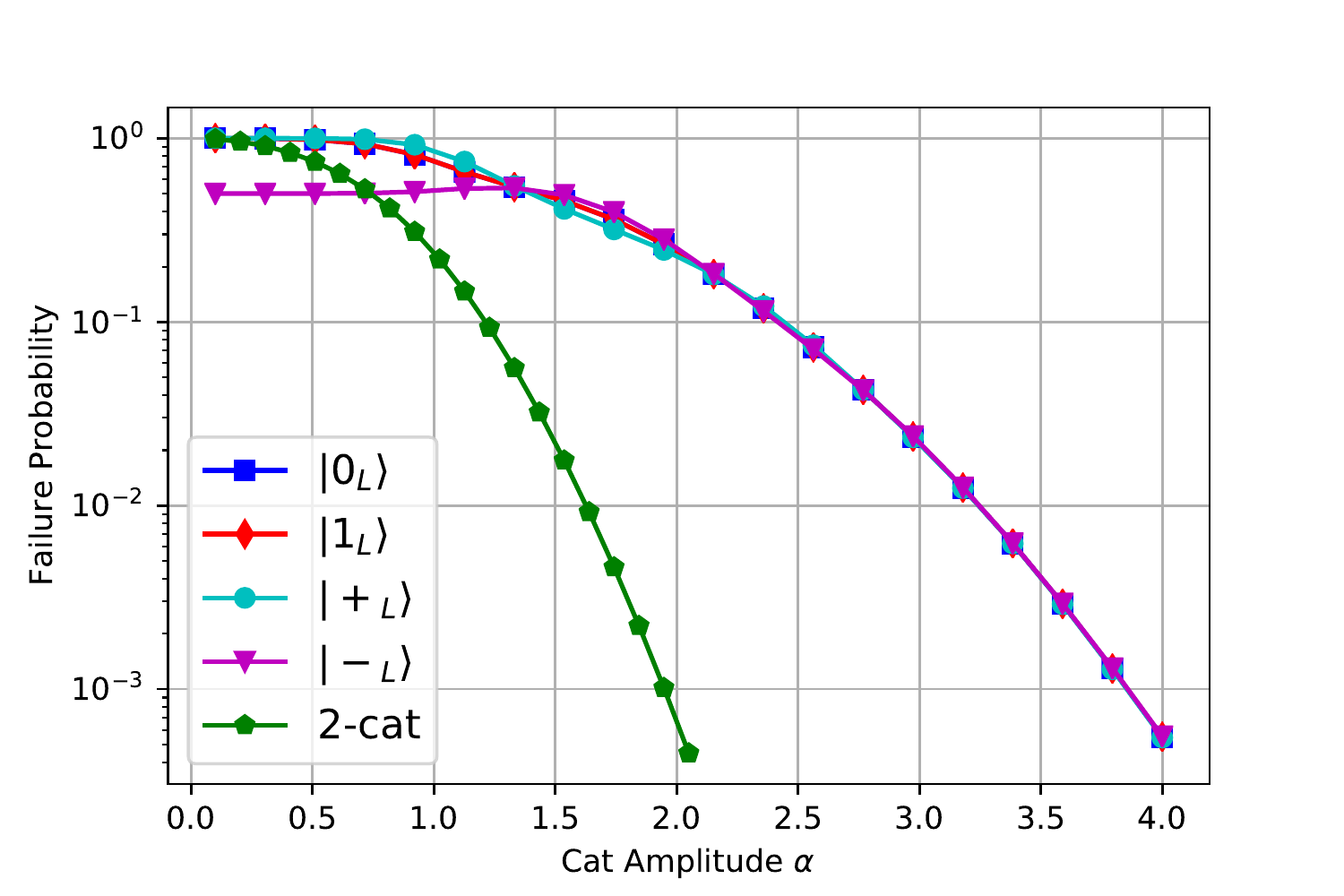}
\caption{ Teleportation failure probability (located error rate).  The upper four curves represent the failure probabilities of four-component cat code for input 
states $\ket{0_L}$ (blue square), $\ket{1_L}$ (red diamond), $\ket{+_L}$ (cyan circle), and $\ket{-_L}$ (magenta triangle), respectively. As a comparison, the failure 
probability for the two-component cat code is also plotted (green pentagon). }
\label{fig:FailProb}
\end{figure}

The photon loss induces a phase flip error for the two-component cat code, and the phase flip error rate is given by \cite{PhysRevLett.100.030503}
\begin{eqnarray}
P_{\rm pf} = \frac{1}{2} \bigg[ 1 + \frac{\sinh (2 \epsilon - 1) \alpha^2 }{\sinh \alpha^2} \bigg]. 
\end{eqnarray} 
Similarly, the photon loss also induces a phase flip error for the four-component cat code, although the single photon loss can be corrected. 
We plot the relation between the phase flip error rate and the photon loss for the two-component and four-component cat codes by requiring that the failure probabilities are at the same level
 in Fig.~\ref{fig:MapLossThreshold}. In particular, Fig.~\ref{fig:MapLossThreshold} establishes a link between the photon loss for the two-component cat code and that 
for the four-component cat code for fixed failure probability $P_{\rm fail}$ and phase flip error rate $P_{\rm pf}$. 
Given the photon loss threshold for two-component cat code, one can thus estimate the fault-tolerant threshold for the four-component cat code 
by using this relation.

\section{Correcting single photon loss in encoded state}\label{sec:PhotonLoss}

In this section, we first briefly review the mechanism of correcting single photon loss using four-component cat code and then describe in detail a concrete procedure to correct single 
photon loss via teleportation. 

The four-component cat code is specifically designed to correct single photon loss. This can be understood as follows. Define four states $\ket{\psi_{\alpha}^{(k)}}$ with $k = 0, 1, 2, 3$ as
\begin{eqnarray}
\ket{\psi_{\alpha}^{(0)}} &=& c_0 \ket{\mathcal{C}^{+}_{\alpha}} + c_1 \ket{\mathcal{C}^{+}_{i \alpha}}, ~~~~~~~~ %\nonumber\\
\ket{\psi_{\alpha}^{(1)}} = c_0 \ket{\mathcal{C}^{-}_{\alpha}} + i c_1 \ket{\mathcal{C}^{-}_{i \alpha}}, \nonumber\\
\ket{\psi_{\alpha}^{(2)}} &=& c_0 \ket{\mathcal{C}^{+}_{\alpha}} - c_1 \ket{\mathcal{C}^{+}_{i \alpha}}, ~~~~~~~~ %\nonumber\\
\ket{\psi_{\alpha}^{(3)}} = c_0 \ket{\mathcal{C}^{-}_{\alpha}} - i c_1 \ket{\mathcal{C}^{-}_{i \alpha}}. 
\end{eqnarray}
Note that $\ket{\psi_{\alpha}^{(0)}}$ and $\ket{\psi_{\alpha}^{(2)}}$ contain even numbers of photon, and therefore they are in the code subspace; 
while $\ket{\psi_{\alpha}^{(1)}}$ and $\ket{\psi_{\alpha}^{(3)}}$ contain only odd numbers of photon, and therefore they are outside the code subspace. 
If a single photon is lost, these states are transformed as \cite{PhysRevLett.111.120501}
\begin{eqnarray}
\ket{\psi_{\alpha}^{(0)}} \rightarrow \ket{\psi_{\alpha}^{(1)}}, ~~~~~~ \ket{\psi_{\alpha}^{(1)}} \rightarrow \ket{\psi_{\alpha}^{(2)}}, ~~~~~~
\ket{\psi_{\alpha}^{(2)}} \rightarrow \ket{\psi_{\alpha}^{(3)}}, ~~~~~~ \ket{\psi_{\alpha}^{(3)}} \rightarrow \ket{\psi_{\alpha}^{(0)}},
\end{eqnarray}
or in a compact form:
\begin{eqnarray}
\ket{\psi_{\alpha}^{(k)}} \rightarrow \frac{ \op{a} \ket{\psi_{\alpha}^{(k)}} }{|| \op{a} \ket{\psi_{\alpha}^{(k)}} ||} = \ket{\psi_{\alpha}^{[(k+1) \, {\rm mod} \, 4]}}. 
\end{eqnarray}
Without loss of generality, we assume that the initial state is $\ket{\psi_{\alpha}^{(0)}}$. 
To check whether there is single photon loss after a lossy channel, we can measure the 
parity operator $\op{\Pi} = e^{i \pi \op{a}^\dag \op{a}}$ because $\langle \psi_{\alpha}^{(k)} | \op{\Pi} \ket{\psi_{\alpha}^{(k)}} = (-1)^k$. Suppose we perform a quantum
nondemolition parity measurement: if the parity is 1, we conclude that there is no single photon loss and no further operations need to be applied; if the parity is $-1$, we conclude 
that there is single photon loss, namely, $\ket{\psi_{\alpha}^{(0)}} \rightarrow \ket{\psi_{\alpha}^{(1)}}$, we then apply a recovery map to bring the state $\ket{\psi_{\alpha}^{(1)}}$
to the initial state $\ket{\psi_{\alpha}^{(0)}}$. The above error correction procedure works only when one photon is lost. If two or more photons are lost, the 
error correction procedure results in a wrong state. This is the intrinsic limitation of the four-component cat code. However, if the photon loss rate is sufficiently low, 
the error correction can significantly suppress the logical error rate.  

To complete the error correction, one needs to overcome two challenges: (1) to check the photon number without destroying the state, and (2) to efficiently implement the recovery
operation. Here we show that these two challenges can be overcome simultaneously by using teleportation. Assume that an encoded qubit with state 
$\ket{\psi_{\alpha}^{(0)}}$ goes through a lossy channel, and then is fed into the teleportation circuit shown in Fig.~\ref{fig:4-port-TP}. After the lossy channel,
the state could either be $\ket{\psi_{\alpha}^{(0)}}$, corresponding to no single photon loss; or $\ket{\psi_{\alpha}^{(1)}}$, corresponding to single photon loss. If there is no 
photon loss, the state is simply teleported to the output, as discussed in Appendix~\ref{sec:GoodTeleportation}; if there is single photon loss, the input of the teleportation circuit becomes 
%\textcolor{red}{\bf [normalization?]}
\begin{eqnarray}
\ket{\psi_{\rm in}^{\, \prime}}_A = \ket{\psi_{\alpha}^{(1)}}_A = c_0 \ket{\mathcal{C}^{-}_{\alpha}}_A + i c_1 \ket{\mathcal{C}^{-}_{i \alpha}}_A,
\end{eqnarray}
where we have explicitly added the subscript ``$A$" to indicate the state at mode $A$. 
The overall input state before teleportation is
\begin{eqnarray}
\ket{\Psi_{\rm in}^{\, \prime}} &=& \ket{\psi_{\rm in}^{\, \prime}}_A \otimes \ket{\Phi_0}_{BC} \otimes \ket{0, 0}_{A' C'}
\nonumber\\
&=&
\mathcal{N} \left[ \left( c_0 \ket{\mathcal{C}^{-}_{\alpha}}_A \ket{\mathcal{C}^{+}_{\alpha}}_C 
+ i c_1 \ket{\mathcal{C}^{-}_{i \alpha}}_A \ket{\mathcal{C}^{+}_{\alpha}}_C \right) \ket{\mathcal{C}^{+}_{\alpha}}_B 
+ \left( c_0 \ket{\mathcal{C}^{-}_{\alpha}}_A \ket{\mathcal{C}^{+}_{i \alpha}}_C 
+ i c_1 \ket{\mathcal{C}^{-}_{i \alpha}}_A \ket{\mathcal{C}^{+}_{i \alpha}}_C \right) \ket{\mathcal{C}^{+}_{i \alpha}}_B \right] \otimes \ket{0, 0}_{A' C'}. 
\end{eqnarray}
The teleportation circuit transforms the four components of the overall input state as follows: 
\begin{eqnarray*}
 \ket{\mathcal{C}^{-}_{\alpha}}_A \ket{\mathcal{C}^{+}_{\alpha}}_C \ket{0}_{A'} \ket{0}_{C'} 
 &\longrightarrow&
\bigg| 0, \alpha, -\frac{\alpha}{\sqrt{2}}, \frac{\alpha}{\sqrt{2}} \bigg\rangle + \bigg| \alpha, 0, \frac{i \alpha}{\sqrt{2}}, \frac{i \alpha}{\sqrt{2}} \bigg\rangle
- \bigg| - \alpha, 0, - \frac{i \alpha}{\sqrt{2}}, - \frac{i \alpha}{\sqrt{2}} \bigg\rangle - \bigg| 0, - \alpha, \frac{\alpha}{\sqrt{2}}, - \frac{\alpha}{\sqrt{2}} \bigg\rangle,
\nonumber\\
 \ket{\mathcal{C}^{-}_{i \alpha}}_A \ket{\mathcal{C}^{+}_{i \alpha}}_C \ket{0}_{A'} \ket{0}_{C'} 
 &\longrightarrow&
\bigg| 0, i \alpha, -\frac{i \alpha}{\sqrt{2}}, \frac{i \alpha}{\sqrt{2}} \bigg\rangle + \bigg| i \alpha, 0, - \frac{\alpha}{\sqrt{2}}, - \frac{\alpha}{\sqrt{2}} \bigg\rangle
- \bigg| - i \alpha, 0, \frac{\alpha}{\sqrt{2}}, \frac{\alpha}{\sqrt{2}} \bigg\rangle - \bigg| 0, -i \alpha, \frac{i \alpha}{\sqrt{2}}, - \frac{i \alpha}{\sqrt{2}} \bigg\rangle,
\nonumber\\
 \ket{\mathcal{C}^{-}_{\alpha}}_A \ket{\mathcal{C}^{+}_{i \alpha}}_C \ket{0}_{A'} \ket{0}_{C'} 
 &\longrightarrow&
\bigg| \frac{\beta^*}{\sqrt{2}}, \frac{\beta}{\sqrt{2}},  0, \beta \bigg\rangle+ \bigg| \frac{\beta}{\sqrt{2}}, \frac{\beta^*}{\sqrt{2}}, - \beta^*, 0 \bigg\rangle
- \bigg| - \frac{\beta}{\sqrt{2}}, - \frac{\beta^*}{\sqrt{2}}, \beta^*, 0 \bigg\rangle - \bigg| - \frac{\beta^*}{\sqrt{2}}, - \frac{\beta}{\sqrt{2}}, 0, - \beta \bigg\rangle,
\nonumber\\
\ket{\mathcal{C}^{-}_{i \alpha}}_A \ket{\mathcal{C}^{+}_{\alpha}}_C \ket{0}_{A'} \ket{0}_{C'} 
 &\longrightarrow&
\bigg| - \frac{\beta^*}{\sqrt{2}}, \frac{\beta}{\sqrt{2}},  - \beta, 0 \bigg\rangle + \bigg| \frac{\beta}{\sqrt{2}}, - \frac{\beta^*}{\sqrt{2}}, 0, - \beta^* \bigg\rangle
- \bigg| - \frac{\beta}{\sqrt{2}}, \frac{\beta^*}{\sqrt{2}}, 0, \beta^* \bigg\rangle - \bigg| \frac{\beta^*}{\sqrt{2}}, - \frac{\beta}{\sqrt{2}}, \beta, 0 \bigg\rangle,
\end{eqnarray*}
where have defined $\beta = \alpha e^{i \pi/4}$, and omitted the subscripts to simplify the notation and chosen the order of the modes as $``ACA'C' "$. 
By combining these results, we find that the output state after the teleportation circuit (before photon number measurement) is 
\begin{eqnarray}\label{eq:OutPutLossState}
&& \ket{\Psi}_{BAA'CC'} = \op{U}_{A'C'} \op{U}_{A'}(\pi/2) \op{U}_{AA'} \op{U}_{CC'} \op{U}_{AC} \ket{\Psi_{\rm in}}
\nonumber\\
&\propto&
\bigg[ \bigg( \bigg| 0, \alpha, - \frac{\alpha}{\sqrt{2}}, \frac{\alpha}{\sqrt{2}} \bigg\rangle 
- \bigg| 0, - \alpha, \frac{\alpha}{\sqrt{2}}, - \frac{\alpha}{\sqrt{2}} \bigg\rangle \bigg) c_0 \ket{\mathcal{C}^{+}_{\alpha}}_B
%\nonumber\\
%&&
+ \bigg( \bigg| 0, i \alpha, - \frac{i \alpha}{\sqrt{2}}, \frac{i \alpha}{\sqrt{2}} \bigg\rangle 
- \bigg|0, - i \alpha, \frac{i \alpha}{\sqrt{2}}, - \frac{i \alpha}{\sqrt{2}} \bigg\rangle \bigg) i c_1 \ket{\mathcal{C}^{+}_{i \alpha}}_B \bigg]
\nonumber\\
&&
+ \bigg[ \bigg( \bigg| \alpha, 0, \frac{i \alpha}{\sqrt{2}}, \frac{i \alpha}{\sqrt{2}} \bigg\rangle 
- \bigg| - \alpha, 0, - \frac{i \alpha}{\sqrt{2}}, - \frac{i \alpha}{\sqrt{2}} \bigg\rangle \bigg) c_0 \ket{\mathcal{C}^{+}_{\alpha}}_B 
%\nonumber\\
%&&
+ \bigg( \bigg| i \alpha, 0, - \frac{\alpha}{\sqrt{2}}, - \frac{\alpha}{\sqrt{2}} \bigg\rangle 
- \bigg| - i \alpha, 0, \frac{\alpha}{\sqrt{2}}, \frac{\alpha}{\sqrt{2}} \bigg\rangle \bigg) i c_1 \ket{\mathcal{C}^{+}_{i \alpha}}_B \bigg]
\nonumber\\
&&
+ \bigg[ \bigg( \bigg| \frac{\beta}{\sqrt{2}}, - \frac{\beta^*}{\sqrt{2}}, 0, - \beta^* \bigg\rangle 
- \bigg| - \frac{\beta}{\sqrt{2}}, \frac{\beta^*}{\sqrt{2}}, 0, \beta^* \bigg\rangle \bigg) i c_1 \ket{\mathcal{C}^{+}_{\alpha}}_B 
%\nonumber\\
%&&
+ \bigg( \bigg| \frac{\beta^*}{\sqrt{2}}, \frac{\beta}{\sqrt{2}},  0, \beta \bigg\rangle 
- \bigg| - \frac{\beta^*}{\sqrt{2}}, - \frac{\beta}{\sqrt{2}}, 0, - \beta \bigg\rangle \bigg) c_0 \ket{\mathcal{C}^{+}_{i \alpha}}_B \bigg]
\nonumber\\
&&
+ \bigg[ \bigg( \bigg| - \frac{\beta^*}{\sqrt{2}}, \frac{\beta}{\sqrt{2}}, - \beta, 0 \bigg\rangle 
- \bigg| \frac{\beta^*}{\sqrt{2}}, - \frac{\beta}{\sqrt{2}}, \beta, 0 \bigg\rangle \bigg) i c_1 \ket{\mathcal{C}^{+}_{\alpha}}_B 
%\nonumber\\
%&&
+ \bigg( \bigg| \frac{\beta}{\sqrt{2}}, \frac{\beta^*}{\sqrt{2}}, - \beta^*, 0 \bigg\rangle 
- \bigg| - \frac{\beta}{\sqrt{2}}, - \frac{\beta^*}{\sqrt{2}}, \beta^*, 0 \bigg\rangle \bigg) c_0 \ket{\mathcal{C}^{+}_{i \alpha}}_B \bigg],
\nonumber\\
\end{eqnarray}
where the proportionality factor is $\mathcal{N} \mathcal{N}_+ \mathcal{N}_-$. 

Four PNR detectors are used to count the number of photons at modes $A, C, A'$ and $C'$, giving click patterns
$\vt{n} = [n_A, n_C, n_{A'}, n_{C'}]$. 
We will derive the conditional output state and its occurring probability for a particular click pattern, from which one can explicitly see whether the input state can be recovered after the single photon loss. 
The results are summarized in Tables~\ref{tab:OneZeroLossState} and \ref{tab:TwoZeroLossState}.

\begin{table}[h]
\caption{ {\bf Pauli corrections for error correction}: click patterns with one vacuum output, single photon loss.  } %title of the table
\centering % centering table
\begin{tabular}{@{} l |c|c|c|c}%\toprule
\hline %\hline
   & ~~~$[0, n_C, n_{A'}, n_{C'}]$~~~ & ~~~$[n_A, 0, n_{A'}, n_{C'}]$~~~  & ~~~$[n_A, n_C, 0, n_{C'}]$~~~ & ~~~$[n_A, n_C, n_{A'}, 0]$~~~ \\
    \hline
   $N = 4k - 1$ & $I$ & $I$ & $X$ & $X$ \\
   \hline
   $N = 4k + 1$ & $Z$ & $Z$ & $ZX$ & $ZX$ \\
\hline %\hline
\end{tabular}
\label{tab:OneZeroLossState}
\end{table}

\begin{table}[h]
\caption{ {\bf Pauli corrections for error correction}: click patterns with two vacuum outputs, single photon loss.   } %title of the table
\centering % centering table
\begin{tabular}{@{} l |c|c|c|c|c|c}%\toprule
\hline %\hline
   & ~~$[0, 0, n_{A'}, n_{C'}]$~~ & ~~$[n_A, n_C, 0, 0]$~~  & ~~$[n_A, 0, n_{A'}, 0]$~~ & ~~$[n_A, 0, 0, n_{C'}]$~~ & ~~$[0, n_C, n_{A'}, 0]$~~ & ~~$[0, n_C, 0, n_{C'}]$~~ \\
    \hline
   $N = 4k - 1$ & $I$ & $X$ & failed & failed & failed & failed \\
   \hline
   $N = 4k + 1$ & $Z$ & $ZX$ & failed & failed & failed & failed \\
\hline %\hline
\end{tabular}
\label{tab:TwoZeroLossState}
\end{table}

\subsection{Absence of photon at one detector}

We first consider situations where only one detector detects no photons and each of the other three detectors registers at least one photon. 
This includes four sorts of click pattern: 
\begin{eqnarray*}
[0, n_C, n_{A'}, n_{C'}], ~~~~~~ [n_A, 0, n_{A'}, n_{C'}], ~~~~~~ [n_A, n_{C}, 0, n_{C'}], ~~~~~~ [n_A, n_{C}, n_{A'}, 0]. 
\end{eqnarray*}
It is found that the initial state can be recovered for all these click patterns by applying appropriate Pauli corrections. This is summarized in Table~\ref{tab:OneZeroLossState}.
The detailed analysis is given as follows. 

(i) $n_{A} =0$, no click at mode $A$.  

The detector at mode $A$ detects no photons, namely, the measurement outcome is $\vt{n} = [0, n_C, n_{A'}, n_{C'}]$. The conditional output state is
\begin{eqnarray}
\langle0, n_C, n_{A'}, n_{C'} \ket{\Psi}_{BAA'CC'} =
\mathcal{N} \mathcal{N}_+ \mathcal{N}_- \frac{e^{- \alpha^2}}{\sqrt{n_C ! \, n_{A'} ! \, n_{C'} !}} \frac{ (-1)^{n_{A'}} \alpha^{N}}{(\sqrt{2})^{n_{A'} + n_{C'}}} 
\big[1 - (-1)^N \big] \bigg( c_0 \ket{\mathcal{C}^{+}_{\alpha}}_B + i^{N+1} c_1 \ket{\mathcal{C}^{+}_{i \alpha}}_B \bigg). 
\end{eqnarray} 
%where $N = n_C + n_{A'}  + n_{C'}$ is the total number of detected photons. 
This implies the (unnormalized) output state at mode $B$ is 
\begin{eqnarray}\label{eq:Case1output}
\ket{\psi_1} =  c_0 \ket{\mathcal{C}^{+}_{\alpha}}_B + i^{N+1} c_1 \ket{\mathcal{C}^{+}_{i \alpha}}_B, 
\end{eqnarray}
and the measurement probability is %\textcolor{red}{ [missing a normalization factor of the output state] }
\begin{eqnarray}
P([0, n_C, n_{A'}, n_{C'}]) = \mathcal{N}^{\, 2} \mathcal{N}_+^{\, 2} \mathcal{N}_-^{\, 2} \frac{e^{- 2 \alpha^2}}{ n_C ! \, n_{A'} ! \, n_{C'} !} \frac{\alpha^{2 N}}{2^{n_{A'} + n_{C'}}} \big[1 - (-1)^N \big]^2 
|\langle \psi_1 \ket{\psi_1}|^2. 
\end{eqnarray}

(ii) $n_C = 0$, no click at mode $C$.

The detector at mode $C$ detects no photons, namely, the measurement outcome is $\vt{n} = [n_A, 0, n_{A'}, n_{C'}]$. The conditional output state is
\begin{eqnarray}
\langle n_A, 0, n_{A'}, n_{C'} \ket{\Psi}_{BAA'CC'} =
\mathcal{N} \mathcal{N}_+  \mathcal{N}_- \frac{e^{- \alpha^2}}{\sqrt{n_A ! \, n_{A'} ! \, n_{C'} !}} \frac{i^{n_{A'} + n_{C'}} \alpha^{N}}{(\sqrt{2})^{n_{A'} + n_{C'}}} 
\big[1 - (-1)^N \big] \bigg( c_0 \ket{\mathcal{C}^{+}_{\alpha}}_B + i^{N+1} c_1 \ket{\mathcal{C}^{+}_{i \alpha}}_B \bigg). 
\end{eqnarray} 
%where $N = n_A + n_{A'}  + n_{C'}$ is the total number of detected photons. 
This implies the (unnormalized) output state at mode $B$ is
\begin{eqnarray}\label{eq:Case1output}
\ket{\psi_2} =  c_0 \ket{\mathcal{C}^{+}_{\alpha}}_B + i^{N+1} c_1 \ket{\mathcal{C}^{+}_{i \alpha}}_B,
\end{eqnarray}
and the measurement probability is 
\begin{eqnarray}
P([n_A, 0, n_{A'}, n_{C'}]) = \mathcal{N}^{\, 2} \mathcal{N}_+^{\, 2} \mathcal{N}_-^{\, 2} \frac{e^{- 2 \alpha^2}}{n_{A} ! \, n_{A'} ! \, n_{C'} !} \frac{\alpha^{2N}}{2^{n_{A'} + n_{C'}}} \big[1 - (-1)^N \big]^2
|\langle \psi_2 \ket{\psi_2}|^2. 
\end{eqnarray}

(iii) $n_{A'} = 0$, no click at mode $A'$.

The detector at mode $A'$ detects no photons, namely, the measurement outcome is $\vt{n} = [n_A, n_C, 0, n_{C'}]$. The conditional output state is
\begin{eqnarray}
\langle n_A, n_C, 0, n_{C'} \ket{\Psi}_{BAA'CC'} =
\mathcal{N} \mathcal{N}_+ \mathcal{N}_- \frac{e^{- \alpha^2}}{\sqrt{n_A ! \, n_{C} ! \, n_{C'} !}} \frac{ e^{- i \pi (n_A - n_{C} - n_{C'})/4 } \alpha^{N}}{(\sqrt{2})^{n_{A} + n_{C}}} 
\big[1 - (-1)^N \big] \bigg( c_0 \ket{\mathcal{C}^{+}_{i \alpha}}_B + i^{N+1} c_1 \ket{\mathcal{C}^{+}_{\alpha}}_B \bigg)
\nonumber\\
\end{eqnarray} 
%where $N = n_A + n_{C}  + n_{C'}$ is the total number of detected photons. 
This implies the (unnormalized) output state at mode $B$ is
\begin{eqnarray}\label{eq:Case1output}
\ket{\psi_3} =  c_0 \ket{\mathcal{C}^{+}_{i \alpha}}_B + i^{N+1} c_1 \ket{\mathcal{C}^{+}_{\alpha}}_B,
\end{eqnarray}
and the measurement probability is 
\begin{eqnarray}
P([n_A, n_{C}, 0, n_{C'}]) = \mathcal{N}^{\, 2} \mathcal{N}_+^{\, 2} \mathcal{N}_-^{\, 2} \frac{e^{- 2 \alpha^2}}{n_{A} ! \, n_{C} ! \, n_{C'} !} \frac{\alpha^{2N}}{2^{n_{A} + n_{C}}} \big[1 - (-1)^N \big]^2 
|\langle \psi_3 \ket{\psi_3}|^2.
\end{eqnarray}

(iv) $n_{C'} = 0$, no click at mode $C'$.

The detector at mode $C'$ detects no photons, namely, the measurement outcome is $\vt{n} = [n_A, n_C, n_{A'}, 0]$. The conditional output state is
\begin{eqnarray}
\langle n_A, n_C, n_{A'}, 0 \ket{\Psi}_{BAA'CC'} &=&
\mathcal{N} \mathcal{N}_+ \mathcal{N}_- \frac{e^{- \alpha^2}}{\sqrt{n_A ! \, n_{C} ! \, n_{A'} !}} \frac{(-1)^{n_{A'}} e^{i \pi (n_A - n_{C} - n_{A'})/4 } \alpha^{N}}{(\sqrt{2})^{n_{A} + n_{C}}} 
\nonumber\\
&&
\times 
\big[1 - (-1)^N \big]  \bigg( c_0 \ket{\mathcal{C}^{+}_{i \alpha}}_B + i^{N+1} c_1 \ket{\mathcal{C}^{+}_{\alpha}}_B \bigg). 
%\nonumber\\
\end{eqnarray} 
%where $N = n_A + n_{C}  + n_{A'}$ is the total number of detected photons. 
This implies the (unnormalized) output state at mode $B$ is
\begin{eqnarray}\label{eq:Case1output}
\ket{\psi_4} =  c_0 \ket{\mathcal{C}^{+}_{i \alpha}}_B + i^{N+1} c_1 \ket{\mathcal{C}^{+}_{\alpha}}_B,
\end{eqnarray}
and the measurement probability is 
\begin{eqnarray}
P([n_A, n_{C}, n_{A'}, 0]) = \mathcal{N}^{\, 2} \mathcal{N}_+^{\, 2} \mathcal{N}_-^{\,2} \frac{e^{- 2 \alpha^2}}{n_{A} ! \, n_{C} ! \, n_{A'} !} \frac{\alpha^{2N}}{2^{n_{A} + n_{C}}} \big[1 - (-1)^N \big]^2
|\langle \psi_4 \ket{\psi_4}|^2.
\end{eqnarray}

It is evident that for cases (i)-(iv) the total number of detected photon is odd when there is single photon loss. However, the total number of detected 
photon is even when there is no single photon loss, as discussed in Appendix~\ref{sec:GoodTeleportation}. Therefore, the total number of detected
photon is an indicator of the parity, namely, the total photon number can tell whether there is single photon loss or not. 
The photon number counting thus completes the first step of error correction: the syndrome measurement. 

In cases (i) and (ii), the output state is
\begin{eqnarray}\label{eq:OutPutEC1}
\ket{\psi_{\rm out}} = c_0 \ket{\mathcal{C}^{+}_{\alpha}}_B + i^{N+1} c_1 \ket{\mathcal{C}^{+}_{i \alpha}}_B. 
\end{eqnarray}
When $N = 4k -1$, the output state $\ket{\psi_{\rm out}} = \ket{\psi_{\alpha}^{(0)}}$, namely, the initial state $\ket{\psi_{\alpha}^{(0)}}$ can be recovered without further operations; when $N = 4k +1$, 
$\ket{\psi_{\rm out}} = c_0 \ket{\mathcal{C}^{+}_{\alpha}}_B - c_1 \ket{\mathcal{C}^{+}_{i \alpha}}_B = Z \ket{\psi_{\alpha}^{(0)}}$, namely, the initial state $\ket{\psi_{\alpha}^{(0)}}$ 
can be recovered by applying an encoded Pauli $Z$ operation. This completes the second step of error correction: converting the faulty state $\ket{\psi_{\alpha}^{(1)}}$ back to the 
initial state $\ket{\psi_{\alpha}^{(0)}}$. 

In cases (iii) and (iv), the output state is 
\begin{eqnarray}\label{eq:OutPutEC2}
\ket{\psi_{\rm out}} = c_0 \ket{\mathcal{C}^{+}_{i \alpha}}_B + i^{N+1} c_1 \ket{\mathcal{C}^{+}_{\alpha}}_B. 
\end{eqnarray}
When $N = 4k -1$, the output state $\ket{\psi_{\rm out}} = X \ket{\psi_{\alpha}^{(0)}}$, namely, the initial state $\ket{\psi_{\alpha}^{(0)}}$ can be recovered by applying an encoded
Pauli $X$ operation, which is simply a $\frac{\pi}{2}$ phase shift; when $N = 4k +1$, 
$\ket{\psi_{\rm out}} = c_0 \ket{\mathcal{C}^{+}_{i \alpha}}_B - c_1 \ket{\mathcal{C}^{+}_{\alpha}}_B = X Z \ket{\psi_{\alpha}^{(0)}}$, namely, the initial state $\ket{\psi_{\alpha}^{(0)}}$ 
can be recovered by first applying an encoded Pauli $X$ operation and then a Pauli $Z$ operation. This completes the second step of error correction: converting the faulty state $\ket{\psi_{\alpha}^{(1)}}$ back to the intial state $\ket{\psi_{\alpha}^{(0)}}$. 

In summary, the teleportation provides a concrete scheme for correcting single photon loss. The PNR detection outcomes tell whether there is
single photon loss. If the total number of detected photon is even, one concludes that there is no photon loss and the circuit simply teleports the initial state to the output; if the total number of 
detected photon is odd, one concludes that there is single photon loss and the circuit recovers the faulty state to the initial state. However, the error correction 
fails if two or more photons are lost, which is the intrinsic limitation of the four-component cat code.

\subsection{Absence of photons at two detectors}

When two detectors detect no photons and each of the other two detectors registers at least one photon, the initial state can be recovered for click patterns:
%It is possible that two detectors detect vacuum and the other two detectors register photons. It is found that the teleportation is still possible when 
%$n_A = n_C = 0$ or $n_{A'} = n_{C'} = 0$, namely, the measurement outcomes are
\begin{eqnarray*}
[0, 0, n_{A'}, n_{C'}], ~~~~~~ [n_A, n_{C}, 0, 0], 
\end{eqnarray*}
and cannot be recovered for click patterns:
\begin{eqnarray*}
[n_A, 0, n_{A'}, 0], ~~~~~~ [n_A, 0, 0, n_{C'}], ~~~~~~ [0, n_C, n_{A'}, 0], ~~~~~~ [0, n_C, 0, n_{C'}]. 
\end{eqnarray*}
The results are summarized in Table~\ref{tab:TwoZeroLossState} and the detailed analysis is given as follows.

(i) $n_A = n_C = 0$.

The detectors at modes $A$ and $C$ detect no photons, namely, the measurement outcome is $\vt{n} = [0, 0, n_{A'}, n_{C'}]$ with $n_{A'} \ne 0$ and $ n_{C'} \ne 0$. The conditional output state is
\begin{eqnarray}
\langle0, 0, n_{A'}, n_{C'} \ket{\Psi}_{BAA'CC'} =
\mathcal{N} \mathcal{N}_+ \mathcal{N}_- \frac{e^{- \alpha^2}}{\sqrt{n_{A'} ! \, n_{C'} !}} \frac{i^N \alpha^{N}}{(\sqrt{2})^{N}} 
\big[1 - (-1)^N \big] \big(1 + i^{n_{A'}  - n_{C'}} \big) \bigg( c_0 \ket{\mathcal{C}^{+}_{\alpha}}_B + i^{N+1} c_1 \ket{\mathcal{C}^{+}_{i \alpha}}_B \bigg). 
\end{eqnarray} 
%where $N = n_{A'}  + n_{C'}$ is the total number of detected photons. 
This implies the (unnormalized) output state at mode $B$ is 
\begin{eqnarray}\label{eq:Case1output}
\ket{\psi_5} =  c_0 \ket{\mathcal{C}^{+}_{\alpha}}_B + i^{N+1} c_1 \ket{\mathcal{C}^{+}_{i \alpha}}_B,
\end{eqnarray}
and the measurement probability is 
\begin{eqnarray}
P([0, 0, n_{A'}, n_{C'}]) = \mathcal{N}^{\, 2} \mathcal{N}_+^{\, 2} \mathcal{N}_-^{\, 2} \frac{e^{- 2 \alpha^2}}{n_{A'} ! \, n_{C'} !} \frac{\alpha^{2N}}{2^{N}} \big[1 - (-1)^N \big]^2 \big|1 + i^{n_{A'}  - n_{C'}} \big|^2
|\langle \psi_5 \ket{\psi_5}|^2.
\end{eqnarray}
The measurement probability is nonzero when $N$ is odd. % and $n_{A'}  - n_{C'} = 4k$ with $k$ an integer. 

(ii) $n_{A'} = n_{C'} = 0$. 

The detectors at modes $A'$ and $C'$ detect no photons, namely, the measurement outcome is $\vt{n} = [n_A, n_C, 0, 0]$ with $n_A \ne 0$ and $ n_{C} \ne 0$. The conditional output state is
\begin{eqnarray}
\langle n_A, n_C, 0, 0 \ket{\Psi}_{BAA'CC'} &=&
\mathcal{N} \mathcal{N}_+ \mathcal{N}_- \frac{e^{- \alpha^2}}{\sqrt{n_A ! \, n_{C} ! }} \frac{e^{- i \pi (n_A - n_{C})/4 }  \alpha^{N}}{(\sqrt{2})^{N}} 
\nonumber\\
&&
\times
\big[1 - (-1)^N \big] \big(1 + i^{n_A - n_C} \big) \bigg( c_0 \ket{\mathcal{C}^{+}_{i \alpha}}_B + i^{N+1} c_1 \ket{\mathcal{C}^{+}_{\alpha}}_B \bigg).
%\nonumber\\
\end{eqnarray} 
%where $N = n_A + n_{C} $ is the total number of detected photons. 
This implies the (unnormalized) output state at mode $B$ is
\begin{eqnarray}\label{eq:Case1output}
\ket{\psi_6} =  c_0 \ket{\mathcal{C}^{+}_{i \alpha}}_B + i^{N+1} c_1 \ket{\mathcal{C}^{+}_{\alpha}}_B.
\end{eqnarray}
and the measurement probability is 
\begin{eqnarray}
P([n_A, n_{C}, 0, 0]) = \mathcal{N}^{\, 2} \mathcal{N}_+^{\, 2} \mathcal{N}_-^{\, 2} \frac{e^{- 2 \alpha^2}}{n_{A} ! \, n_{C} ! } \frac{\alpha^{2N}}{2^{N}} \big[1 - (-1)^N \big]^2 \big|1 + i^{n_A - n_C} \big|^2
|\langle \psi_6 \ket{\psi_6}|^2.
\end{eqnarray}
The measurement probability is nonzero when $N$ is odd. %even and $ n_A - n_C= 4k$ with $k$ an integer.

(iii) $n_{C} = n_{C'} = 0$.

The detectors at modes $C$ and $C'$ detect no photons, namely, the measurement outcome is $\vt{n} = [n_A, 0, n_{A'}, 0]$ with $n_A \ne 0$ and $ n_{A'} \ne 0$. The conditional output state is
\begin{eqnarray}
&& \langle n_A, 0, n_{A'}, 0 \ket{\Psi}_{BAA'CC'} 
\nonumber\\
&=&
\mathcal{N} \mathcal{N}_+ \mathcal{N}_- \frac{e^{-\alpha^2}}{\sqrt{n_A ! \, n_{A'} !}} \alpha^{N} \big[ 1 - (-1)^{N} \big] i^{n_{A'}}
\bigg[ \frac{1}{(\sqrt{2})^{n_{A'}}} \bigg( c_0 \ket{\mathcal{C}^{+}_{\alpha}}_B + i^{N + 1} c_1 \ket{\mathcal{C}^{+}_{i \alpha}}_B, \bigg)
\nonumber\\
&&
+ \frac{ e^{i \pi N/4}}{(\sqrt{2})^{n_{A}}} \bigg( c_0 \ket{\mathcal{C}^{+}_{i \alpha}}_B + i^{N + 1} c_1 \ket{\mathcal{C}^{+}_{\alpha}}_B, \bigg)
\bigg]
\nonumber\\
&=&
\mathcal{N} \mathcal{N}_+ \mathcal{N}_- \big[ 1 - (-1)^N \big] e^{-\alpha^2} \frac{\alpha^N}{\sqrt{n_A ! n_{A'} !}} i^{n_{A'}}
\bigg\{ \bigg[ \frac{1}{(\sqrt{2})^{n_{A'}}} c_0 + \frac{ e^{i N \pi/4}}{(\sqrt{2})^{n_{A}}} i^{N+1} c_1 \bigg]  \ket{\mathcal{C}^{+}_{\alpha}}_B
\nonumber\\
&&
+ \bigg[ \frac{1}{(\sqrt{2})^{n_{A'}}} i^{N+1} c_1 + \frac{ e^{i N \pi/4}}{(\sqrt{2})^{n_{A}}} c_0 \bigg]  \ket{\mathcal{C}^{+}_{i \alpha}}_B
\bigg\}.
\end{eqnarray}
%and the measurement probability is
%\begin{eqnarray}
%P([n_A, 0, n_{A'}, 0]) = | \langle n_A, 0, n_{A'}, 0 \ket{\Psi}_{BAA'CC'} |^2. 
%\end{eqnarray}
The initial state cannot be recovered. 

(iv) $n_{C} = n_{A'} = 0$.

The detectors at modes $C$ and $A'$ detect no photons, namely, 
the measurement outcome is $\vt{n} = [n_A, 0, 0, n_{C'}]$ with $n_A \ne 0$ and $ n_{C'} \ne 0$. The conditional output state is
\begin{eqnarray}
&& \langle n_A, 0, 0, n_{C'} \ket{\Psi}_{BAA'CC'} 
\nonumber\\
&=&
\mathcal{N} \mathcal{N}_+ \mathcal{N}_- \frac{e^{-\alpha^2}}{\sqrt{n_A ! \, n_{C'} !}} \alpha^{N} \big[ 1 - (-1)^{N} \big] i^{n_{C'}}
\bigg[ \frac{1}{(\sqrt{2})^{n_{C'}}} \bigg( c_0 \ket{\mathcal{C}^{+}_{\alpha}}_B + i^{N + 1} c_1 \ket{\mathcal{C}^{+}_{i \alpha}}_B, \bigg)
\nonumber\\
&&
+ \frac{ e^{- i N \pi/4}}{(\sqrt{2})^{n_{A}}} \bigg( c_0 \ket{\mathcal{C}^{+}_{i \alpha}}_B + i^{N + 1} c_1 \ket{\mathcal{C}^{+}_{\alpha}}_B, \bigg)
\bigg]
\nonumber\\
&=&
\mathcal{N} \mathcal{N}_+ \mathcal{N}_- \frac{e^{-\alpha^2}}{\sqrt{n_A ! \, n_{C'} !}} \alpha^N \big[ 1 - (-1)^N \big] i^{n_{C'}}
\bigg\{ \bigg[ \frac{1}{(\sqrt{2})^{n_{C'}}} c_0 + \frac{ e^{- i N \pi/4}}{(\sqrt{2})^{n_{A}}} i^{N + 1} c_1 \bigg]  \ket{\mathcal{C}^{+}_{\alpha}}_B
\nonumber\\
&&
+ \bigg[ \frac{1}{(\sqrt{2})^{n_{C'}}} i^{N + 1} c_1 + \frac{e^{- i N \pi/4}}{(\sqrt{2})^{n_{A}}} c_0 \bigg]  \ket{\mathcal{C}^{+}_{i \alpha}}_B
\bigg\}.
\end{eqnarray} 
%and the measurement probability is
%\begin{eqnarray}
%P([n_A, 0, 0, n_{C'}]) = | \langle n_A, 0, 0, n_{C'} \ket{\Psi}_{BAA'CC'} |^2. 
%\end{eqnarray}
The initial state cannot be recovered.

(v) $n_{A} = n_{C'} = 0$.

The detectors at modes $A$ and $C'$ detect no photons, namely, 
the measurement outcome is $\vt{n} = [0, n_C, n_{A'}, 0]$ with $n_C \ne 0$ and $ n_{A'} \ne 0$. The conditional output state is
\begin{eqnarray}
&& \langle 0, n_C, n_{A'}, 0 \ket{\Psi}_{BAA'CC'} 
\nonumber\\
&=&
\mathcal{N} \mathcal{N}_+ \mathcal{N}_- \frac{e^{-\alpha^2}}{\sqrt{n_C ! \, n_{A'} !}} \alpha^{N} \big[ 1 - (-1)^{N} \big] (-1)^{n_{A'}}
\bigg[ \frac{1}{(\sqrt{2})^{n_{A'}}} \bigg( c_0 \ket{\mathcal{C}^{+}_{\alpha}}_B + i^{N + 1} c_1 \ket{\mathcal{C}^{+}_{i \alpha}}_B, \bigg)
\nonumber\\
&&
+ \frac{e^{- i N \pi/4}}{(\sqrt{2})^{n_{C}}} \bigg( c_0 \ket{\mathcal{C}^{+}_{i \alpha}}_B + i^{N + 1} c_1 \ket{\mathcal{C}^{+}_{\alpha}}_B, \bigg)
\bigg]
\nonumber\\
&=&
\mathcal{N} \mathcal{N}_+ \mathcal{N}_- \frac{e^{-\alpha^2}}{\sqrt{n_C ! \, n_{A'} !}} \alpha^{N} \big[ 1 - (-1)^{N} \big] (-1)^{n_{A'}}
\bigg\{ \bigg[ \frac{1}{(\sqrt{2})^{n_{A'}}} c_0 + \frac{e^{- i N \pi /4}}{(\sqrt{2})^{n_{C}}} i^{N + 1} c_1 \bigg]  \ket{\mathcal{C}^{+}_{\alpha}}_B
\nonumber\\
&&
+ \bigg[ \frac{1}{(\sqrt{2})^{n_{A'}}} i^{N + 1} c_1 + \frac{ e^{- i N \pi/4}}{(\sqrt{2})^{n_{C}}} c_0 \bigg]  \ket{\mathcal{C}^{+}_{i \alpha}}_B
\bigg\}.
\end{eqnarray}
%and the measurement probability is
%\begin{eqnarray}
%P([0, n_C, n_{A'}, 0]) = | \langle 0, n_C, n_{A'}, 0 \ket{\Psi}_{BAA'CC'} |^2. 
%\end{eqnarray}
The initial state cannot be recovered.

(vi) $n_{A} = n_{A'} = 0$.

The detectors at modes $A$ and $A'$ detect no photons, namely, 
the measurement outcome is $\vt{n} = [0, n_C, 0, n_{C'}]$ with $n_C \ne 0$ and $ n_{C'} \ne 0$. The conditional output state is
\begin{eqnarray}
&& \langle 0, n_C, 0, n_{C'} \ket{\Psi}_{BAA'CC'} 
\nonumber\\
&=&
\mathcal{N} \mathcal{N}_+ \mathcal{N}_- \frac{e^{-\alpha^2}}{\sqrt{n_C ! \, n_{C'} !}} \alpha^{N} \big[ 1 - (-1)^{N} \big] 
\bigg[ \frac{1}{(\sqrt{2})^{n_{C'}}} \bigg( c_0 \ket{\mathcal{C}^{+}_{\alpha}}_B + i^{N + 1} c_1 \ket{\mathcal{C}^{+}_{i \alpha}}_B, \bigg)
\nonumber\\
&&
+ \frac{ e^{i N \pi /4}}{(\sqrt{2})^{n_{C}}} \bigg( c_0 \ket{\mathcal{C}^{+}_{i \alpha}}_B + i^{N + 1} c_1 \ket{\mathcal{C}^{+}_{\alpha}}_B, \bigg)
\bigg]
\nonumber\\
&=&
\mathcal{N} \mathcal{N}_+ \mathcal{N}_- \frac{e^{-\alpha^2}}{\sqrt{n_C ! \, n_{C'} !}} \alpha^{N} \big[ 1 - (-1)^{N} \big] 
\bigg\{ \bigg[ \frac{1}{(\sqrt{2})^{n_{C'}}} c_0 + \frac{ e^{i N \pi /4}}{(\sqrt{2})^{n_{C}}} i^{N + 1} c_1 \bigg]  \ket{\mathcal{C}^{+}_{\alpha}}_B
\nonumber\\
&&
+ \bigg[ \frac{1}{(\sqrt{2})^{n_{C'}}} i^{N + 1} c_1 + \frac{ e^{i N \pi/4}}{(\sqrt{2})^{n_{C}}} c_0 \bigg]  \ket{\mathcal{C}^{+}_{i \alpha}}_B
\bigg\}.
\end{eqnarray}
%and the measurement probability is
%\begin{eqnarray}
%P([0, n_C, 0, n_{C'}]) = | \langle 0, n_C, 0, n_{C'} \ket{\Psi}_{BAA'CC'} |^2. 
%\end{eqnarray}
The initial state cannot be recovered.

\subsection{Absence of photons at three detectors}

When only one detector registers photons, the teleportation fails and the initial state cannot be recovered. These click patterns include: 
\begin{eqnarray*}
[n_A, 0, 0, 0], ~~~~~~ [0, n_C,  0, 0], ~~~~~~ [0, 0, n_{A'}, 0], ~~~~~~ [0, 0, 0, n_{C'}]. 
\end{eqnarray*}
We explicitly write down the conditional output states for these click patterns, from which one can see clearly why the error correction fails.

(i) $n_A = N \ne 0$.

The detector at mode $A$ detects photons, namely, 
the measurement outcome is $\vt{n} = [N, 0, 0, 0]$. The conditional output state is
\begin{eqnarray}
&& \langle N, 0, 0, 0 \ket{\Psi}_{BAA'CC'} 
\nonumber\\
&=&
\mathcal{N} \mathcal{N}_+ \mathcal{N}_- \big[ 1 - (-1)^N \big] e^{-\alpha^2} \frac{\alpha^N}{\sqrt{N!}}
\bigg[ \bigg( c_0 \ket{\mathcal{C}^{+}_{\alpha}}_B + i^{N + 1} c_1 \ket{\mathcal{C}^{+}_{i \alpha}}_B \bigg)
+ \frac{e^{i N \pi/4} + e^{- i N \pi/4}} {(\sqrt{2})^N} \bigg( c_0 \ket{\mathcal{C}^{+}_{i \alpha}}_B + i^{N + 1} c_1 \ket{\mathcal{C}^{+}_{\alpha}}_B \bigg)
\bigg]
\nonumber\\
&=&
\mathcal{N} \mathcal{N}_+ \mathcal{N}_- \big[ 1 - (-1)^N \big] e^{-\alpha^2} \frac{\alpha^N}{\sqrt{N!}}
\bigg\{ \bigg[c_0 + \frac{2 \cos(N \pi/4)}{(\sqrt{2})^N} i^{N + 1} c_1 \bigg] \ket{\mathcal{C}^{+}_{\alpha}}_B
+ \bigg[ i^{N + 1} c_1 + \frac{2 \cos(N \pi/4)}{(\sqrt{2})^N} c_0 \bigg] \ket{\mathcal{C}^{+}_{i \alpha}}_B
\bigg\}. 
\end{eqnarray}
%and the measurement probability is
%\begin{eqnarray}
%P([N, 0, 0, 0]) = | \langle N, 0, 0, 0 \ket{\Psi}_{BAA'CC'} |^2. 
%\end{eqnarray}
The initial state cannot be recovered. 

(ii) $n_C = N \ne 0$. 

The detector at mode $C$ detects photons, namely,
the measurement outcome is $\vt{n} = [0, N, 0, 0]$. The conditional output state is the same as case (i) and the initial state cannot be recovered. 

(iii) $n_{A'} = N \ne 0$.

The detector at mode $A'$ detects photons, namely, the measurement outcome is $\vt{n} = [0, 0, N, 0]$. The conditional output state is
\begin{eqnarray}
&& \langle 0, 0, N, 0 \ket{\Psi}_{BAA'CC'} 
\nonumber\\
&=&
\mathcal{N} \mathcal{N}_+ \mathcal{N}_- \big[ 1 - (-1)^N \big] e^{-\alpha^2}  \frac{\alpha^N}{\sqrt{N!}} i^N
\bigg[ \frac{1 + i^N}{(\sqrt{2})^N} \bigg( c_0 \ket{\mathcal{C}^{+}_{\alpha}}_B + i^{N + 1} c_1 \ket{\mathcal{C}^{+}_{i \alpha}}_B \bigg)
+ e^{ i N \pi/4} \bigg( c_0 \ket{\mathcal{C}^{+}_{i \alpha}}_B + i^{N + 1} c_1 \ket{\mathcal{C}^{+}_{\alpha}}_B \bigg)
\bigg]
\nonumber\\
&=&
\mathcal{N} \mathcal{N}_+ \mathcal{N}_- \big[ 1 - (-1)^N \big] e^{-\alpha^2}  \frac{\alpha^N}{\sqrt{N!}} i^N
\bigg\{ \bigg[ e^{ i N \pi/4} i^{N + 1} c_1 +  \frac{1 + i^N}{(\sqrt{2})^N} c_0 \bigg]  \ket{\mathcal{C}^{+}_{\alpha}}_B
+ \bigg[ e^{ i N \pi/4} c_0 +  \frac{1 + i^N}{(\sqrt{2})^N} i^{N + 1} c_1 \bigg]  \ket{\mathcal{C}^{+}_{i \alpha}}_B
\bigg\}. 
\nonumber\\
\end{eqnarray}
%and the measurement probability is
%\begin{eqnarray}
%P([0, 0, N, 0]) = | \langle 0, 0, N, 0 \ket{\Psi}_{BAA'CC'} |^2. 
%\end{eqnarray}
The initial state cannot be recovered.

(iv) $n_{C'} = N \ne 0$.

The detector at mode $C'$ detects photons, namely,
the measurement outcome is $\vt{n} = [0, 0, 0, N]$. The conditional output state is the same as case (iii) except a global phase $i^N$, and the initial state cannot be recovered. 

\section{Correcting single photon loss in Bell state}\label{ap:LossBellState}

In this section, we discuss the correction of single photon loss in the entangled Bell state. 

Ideally, the entangled resource state used for teleportation is given by Eq.~\eqref{eq:bell}. If a single photon is lost in mode $C$ before coupling with the encoded input state, then the entangled 
resource becomes %\textcolor{red}{[{\bf normalization?}] }
\begin{eqnarray}
\ket{\Phi_0^{\, \prime}}_{BC} % &=& \mathcal{N} \left(\ket{0_L}_{B} \ket{0_L}_{C} + \ket{1_L}_{B} \ket{1_L}_{C} \right) 
= \mathcal{N} \left(\ket{\mathcal{C}^{+}_{\alpha}}_{B} \ket{\mathcal{C}^{-}_{\alpha}}_{C} + i \, \ket{\mathcal{C}^{+}_{i \alpha}}_{B} \ket{\mathcal{C}^{-}_{i \alpha}}_{C} \right).
\end{eqnarray}
Assume that there is no photon loss in mode $A$, then the overall input state is
\begin{eqnarray}
\ket{\Psi_{\rm in}^{\, \prime}} &=& \ket{\psi_{\rm in}}_A \otimes \ket{\Phi_0^{\, \prime} }_{BC} \otimes \ket{0, 0}_{A' C'}
\nonumber\\
&=&
\mathcal{N} \left[ \left( c_0 \ket{\mathcal{C}^{+}_{\alpha}}_A \ket{\mathcal{C}^{-}_{\alpha}}_C 
+ c_1 \ket{\mathcal{C}^{+}_{i \alpha}}_A \ket{\mathcal{C}^{-}_{\alpha}}_C \right) \ket{\mathcal{C}^{+}_{\alpha}}_B 
+ i \left( c_0 \ket{\mathcal{C}^{+}_{\alpha}}_A \ket{\mathcal{C}^{-}_{i \alpha}}_C 
+ c_1 \ket{\mathcal{C}^{+}_{i \alpha}}_A \ket{\mathcal{C}^{-}_{i \alpha}}_C \right) \ket{\mathcal{C}^{+}_{i \alpha}}_B \right] \otimes \ket{0, 0}_{A' C'}. 
\end{eqnarray}
The teleportation circuit transforms the four components of the overall input state as follows: 
\begin{eqnarray*}
 \ket{\mathcal{C}^{+}_{\alpha}}_A \ket{\mathcal{C}^{-}_{\alpha}}_C \ket{0}_{A'} \ket{0}_{C'} 
 &\longrightarrow&
\bigg| 0, \alpha, -\frac{\alpha}{\sqrt{2}}, \frac{\alpha}{\sqrt{2}} \bigg\rangle - \bigg| \alpha, 0, \frac{i \alpha}{\sqrt{2}}, \frac{i \alpha}{\sqrt{2}} \bigg\rangle
+ \bigg| - \alpha, 0, - \frac{i \alpha}{\sqrt{2}}, - \frac{i \alpha}{\sqrt{2}} \bigg\rangle - \bigg| 0, - \alpha, \frac{\alpha}{\sqrt{2}}, - \frac{\alpha}{\sqrt{2}} \bigg\rangle,
\nonumber\\
 \ket{\mathcal{C}^{+}_{i \alpha}}_A \ket{\mathcal{C}^{-}_{i \alpha}}_C \ket{0}_{A'} \ket{0}_{C'} 
 &\longrightarrow&
\bigg| 0, i \alpha, -\frac{i \alpha}{\sqrt{2}}, \frac{i \alpha}{\sqrt{2}} \bigg\rangle - \bigg| i \alpha, 0, - \frac{\alpha}{\sqrt{2}}, - \frac{\alpha}{\sqrt{2}} \bigg\rangle
+ \bigg| - i \alpha, 0, \frac{\alpha}{\sqrt{2}}, \frac{\alpha}{\sqrt{2}} \bigg\rangle - \bigg| 0, -i \alpha, \frac{i \alpha}{\sqrt{2}}, - \frac{i \alpha}{\sqrt{2}} \bigg\rangle,
\nonumber\\
 \ket{\mathcal{C}^{+}_{\alpha}}_A \ket{\mathcal{C}^{-}_{i \alpha}}_C \ket{0}_{A'} \ket{0}_{C'} 
 &\longrightarrow&
\bigg| \frac{\beta^*}{\sqrt{2}}, \frac{\beta}{\sqrt{2}},  0, \beta \bigg\rangle - \bigg| \frac{\beta}{\sqrt{2}}, \frac{\beta^*}{\sqrt{2}}, - \beta^*, 0 \bigg\rangle
+ \bigg| - \frac{\beta}{\sqrt{2}}, - \frac{\beta^*}{\sqrt{2}}, \beta^*, 0 \bigg\rangle - \bigg| - \frac{\beta^*}{\sqrt{2}}, - \frac{\beta}{\sqrt{2}}, 0, - \beta \bigg\rangle,
\nonumber\\
\ket{\mathcal{C}^{+}_{i \alpha}}_A \ket{\mathcal{C}^{-}_{\alpha}}_C \ket{0}_{A'} \ket{0}_{C'} 
 &\longrightarrow&
\bigg| - \frac{\beta^*}{\sqrt{2}}, \frac{\beta}{\sqrt{2}},  - \beta, 0 \bigg\rangle - \bigg| \frac{\beta}{\sqrt{2}}, - \frac{\beta^*}{\sqrt{2}}, 0, - \beta^* \bigg\rangle
+ \bigg| - \frac{\beta}{\sqrt{2}}, \frac{\beta^*}{\sqrt{2}}, 0, \beta^* \bigg\rangle - \bigg| \frac{\beta^*}{\sqrt{2}}, - \frac{\beta}{\sqrt{2}}, \beta, 0 \bigg\rangle,
\end{eqnarray*}
where have defined $\beta = \alpha e^{i \pi/4}$, and omitted the subscripts to simplify the notation and chosen the order of the modes as $``ACA'C' "$. 
By combining these results, we find that the output state after the teleportation circuit (before photon number measurement) is 
\begin{eqnarray}\label{eq:OutPutState2}
&& \ket{\Psi}_{BAA'CC'} = \op{U}_{A'C'} \op{U}_{A'}(\pi/2) \op{U}_{AA'} \op{U}_{CC'} \op{U}_{AC} \ket{\Psi_{\rm in}}
\nonumber\\
&\propto&
\bigg[ \bigg( \bigg| 0, \alpha, - \frac{\alpha}{\sqrt{2}}, \frac{\alpha}{\sqrt{2}} \bigg\rangle 
- \bigg| 0, - \alpha, \frac{\alpha}{\sqrt{2}}, - \frac{\alpha}{\sqrt{2}} \bigg\rangle \bigg) c_0 \ket{\mathcal{C}^{+}_{\alpha}}_B
%\nonumber\\
%&&
+ \bigg( \bigg| 0, i \alpha, - \frac{i \alpha}{\sqrt{2}}, \frac{i \alpha}{\sqrt{2}} \bigg\rangle 
- \bigg|0, - i \alpha, \frac{i \alpha}{\sqrt{2}}, - \frac{i \alpha}{\sqrt{2}} \bigg\rangle \bigg) i c_1 \ket{\mathcal{C}^{+}_{i \alpha}}_B \bigg]
\nonumber\\
&&
- \bigg[ \bigg( \bigg| \alpha, 0, \frac{i \alpha}{\sqrt{2}}, \frac{i \alpha}{\sqrt{2}} \bigg\rangle 
- \bigg| - \alpha, 0, - \frac{i \alpha}{\sqrt{2}}, - \frac{i \alpha}{\sqrt{2}} \bigg\rangle \bigg) c_0 \ket{\mathcal{C}^{+}_{\alpha}}_B 
%\nonumber\\
%&&
+ \bigg( \bigg| i \alpha, 0, - \frac{\alpha}{\sqrt{2}}, - \frac{\alpha}{\sqrt{2}} \bigg\rangle 
- \bigg| - i \alpha, 0, \frac{\alpha}{\sqrt{2}}, \frac{\alpha}{\sqrt{2}} \bigg\rangle \bigg) i c_1 \ket{\mathcal{C}^{+}_{i \alpha}}_B \bigg]
\nonumber\\
&&
+ \bigg[ - \bigg( \bigg| \frac{\beta}{\sqrt{2}}, - \frac{\beta^*}{\sqrt{2}}, 0, - \beta^* \bigg\rangle 
- \bigg| - \frac{\beta}{\sqrt{2}}, \frac{\beta^*}{\sqrt{2}}, 0, \beta^* \bigg\rangle \bigg) c_1 \ket{\mathcal{C}^{+}_{\alpha}}_B 
%\nonumber\\
%&&
+ \bigg( \bigg| \frac{\beta^*}{\sqrt{2}}, \frac{\beta}{\sqrt{2}},  0, \beta \bigg\rangle 
- \bigg| - \frac{\beta^*}{\sqrt{2}}, - \frac{\beta}{\sqrt{2}}, 0, - \beta \bigg\rangle \bigg) i c_0 \ket{\mathcal{C}^{+}_{i \alpha}}_B \bigg]
\nonumber\\
&&
+ \bigg[ \bigg( \bigg| - \frac{\beta^*}{\sqrt{2}}, \frac{\beta}{\sqrt{2}}, - \beta, 0 \bigg\rangle 
- \bigg| \frac{\beta^*}{\sqrt{2}}, - \frac{\beta}{\sqrt{2}}, \beta, 0 \bigg\rangle \bigg) c_1 \ket{\mathcal{C}^{+}_{\alpha}}_B 
%\nonumber\\
%&&
- \bigg( \bigg| \frac{\beta}{\sqrt{2}}, \frac{\beta^*}{\sqrt{2}}, - \beta^*, 0 \bigg\rangle 
- \bigg| - \frac{\beta}{\sqrt{2}}, - \frac{\beta^*}{\sqrt{2}}, \beta^*, 0 \bigg\rangle \bigg) i c_0 \ket{\mathcal{C}^{+}_{i \alpha}}_B \bigg],
\nonumber\\
\end{eqnarray}
where the proportionality factor is $\mathcal{N} \mathcal{N}_+ \mathcal{N}_-$.

\subsection{Absence of photon at one detector}

We first consider situations where only one detector detects no photons and each of the other three detectors registers at least one photon. 
This includes four sorts of click pattern: 
\begin{eqnarray*}
[0, n_C, n_{A'}, n_{C'}], ~~~~~~ [n_A, 0, n_{A'}, n_{C'}], ~~~~~~ [n_A, n_{C}, 0, n_{C'}], ~~~~~~ [n_A, n_{C}, n_{A'}, 0]. 
\end{eqnarray*}
%It is found that the initial state can be recovered for all these measurement patterns by applying appropriate Pauli corrections. This is summarized in Table~\ref{tab:OneZeroLossState}.
We find that the output states are the same as those when a single photon is lost in the input mode, except a possible global phase. Therefore, the initial state can be recovered by 
applying appropriate Pauli corrections, which is summarized in Table~\ref{tab:OneZeroLossState}. 
The detailed analysis is given as follows. 

(i) $n_{A} =0$, no click at mode $A$.  

The detector at mode $A$ detects no photons, namely, the measurement outcome is $\vt{n} = [0, n_C, n_{A'}, n_{C'}]$. The conditional output state is
\begin{eqnarray}
\langle0, n_C, n_{A'}, n_{C'} \ket{\Psi}_{BAA'CC'} =
\mathcal{N} \mathcal{N}_+ \mathcal{N}_- \frac{e^{- \alpha^2}}{\sqrt{n_C ! \, n_{A'} ! \, n_{C'} !}} \frac{ (-1)^{n_{A'}} \alpha^{N}}{(\sqrt{2})^{n_{A'} + n_{C'}}} 
\big[1 - (-1)^N \big] \bigg( c_0 \ket{\mathcal{C}^{+}_{\alpha}}_B + i^{N+1} c_1 \ket{\mathcal{C}^{+}_{i \alpha}}_B \bigg).
\end{eqnarray} 
%where $N = n_C + n_{A'}  + n_{C'}$ is the total number of detected photons. 
This implies the (unnormalized) output state is 
\begin{eqnarray}\label{eq:Case1output}
\ket{\psi_1} =  c_0 \ket{\mathcal{C}^{+}_{\alpha}}_B + i^{N+1} c_1 \ket{\mathcal{C}^{+}_{i \alpha}}_B, 
\end{eqnarray}
and the measurement probability is %\textcolor{red}{ [missing a normalization factor of the output state] }
\begin{eqnarray}
P([0, n_C, n_{A'}, n_{C'}]) = \mathcal{N}^{\, 2} \mathcal{N}_+^{\, 2} \mathcal{N}_-^{\, 2} \frac{e^{- 2 \alpha^2}}{ n_C ! \, n_{A'} ! \, n_{C'} !} \frac{\alpha^{2 N}}{2^{n_{A'} + n_{C'}}} \big[1 - (-1)^N \big]^2 
|\langle \psi_1 \ket{\psi_1}|^2. 
\end{eqnarray}

(ii) $n_C = 0$, no click at mode $C$. 

The detector at mode $C$ detects no photons, namely, the measurement outcome is $\vt{n} = [n_A, 0, n_{A'}, n_{C'}]$. The conditional output state is
\begin{eqnarray}
\langle n_A, 0, n_{A'}, n_{C'} \ket{\Psi}_{BAA'CC'} =
\mathcal{N} \mathcal{N}_+  \mathcal{N}_- \frac{e^{- \alpha^2}}{\sqrt{n_A ! \, n_{A'} ! \, n_{C'} !}} \frac{i^{n_{A'} + n_{C'} + 2} \alpha^{N}}{(\sqrt{2})^{n_{A'} + n_{C'}}} 
\big[1 - (-1)^N \big] \bigg( c_0 \ket{\mathcal{C}^{+}_{\alpha}}_B + i^{N+1} c_1 \ket{\mathcal{C}^{+}_{i \alpha}}_B \bigg). 
\end{eqnarray} 
%where $N = n_A + n_{A'}  + n_{C'}$ is the total number of detected photons. 
This implies the (unnormalized) output state is
\begin{eqnarray}\label{eq:Case1output}
\ket{\psi_2} =  c_0 \ket{\mathcal{C}^{+}_{\alpha}}_B + i^{N+1} c_1 \ket{\mathcal{C}^{+}_{i \alpha}}_B.
\end{eqnarray}
and the measurement probability is 
\begin{eqnarray}
P([n_A, 0, n_{A'}, n_{C'}]) = \mathcal{N}^{\, 2} \mathcal{N}_+^{\, 2} \mathcal{N}_-^{\, 2} \frac{e^{- 2 \alpha^2}}{n_{A} ! \, n_{A'} ! \, n_{C'} !} \frac{\alpha^{2N}}{2^{n_{A'} + n_{C'}}} \big[1 - (-1)^N \big]^2
|\langle \psi_2 \ket{\psi_2}|^2. 
\end{eqnarray}

(iii) $n_{A'} = 0$, no click at mode $A'$.

The detector at mode $A'$ detects no photons, namely, the measurement outcome is $\vt{n} = [n_A, n_C, 0, n_{C'}]$. The conditional output state is
\begin{eqnarray}
\langle n_A, n_C, 0, n_{C'} \ket{\Psi}_{BAA'CC'} =
\mathcal{N} \mathcal{N}_+ \mathcal{N}_- \frac{e^{- \alpha^2}}{\sqrt{n_A ! \, n_{C} ! \, n_{C'} !}} \frac{ i\, e^{- i \pi (n_A - n_{C} - n_{C'})/4 } \alpha^{N}}{(\sqrt{2})^{n_{A} + n_{C}}} 
\big[1 - (-1)^N \big] \bigg( c_0 \ket{\mathcal{C}^{+}_{i \alpha}}_B + i^{N+1} c_1 \ket{\mathcal{C}^{+}_{\alpha}}_B \bigg).
\nonumber\\
\end{eqnarray} 
%where $N = n_A + n_{C}  + n_{C'}$ is the total number of detected photons. 
This implies the (unnormalized) output state is
\begin{eqnarray}\label{eq:Case1output}
\ket{\psi_3} =  c_0 \ket{\mathcal{C}^{+}_{i \alpha}}_B + i^{N+1} c_1 \ket{\mathcal{C}^{+}_{\alpha}}_B.
\end{eqnarray}
and the measurement probability is 
\begin{eqnarray}
P([n_A, n_{C}, 0, n_{C'}]) = \mathcal{N}^{\, 2} \mathcal{N}_+^{\, 2} \mathcal{N}_-^{\, 2} \frac{e^{- 2 \alpha^2}}{n_{A} ! \, n_{C} ! \, n_{C'} !} \frac{\alpha^{2N}}{2^{n_{A} + n_{C}}} \big[1 - (-1)^N \big]^2 
|\langle \psi_3 \ket{\psi_3}|^2.
\end{eqnarray}

(iv) $n_{C'} = 0$, no click at mode $C'$.

The detector at mode $C'$ detects no photons, namely, the measurement outcome is $\vt{n} = [n_A, n_C, n_{A'}, 0]$. The conditional output state is
\begin{eqnarray}
\langle n_A, n_C, n_{A'}, 0 \ket{\Psi}_{BAA'CC'} &=&
\mathcal{N} \mathcal{N}_+ \mathcal{N}_- \frac{e^{- \alpha^2}}{\sqrt{n_A ! \, n_{C} ! \, n_{A'} !}} \frac{(-1)^{n_{A'} + 1} i\, e^{i \pi (n_A - n_{C} - n_{A'})/4 } \alpha^{N}}{(\sqrt{2})^{n_{A} + n_{C}}} 
\nonumber\\
&&
\times 
\big[1 - (-1)^N \big]  \bigg( c_0 \ket{\mathcal{C}^{+}_{i \alpha}}_B + i^{N+1} c_1 \ket{\mathcal{C}^{+}_{\alpha}}_B \bigg).
%\nonumber\\
\end{eqnarray} 
%where $N = n_A + n_{C}  + n_{A'}$ is the total number of detected photons. 
This implies the (unnormalized) output state is
\begin{eqnarray}\label{eq:Case1output}
\ket{\psi_4} =  c_0 \ket{\mathcal{C}^{+}_{i \alpha}}_B + i^{N+1} c_1 \ket{\mathcal{C}^{+}_{\alpha}}_B.
\end{eqnarray}
and the measurement probability is 
\begin{eqnarray}
P([n_A, n_{C}, n_{A'}, 0]) = \mathcal{N}^{\, 2} \mathcal{N}_+^{\, 2} \mathcal{N}_-^{\,2} \frac{e^{- 2 \alpha^2}}{n_{A} ! \, n_{C} ! \, n_{A'} !} \frac{\alpha^{2N}}{2^{n_{A} + n_{C}}} \big[1 - (-1)^N \big]^2
|\langle \psi_4 \ket{\psi_4}|^2.
\end{eqnarray}

\subsection{Absence of photons at two detectors}

When two detectors detect no photons and each of the other two detectors registers at least one photon, the initial state can be recovered for click patterns:
\begin{eqnarray*}
[0, 0, n_{A'}, n_{C'}], ~~~~~~ [n_A, n_{C}, 0, 0], 
\end{eqnarray*}
and cannot be recovered for click patterns:
\begin{eqnarray*}
[n_A, 0, n_{A'}, 0], ~~~~~~ [n_A, 0, 0, n_{C'}], ~~~~~~ [0, n_C, n_{A'}, 0], ~~~~~~ [0, n_C, 0, n_{C'}]. 
\end{eqnarray*}
This is the same as the case where a single photon is lost in the input mode. 
The results are summarized in Table~\ref{tab:TwoZeroLossState} and the detailed analysis is given as follows.

(i) $n_A = n_C = 0$.

The detectors at modes $A$ and $C$ detect no photons, namely, the measurement outcome is $\vt{n} = [0, 0, n_{A'}, n_{C'}]$ with $n_{A'} \ne 0$ and $ n_{C'} \ne 0$. The conditional output state is
\begin{eqnarray}
\langle0, 0, n_{A'}, n_{C'} \ket{\Psi}_{BAA'CC'} =
\mathcal{N} \mathcal{N}_+ \mathcal{N}_- \frac{e^{- \alpha^2}}{\sqrt{n_{A'} ! \, n_{C'} !}} \frac{i^N \alpha^{N}}{(\sqrt{2})^{N}} 
\big[1 - (-1)^N \big] \big(i^{n_{A'}  - n_{C'}} - 1 \big) \bigg( c_0 \ket{\mathcal{C}^{+}_{\alpha}}_B + i^{N+1} c_1 \ket{\mathcal{C}^{+}_{i \alpha}}_B \bigg).
\end{eqnarray} 
%where $N = n_{A'}  + n_{C'}$ is the total number of detected photons. 
This implies the (unnormalized) output state is 
\begin{eqnarray}\label{eq:Case1output}
\ket{\psi_5} =  c_0 \ket{\mathcal{C}^{+}_{\alpha}}_B + i^{N+1} c_1 \ket{\mathcal{C}^{+}_{i \alpha}}_B,
\end{eqnarray}
and the measurement probability is 
\begin{eqnarray}
P([0, 0, n_{A'}, n_{C'}]) = \mathcal{N}^{\, 2} \mathcal{N}_+^{\, 2} \mathcal{N}_-^{\, 2} \frac{e^{- 2 \alpha^2}}{n_{A'} ! \, n_{C'} !} \frac{\alpha^{2N}}{2^{N}} \big[1 - (-1)^N \big]^2 \big|1 - i^{n_{A'}  - n_{C'}} \big|^2
|\langle \psi_5 \ket{\psi_5}|^2.
\end{eqnarray}
The measurement probability is nonzero when $N$ is odd. % and $n_{A'}  - n_{C'} = 4k$ with $k$ an integer. 

(ii) $n_{A'} = n_{C'} = 0$. 

The detectors at mode $A'$ and $C'$ detect no photons, namely, the measurement outcome is $\vt{n} = [n_A, n_C, 0, 0]$ with $n_A \ne 0$ and $ n_{C} \ne 0$. The conditional output state is
\begin{eqnarray}
\langle n_A, n_C, 0, 0 \ket{\Psi}_{BAA'CC'} &=&
\mathcal{N} \mathcal{N}_+ \mathcal{N}_- \frac{e^{- \alpha^2}}{\sqrt{n_A ! \, n_{C} ! }} \frac{i \, e^{- i \pi (n_A - n_{C})/4 }  \alpha^{N}}{(\sqrt{2})^{N}} 
\nonumber\\
&&
\times
\big[1 - (-1)^N \big] \big(1 - i^{n_A - n_C} \big) \bigg( c_0 \ket{\mathcal{C}^{+}_{i \alpha}}_B + i^{N+1} c_1 \ket{\mathcal{C}^{+}_{\alpha}}_B \bigg). 
%\nonumber\\
\end{eqnarray} 
%where $N = n_A + n_{C} $ is the total number of detected photons. 
This implies the (unnormalized) output state is
\begin{eqnarray}\label{eq:Case1output}
\ket{\psi_6} =  c_0 \ket{\mathcal{C}^{+}_{i \alpha}}_B + i^{N+1} c_1 \ket{\mathcal{C}^{+}_{\alpha}}_B,
\end{eqnarray}
and the measurement probability is 
\begin{eqnarray}
P([n_A, n_{C}, 0, 0]) = \mathcal{N}^{\, 2} \mathcal{N}_+^{\, 2} \mathcal{N}_-^{\, 2} \frac{e^{- 2 \alpha^2}}{n_{A} ! \, n_{C} ! } \frac{\alpha^{2N}}{2^{N}} \big[1 - (-1)^N \big]^2 \big|1 - i^{n_A - n_C} \big|^2
|\langle \psi_6 \ket{\psi_6}|^2.
\end{eqnarray}
The measurement probability is nonzero only when $N$ is odd. %even and $ n_A - n_C= 4k$ with $k$ an integer.

(iii) $n_{C} = n_{C'} = 0$.

The detectors at modes $C$ and $C'$ detect no photons, namely, the measurement outcome is $\vt{n} = [n_A, 0, n_{A'}, 0]$ with $n_A \ne 0$ and $ n_{A'} \ne 0$. 
The conditional output state is
\begin{eqnarray}
&& \langle n_A, 0, n_{A'}, 0 \ket{\Psi}_{BAA'CC'} 
\nonumber\\
&=&
\mathcal{N} \mathcal{N}_+ \mathcal{N}_- \frac{e^{-\alpha^2}}{\sqrt{n_A ! \, n_{A'} !}} \alpha^{N} \big[ 1 - (-1)^{N} \big] i^{n_{A'} + 2}
\bigg[ \frac{1}{(\sqrt{2})^{n_{A'}}} \bigg( c_0 \ket{\mathcal{C}^{+}_{\alpha}}_B + i^{N + 1} c_1 \ket{\mathcal{C}^{+}_{i \alpha}}_B, \bigg)
\nonumber\\
&&
+ \frac{ i \, e^{i \pi N/4}}{(\sqrt{2})^{n_{A}}} \bigg( c_0 \ket{\mathcal{C}^{+}_{i \alpha}}_B + i^{N + 1} c_1 \ket{\mathcal{C}^{+}_{\alpha}}_B, \bigg)
\bigg]
\nonumber\\
&=&
\mathcal{N} \mathcal{N}_+ \mathcal{N}_- \big[ 1 - (-1)^N \big] e^{-\alpha^2} \frac{\alpha^N}{\sqrt{n_A ! n_{A'} !}} i^{n_{A'} + 2}
\bigg\{ \bigg[ \frac{1}{(\sqrt{2})^{n_{A'}}} c_0 + \frac{ e^{i N \pi/4}}{(\sqrt{2})^{n_{A}}} i^{N+2} c_1 \bigg]  \ket{\mathcal{C}^{+}_{\alpha}}_B
\nonumber\\
&&
+ \bigg[ \frac{1}{(\sqrt{2})^{n_{A'}}} i^{N+1} c_1 + \frac{ i \, e^{i N \pi/4}}{(\sqrt{2})^{n_{A}}} c_0 \bigg]  \ket{\mathcal{C}^{+}_{i \alpha}}_B
\bigg\}.
\end{eqnarray} 
%and the measurement probability is
%\begin{eqnarray}
%P([n_A, 0, n_{A'}, 0]) = | \langle n_A, 0, n_{A'}, 0 \ket{\Psi}_{BAA'CC'} |^2. 
%\end{eqnarray}
The initial state cannot be recovered.

(iv) $n_{C} = n_{A'} = 0$.

The detectors at modes $C$ and $A'$ detect no photons, namely,
the measurement outcome is $\vt{n} = [n_A, 0, 0, n_{C'}]$ with $n_A \ne 0$ and $ n_{C'} \ne 0$. The conditional output state is
\begin{eqnarray}
&& \langle n_A, 0, 0, n_{C'} \ket{\Psi}_{BAA'CC'} 
\nonumber\\
&=&
\mathcal{N} \mathcal{N}_+ \mathcal{N}_- \frac{e^{-\alpha^2}}{\sqrt{n_A ! \, n_{C'} !}} \alpha^{N} \big[ 1 - (-1)^{N} \big] i^{n_{C'} + 2}
\bigg[ \frac{1}{(\sqrt{2})^{n_{C'}}} \bigg( c_0 \ket{\mathcal{C}^{+}_{\alpha}}_B + i^{N + 1} c_1 \ket{\mathcal{C}^{+}_{i \alpha}}_B, \bigg)
\nonumber\\
&&
- \frac{ i \, e^{- i N \pi/4}}{(\sqrt{2})^{n_{A}}} \bigg( c_0 \ket{\mathcal{C}^{+}_{i \alpha}}_B + i^{N + 1} c_1 \ket{\mathcal{C}^{+}_{\alpha}}_B, \bigg)
\bigg]
\nonumber\\
&=&
\mathcal{N} \mathcal{N}_+ \mathcal{N}_- \frac{e^{-\alpha^2}}{\sqrt{n_A ! \, n_{C'} !}} \alpha^N \big[ 1 - (-1)^N \big] i^{n_{C'} + 2}
\bigg\{ \bigg[ \frac{1}{(\sqrt{2})^{n_{C'}}} c_0 - \frac{ e^{- i N \pi/4}}{(\sqrt{2})^{n_{A}}} i^{N + 2} c_1 \bigg]  \ket{\mathcal{C}^{+}_{\alpha}}_B
\nonumber\\
&&
+ \bigg[ \frac{1}{(\sqrt{2})^{n_{C'}}} i^{N + 1} c_1 - \frac{ i \, e^{- i N \pi/4}}{(\sqrt{2})^{n_{A}}} c_0 \bigg]  \ket{\mathcal{C}^{+}_{i \alpha}}_B
\bigg\}.
\end{eqnarray}
%and the measurement probability is
%\begin{eqnarray}
%P([n_A, 0, 0, n_{C'}]) = | \langle n_A, 0, 0, n_{C'} \ket{\Psi}_{BAA'CC'} |^2. 
%\end{eqnarray}
The initial state cannot be recovered.

(v) $n_{A} = n_{C'} = 0$.

The detectors at modes $A$ and $C'$ detect no photons, namely, 
the measurement outcome is $\vt{n} = [0, n_C, n_{A'}, 0]$ with $n_C \ne 0$ and $ n_{A'} \ne 0$. The conditional output state is
\begin{eqnarray}
&& \langle 0, n_C, n_{A'}, 0 \ket{\Psi}_{BAA'CC'} 
\nonumber\\
&=&
\mathcal{N} \mathcal{N}_+ \mathcal{N}_- \frac{e^{-\alpha^2}}{\sqrt{n_C ! \, n_{A'} !}} \alpha^{N} \big[ 1 - (-1)^{N} \big] (-1)^{n_{A'}}
\bigg[ \frac{1}{(\sqrt{2})^{n_{A'}}} \bigg( c_0 \ket{\mathcal{C}^{+}_{\alpha}}_B + i^{N + 1} c_1 \ket{\mathcal{C}^{+}_{i \alpha}}_B, \bigg)
\nonumber\\
&&
- \frac{ i \, e^{- i N \pi/4}}{(\sqrt{2})^{n_{C}}} \bigg( c_0 \ket{\mathcal{C}^{+}_{i \alpha}}_B + i^{N + 1} c_1 \ket{\mathcal{C}^{+}_{\alpha}}_B, \bigg)
\bigg]
\nonumber\\
&=&
\mathcal{N} \mathcal{N}_+ \mathcal{N}_- \frac{e^{-\alpha^2}}{\sqrt{n_C ! \, n_{A'} !}} \alpha^{N} \big[ 1 - (-1)^{N} \big] (-1)^{n_{A'}}
\bigg\{ \bigg[ \frac{1}{(\sqrt{2})^{n_{A'}}} c_0 - \frac{e^{- i N \pi /4}}{(\sqrt{2})^{n_{C}}} i^{N + 2} c_1 \bigg]  \ket{\mathcal{C}^{+}_{\alpha}}_B
\nonumber\\
&&
+ \bigg[ \frac{1}{(\sqrt{2})^{n_{A'}}} i^{N + 1} c_1 - \frac{ i \, e^{- i N \pi/4}}{(\sqrt{2})^{n_{C}}} c_0 \bigg]  \ket{\mathcal{C}^{+}_{i \alpha}}_B
\bigg\}.
\end{eqnarray}
%and the measurement probability is
%\begin{eqnarray}
%P([0, n_C, n_{A'}, 0]) = | \langle 0, n_C, n_{A'}, 0 \ket{\Psi}_{BAA'CC'} |^2. 
%\end{eqnarray}
The initial state cannot be recovered.

(vi) $n_{A} = n_{A'} = 0$.

The detectors at modes $A$ and $A'$ detect no photons, namely, 
the measurement outcome is $\vt{n} = [0, n_C, 0, n_{C'}]$ with $n_C \ne 0$ and $ n_{C'} \ne 0$. The conditional state is
\begin{eqnarray}
&& \langle 0, n_C, 0, n_{C'} \ket{\Psi}_{BAA'CC'} 
\nonumber\\
&=&
\mathcal{N} \mathcal{N}_+ \mathcal{N}_- \frac{e^{-\alpha^2}}{\sqrt{n_C ! \, n_{C'} !}} \alpha^{N} \big[ 1 - (-1)^{N} \big] 
\bigg[ \frac{1}{(\sqrt{2})^{n_{C'}}} \bigg( c_0 \ket{\mathcal{C}^{+}_{\alpha}}_B + i^{N + 1} c_1 \ket{\mathcal{C}^{+}_{i \alpha}}_B, \bigg)
\nonumber\\
&&
+ \frac{ i \, e^{i N \pi /4}}{(\sqrt{2})^{n_{C}}} \bigg( c_0 \ket{\mathcal{C}^{+}_{i \alpha}}_B + i^{N + 1} c_1 \ket{\mathcal{C}^{+}_{\alpha}}_B, \bigg)
\bigg]
\nonumber\\
&=&
\mathcal{N} \mathcal{N}_+ \mathcal{N}_- \frac{e^{-\alpha^2}}{\sqrt{n_C ! \, n_{C'} !}} \alpha^{N} \big[ 1 - (-1)^{N} \big] 
\bigg\{ \bigg[ \frac{1}{(\sqrt{2})^{n_{C'}}} c_0 + \frac{ e^{i N \pi /4}}{(\sqrt{2})^{n_{C}}} i^{N + 2} c_1 \bigg]  \ket{\mathcal{C}^{+}_{\alpha}}_B
\nonumber\\
&&
+ \bigg[ \frac{1}{(\sqrt{2})^{n_{C'}}} i^{N + 1} c_1 + \frac{ i \, e^{i N \pi/4}}{(\sqrt{2})^{n_{C}}} c_0 \bigg]  \ket{\mathcal{C}^{+}_{i \alpha}}_B
\bigg\}.
\end{eqnarray}
%and the measurement probability is
%\begin{eqnarray}
%P([0, n_C, 0, n_{C'}]) = | \langle 0, n_C, 0, n_{C'} \ket{\Psi}_{BAA'CC'} |^2. 
%\end{eqnarray}
The initial state cannot be recovered.

\subsection{Absence of photons at three detectors}

When only one detector registers photons, the teleportation fails and the initial state cannot be recovered. These click patterns include: 
\begin{eqnarray*}
[n_A, 0, 0, 0], ~~~~~~ [0, n_C,  0, 0], ~~~~~~ [0, 0, n_{A'}, 0], ~~~~~~ [0, 0, 0, n_{C'}]. 
\end{eqnarray*}
We explicitly write down the conditional output states for these click patterns, from which one can see clearly why the error correction fails.

(i) $n_A = N \neq 0$. 

The detector at mode $A$ detects photons, namely, the measurement outcome is $\vt{n} = [N, 0, 0, 0]$. The conditional output state is
\begin{eqnarray}
&& \langle N, 0, 0, 0 \ket{\Psi}_{BAA'CC'} 
\nonumber\\
&=&
\mathcal{N} \mathcal{N}_+ \mathcal{N}_- \big[ 1 - (-1)^N \big] e^{-\alpha^2} \frac{\alpha^N}{\sqrt{N!}}
\bigg[ - \bigg( c_0 \ket{\mathcal{C}^{+}_{\alpha}}_B + i^{N + 1} c_1 \ket{\mathcal{C}^{+}_{i \alpha}}_B \bigg)
+ \frac{i \, e^{- i N \pi/4} - i \, e^{i N \pi/4}} {(\sqrt{2})^N} \bigg( c_0 \ket{\mathcal{C}^{+}_{i \alpha}}_B + i^{N + 1} c_1 \ket{\mathcal{C}^{+}_{\alpha}}_B \bigg)
\bigg]
\nonumber\\
&=&
\mathcal{N} \mathcal{N}_+ \mathcal{N}_- \big[ 1 - (-1)^N \big] e^{-\alpha^2} \frac{\alpha^N}{\sqrt{N!}}
\bigg\{ \bigg[- c_0 + \frac{2 \sin(N \pi/4)}{(\sqrt{2})^N} i^{N + 1} c_1 \bigg] \ket{\mathcal{C}^{+}_{\alpha}}_B
+ \bigg[ - i^{N + 1} c_1 + \frac{2 \sin(N \pi/4)}{(\sqrt{2})^N} c_0 \bigg] \ket{\mathcal{C}^{+}_{i \alpha}}_B
\bigg\}. 
\end{eqnarray}
%The measurement probability is
%\begin{eqnarray}
%P([N, 0, 0, 0]) = | \langle N, 0, 0, 0 \ket{\Psi}_{BAA'CC'} |^2. 
%\end{eqnarray}
The initial state cannot be recovered since $N$ should be odd. 

(ii) $n_C = N \ne 0$. 

The detector at mode $C$ detects photons, namely,
the measurement outcome is $\vt{n} = [0, N, 0, 0]$. The conditional output state is the same as case (i) except a global phase $e^{i \pi}$ and the initial state cannot be recovered. 

(iii) $n_{A'} = N \ne 0$. 

The detector at mode $A'$ detects photons, namely, 
the measurement outcome is $\vt{n} = [0, 0, N, 0]$. The conditional output state is
\begin{eqnarray}
&& \langle 0, 0, N, 0 \ket{\Psi}_{BAA'CC'} 
\nonumber\\
&=&
\mathcal{N} \mathcal{N}_+ \mathcal{N}_- \big[ 1 - (-1)^N \big] e^{-\alpha^2}  \frac{\alpha^N}{\sqrt{N!}} (-i^N)
\bigg[ \frac{1 - i^N}{(\sqrt{2})^N} \bigg( c_0 \ket{\mathcal{C}^{+}_{\alpha}}_B + i^{N + 1} c_1 \ket{\mathcal{C}^{+}_{i \alpha}}_B \bigg)
+ i\, e^{ i N \pi/4} \bigg( c_0 \ket{\mathcal{C}^{+}_{i \alpha}}_B + i^{N + 1} c_1 \ket{\mathcal{C}^{+}_{\alpha}}_B \bigg)
\bigg]
\nonumber\\
&=&
\mathcal{N} \mathcal{N}_+ \mathcal{N}_- \big[ 1 - (-1)^N \big] e^{-\alpha^2}  \frac{\alpha^N}{\sqrt{N!}} (-i^N)
\bigg\{ \bigg[ e^{ i N \pi/4} i^{N + 2} c_1 +  \frac{1 - i^N}{(\sqrt{2})^N} c_0 \bigg]  \ket{\mathcal{C}^{+}_{\alpha}}_B
+ \bigg[ i \,e^{ i N \pi/4} c_0 +  \frac{1 - i^N}{(\sqrt{2})^N} i^{N + 1} c_1 \bigg]  \ket{\mathcal{C}^{+}_{i \alpha}}_B
\bigg\}. 
\nonumber\\
\end{eqnarray}
%The measurement probability is
%\begin{eqnarray}
%P([0, 0, N, 0]) = | \langle 0, 0, N, 0 \ket{\Psi}_{BAA'CC'} |^2. 
%\end{eqnarray}
The initial state cannot be recovered since $N$ should be odd. 

(iv) $n_{C'} = N \ne 0$.

The detector at mode $C'$ detects photons, namely, the measurement outcome is $\vt{n} = [0, 0, 0, N]$. The conditional output state is the same as case (iii) except a global phase $- i^N$ and 
the initial state cannot be recovered.

In summary, when there is single photon loss only in one of the arms of the encoded Bell state, the output states are the same as those when there is single photon loss only 
in the encoded input state, except a possible global phase that is irrelevant. This implies that the proposed error correction procedure not only can mitigate single photon loss in the 
encoded state, but also can correct single photon loss in the entangled Bell state. In the experiment, one does not have to know whether a single photon is lost is the 
encoded state or in the Bell state. Actually, one cannot discriminate these two cases. What one needs to do is to follow the error correction procedure, which then corrects single photon loss in 
both cases.

\section{Correcting single photon loss in imperfect teleportation circuit}

When discussing state teleportation, in Appendix~\ref{sec:GoodTeleportation}, and error correction of single photon loss in the encoded state, in Appendix~\ref{sec:PhotonLoss}, we assume that the teleportation circuit (including the entangled resource state $\ket{\Phi_0}_B$,  four beam splitters, one phase shifter and the PNR detectors) is perfect. In an actual experiment, they are all imperfect and suffer from photon loss. An important question is whether the proposed error correction procedure can also correct single photon loss in the teleportation circuit. There are several possible sources of photon loss in the teleportation circuit. Interestingly, we show in Appendix~\ref{ap:LossBellState} that the single photon loss in one of the arms of the Bell state can be corrected. This implies that the single photon loss in the teleportation circuit can be in principle corrected. A complete analysis is needed to take into account all other sources of photon loss in the teleportation circuit, which we leave for future work. 

\begin{figure}[ht]
  \subfigure[]{
	   \includegraphics[width=0.42\textwidth]{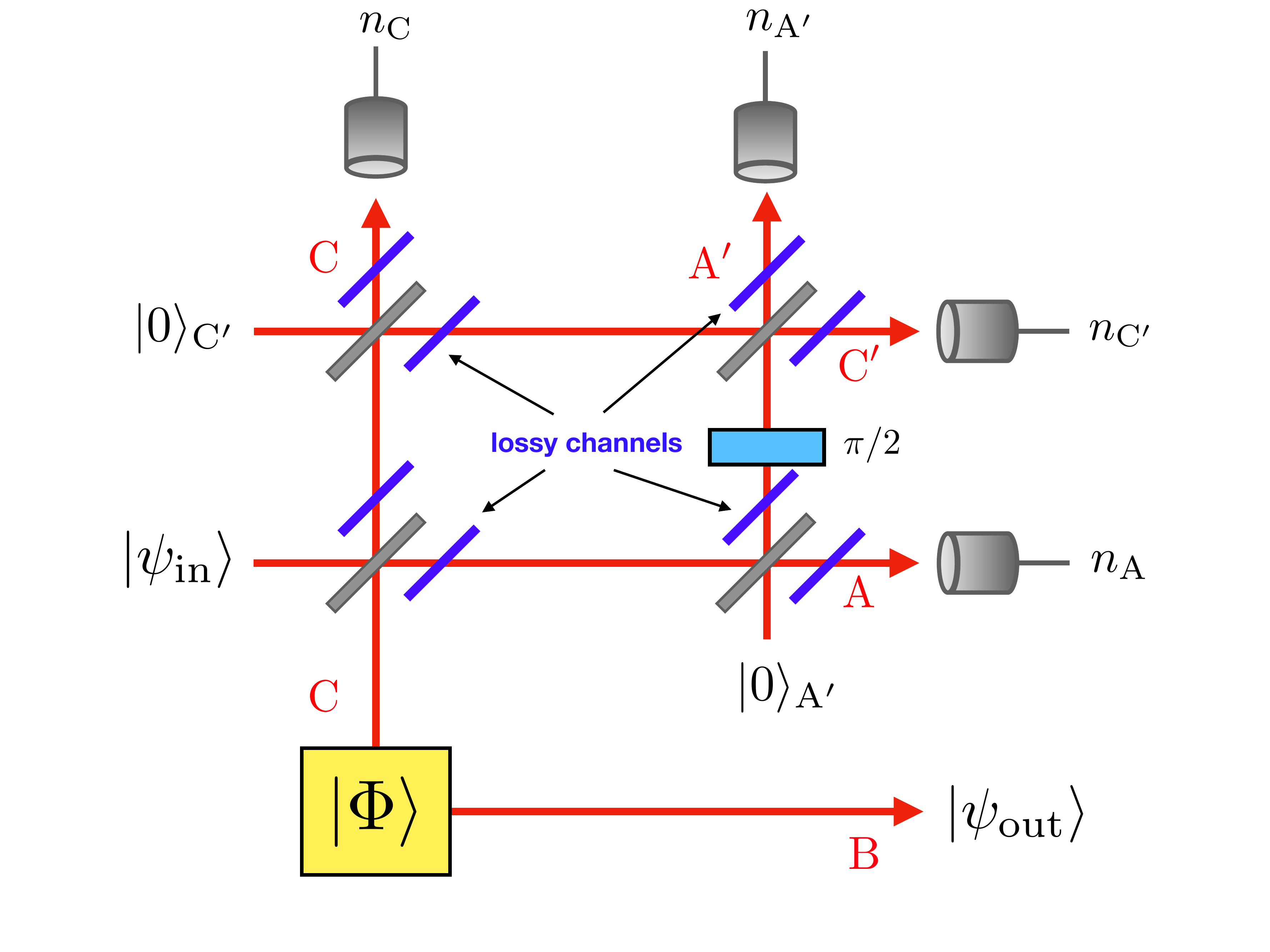}
	}	
  \subfigure[]{
	   \centering
	   \includegraphics[width=0.42\textwidth]{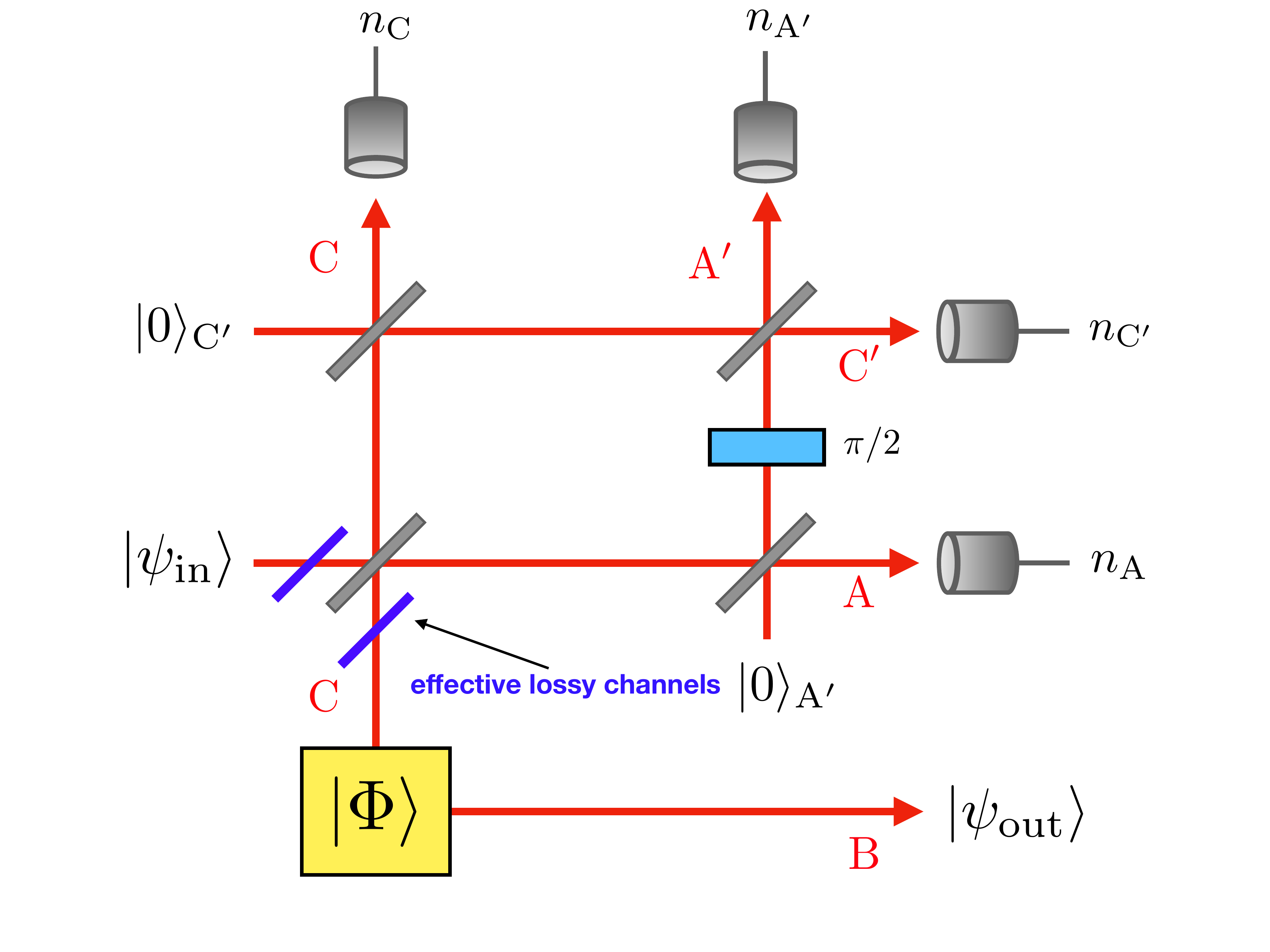}
	}
\caption{Uniform loss model. (a) The imperfection (photon loss) of the teleportation circuit is modeled by various lossy channels (blue beam splitters) after or before the optical components, including the beam splitters, phase shifter and PNR detectors. (b) When the photon loss rate along each path is the same, then four effective lossy channels with the same loss rate can be formed and further moved to the front of the teleportation circuit. The effective lossy channels in the input modes $A'$ and $C'$ are neglected because they have no effect on the vacuum inputs. }
\label{fig:uniform-loss}
\end{figure}

However, based on the results in Appendix~\ref{ap:LossBellState}, it can be simply shown that single photon loss in the beam splitters, phase shifter and photon number detectors can be mitigated in some special cases. Assume that the optical elements and the PNR detectors are not perfect, but the photon loss rate along each path (the modes $A, C, A'$ and $C'$) is the same. Under this balanced-loss assumption, one can combine all lossy channels along each path into one effective lossy channel. Since the photon loss rate of the effective lossy channel in each path is the same, the effective lossy channels commute with the optical elements. One therefore can move them to the front of the teleportation circuit, resulting in a lossy channel in the input mode $A$, which may cause single photon loss in the encoded input state, and another lossy channel in one of the arms of the Bell state, which may cause singe photon loss in the Bell state, as shown in Fig.~\ref{fig:uniform-loss}. As a result, the photon loss from the optical elements and PNR detectors can be mitigated via the proposed error correction procedure.

\section{Generate resource state}

The entangled resource state $\ket{\Phi_0}$ can be generated by the circuit shown in Fig.~\ref{fig:resoure-state} with the state 
\begin{eqnarray}
\ket{+_L}_{\sqrt{2} \alpha} \sim \ket{0_L}_{\sqrt{2} \alpha} + \ket{1_L}_{\sqrt{2} \alpha} \sim \ket{\sqrt{2} \alpha} + \ket{-\sqrt{2} \alpha} + \ket{\sqrt{2} i \alpha} + \ket{- \sqrt{2} i \alpha}
\end{eqnarray}
as an initial resource state. The input of modes $A$ and $C$ is prepared in state $\ket{+_L}_{\sqrt{2} \alpha}$, and the input of modes $A'$ and $C'$ is prepared in vacuum. The output state before photon number detection from this circuit is 
\begin{eqnarray}
&& \ket{\alpha, \alpha, -\beta^*, \beta} + \ket{\alpha, - \alpha, \beta, - \beta^*} + \ket{\alpha, i \alpha, 0, \sqrt{2} i \alpha} + \ket{\alpha, - i \alpha, \sqrt{2} i \alpha, 0} + 
\nonumber\\
&& \ket{- \alpha, \alpha, -\beta, \beta^*} + \ket{- \alpha, - \alpha, \beta^*, - \beta} + \ket{- \alpha, i \alpha, - \sqrt{2} i \alpha, 0} + \ket{- \alpha, - i \alpha, 0, - \sqrt{2} i \alpha} + 
\nonumber\\
&& \ket{i \alpha, \alpha, - \sqrt{2} \alpha, 0} + \ket{i \alpha, - \alpha, 0, - \sqrt{2} \alpha} + \ket{i \alpha, i \alpha, - \beta,- \beta^*} + \ket{i \alpha, - i \alpha, - \beta^*, - \beta} +
\nonumber\\
&& \ket{ - i \alpha, \alpha, 0, \sqrt{2} \alpha} + \ket{- i \alpha, - \alpha, \sqrt{2} \alpha, 0} + \ket{- i \alpha, i \alpha, \beta^*, \beta} + \ket{- i \alpha, - i \alpha, \beta, \beta^*},
\end{eqnarray}
where $\beta = \alpha e^{i \pi/4}$ and the normalization factor $e^{\alpha^2}/\sqrt{8(\cosh 2 \alpha^2 + \cos 2 \alpha^2) }$ has been neglected. 
We now consider the conditional state obtained when both detectors click. This removes all the zero photon components. 
%Performing the final rotation by $-i$ of the lower (final) mode 
When the detectors at modes $A'$ and $C'$ record $n$ and $m$ photon counts respectively, the projected (unnormalized) state is then given by:
\begin{eqnarray}
&& e^{-{{\alpha^2}}} {{\alpha^{n+m}}\over{\sqrt{n! m!}}} \big[ (-1)^n f_{nm}^* \ket{\alpha, \alpha} + (-1)^m f_{nm} \ket{\alpha, - \alpha} 
+ (-1)^n f_{nm} \ket{- \alpha, \alpha} + (-1)^m f_{nm}^* \ket{ - \alpha, - \alpha} 
\nonumber\\
&& + (-1)^{n+m} f_{nm} \ket{ i \alpha, i \alpha} + (-1)^{n+m} f_{nm}^* \ket{ i \alpha, - i \alpha} 
+ f_{nm}^* \ket{- i \alpha, i \alpha} + f_{nm} \ket{ - i \alpha, - i \alpha} \big],
\end{eqnarray}
where $f_{nm} = e^{i \pi (n - m)/4}$. 
We see that if we obtain equal counts on the two detectors, $n=m$, and $n$ is even then the projected state is the teleportation resource state given in Eq.~\eqref{eq:bell}. If $n=m$, but $n$ is odd we obtain the phase flipped Bell state
$\mathcal{N} \big(\ket{0_L} \ket{0_L} - \ket{1_L} \ket{1_L} \big)$, which is still useful for teleportation, just requiring a different correction syndrome.
The probability of obtaining equal counts on the two detectors is 
\begin{eqnarray}
P_{nn} =  \frac{ \cosh^2 \alpha^2 + \cos^2 \alpha^2}{4 \cosh^2 \alpha^2 \, (\cosh 2 \alpha^2 + \cos 2 \alpha^2)} \frac{\alpha^{4n}}{(n!)^2},
\end{eqnarray}
with $n \ge 1$. 

\end{widetext}

\section{Probability of losing different number of photons}

In this section, we briefly review the pure lossy channel and evaluate the probability of losing a particular number of photon. The pure lossy channel can be described by a set of Krauss operators $\{ \op{E}_\ell \}$, so that $\op{\rho}^{\, \prime} = \sum_{\ell = 0}^{\infty} \op{E}_\ell \op{\rho} \op{E}_\ell^\dag$, where $\op{\rho}$ and $\op{\rho}^{\, \prime}$ are the density operators before and after going through the lossy channel, respectively. If we denote the loss rate as $\gamma = \epsilon$, then the Krauss operators can be written as
\begin{eqnarray}
\op{E}_\ell = \bigg( \frac{\gamma}{1 - \gamma} \bigg)^{\ell/2} \frac{\op{a}^\ell}{\sqrt{\ell !}} (1 - \gamma)^{\op{n}/2},
\end{eqnarray}
where $\op{a}$ $(\op{a}^\dag)$ is the annihilation (creation) operator and $\op{n} = \op{a}^\dag \op{a}$ is the photon number operator. 
Define an unnormalized density operator $\op{\bar \rho}_\ell$ as
\begin{eqnarray}
\op{\bar \rho}_\ell = \op{E}_\ell \op{\rho} \op{E}_\ell^\dag, 
\end{eqnarray}
which represents the output density operator after the initial state experiences a quantum jump $ \op{E}_\ell $. The quantum jump $ \op{E}_\ell $ occurs with probability
\begin{eqnarray}
P_\ell(\gamma) = \text{tr} (\op{\bar \rho}_\ell),
\end{eqnarray}
and the normalized output density operator is
\begin{eqnarray}
\op{\rho}_\ell = \frac{\op{\bar \rho}_\ell}{P_\ell(\gamma)} = \frac{\op{\bar \rho}_\ell}{\text{tr} (\op{\bar \rho}_\ell)}. 
\end{eqnarray}
Consider the case where the input state is a pure state $\ket{\psi}$, which can be expanded in the Fock basis as
\begin{eqnarray}
\ket{\psi} = \sum_{n = 0}^{\infty} d_n \ket{n}
\end{eqnarray}
with $d_n$ the complex coefficients. 
The probability $P_\ell (\gamma)$ becomes
\begin{eqnarray}\label{eq:JumpProb}
P_\ell (\gamma) &=& \text{tr} \big( \op{E}_\ell \ket{\psi} \bra{\psi} \op{E}_\ell^\dag \big) 
\nonumber\\
&=&
\sum_{n = \ell}^{\infty} |d_n|^2 \bigg( \frac{\gamma}{1 - \gamma} \bigg)^{\ell} \frac{(1 - \gamma)^{n}}{ \ell ! } \bra{n} (\op{a}^\dag)^\ell \op{a}^\ell \ket{n}
\nonumber\\
&=&
\sum_{n = \ell}^{\infty} |d_n|^2 
\begin{pmatrix} n \\ \ell \end{pmatrix} 
\gamma^\ell (1 - \gamma)^{n - \ell}. 
\end{eqnarray}

In this work, we only consider input states in the code subspace. In the Fock basis, the logical $\ket{0_L}$ and $\ket{1_L}$ can be 
expressed as
\begin{eqnarray}\label{eq:LogicalFockBasis}
\ket{0_L} &=& \ket{\mathcal{C}^{+}_{\alpha}} = \frac{1}{\sqrt{2(1 + e^{-2 \alpha^2})}} \big( \ket{\alpha} + \ket{- \alpha} \big) 
\nonumber\\
&=& 
\frac{1}{\sqrt{\cosh \alpha^2}} \sum_{n = 0}^{\infty} \frac{\alpha^{2n}}{\sqrt{(2n)!}} \ket{2 n}, \nonumber\\
\ket{1_L} &=& \ket{\mathcal{C}^{+}_{i \alpha}} = \frac{1}{\sqrt{2(1 + e^{-2 \alpha^2})}} \big( \ket{i \alpha} + \ket{- i \alpha} \big)
\nonumber\\
&=&
\frac{1}{\sqrt{\cosh \alpha^2}} \sum_{n = 0}^{\infty} (-1)^n \frac{ \alpha^{2n}}{\sqrt{(2n)!}} \ket{2 n}. 
\end{eqnarray}
Therefore an arbitrary state in the code subspace can be expressed as
\begin{eqnarray}
\ket{\psi_{\rm code}} &=& c_0 \ket{0_L} + c_1 \ket{1_L} 
\nonumber\\
&=&
\frac{1}{\sqrt{\cosh \alpha^2}} \sum_{n = 0}^{\infty} \big[ c_0 + (-1)^n c_1 \big] \frac{ \alpha^{2n}}{\sqrt{(2n)!}} \ket{2 n},
\nonumber\\
\end{eqnarray}
which gives
\begin{eqnarray*}
d_n = \left\{ 
\begin{array}{lc} 
0, & n ~ \text{is odd}; \\ 
\frac{1}{\sqrt{\cosh \alpha^2}} \big[ c_0 + (-1)^{n/2} c_1 \big] \frac{ \alpha^{n}}{\sqrt{n !}}, & n ~ \text{is even}. 
\end{array} \right. 
\end{eqnarray*}
Therefore, for a general encoded input state, the quantum jump  $ \op{E}_\ell $ occurs with probability
\begin{eqnarray}\label{eq:JumpProbCode}
P_\ell (\gamma) &=& \frac{1}{\cosh \alpha^2} \sum_{n \ge \ell/2} \big| c_0 + (-1)^n c_1 \big|^2 \frac{ \alpha^{4n}}{(2n)!}
\nonumber\\
&&
\times \begin{pmatrix} 2n \\ \ell \end{pmatrix} 
\gamma^\ell (1 - \gamma)^{2 n - \ell}. 
\end{eqnarray}

\vspace{0.5cm}

\vspace{10 mm}

%\bibliography{ref_4CatCode}

\end{document}